\documentclass[aps,prb,groupedaddress,twocolumn]{revtex4}
\usepackage{epsfig}
\usepackage[dvipsnames,usenames]{color}
\usepackage{color}
\usepackage{amsmath}
\usepackage{comment}
\usepackage{array}
\usepackage{multirow}
\usepackage{tabularx}
\usepackage{float}

\emergencystretch=\maxdimen
\begin{document}
\title{Tunable magnetism of hexagonal Anderson droplet on the triangular lattice}
\author{Mi Jiang}
\affiliation{Stewart Blusson Quantum Matter Institute, University of British Columbia, Vancouver, BC, Canada}

\begin{abstract}
Motivated by recent progress on quantum engineered Kondo lattices, we numerically investigated the local magnetic properties of a hexagonal Anderson droplet consisting of multiple rings of magnetic atoms periodically arrayed on a triangular lattice. 
We demonstrated the tunability of the magnetic properties via their evolution with the droplet geometry for two types of systems with distinct local orbital occupancy profile.
We found that the local susceptibility of the droplet center of some types of droplets can be remarkably enhanced in contrast to the conventionally rapid decrease due to spin correlations of surrounding droplet rings. The tunability of the magnetic properties is attributed to the charge redistribution with varying the droplet geometry enforced by the confined lattice with open boundary. Our simulations complements the exploration of the novel artificial tunability of engineered lattice systems.
\end{abstract}

\maketitle

\section{Introduction}
There has been a recent paradigm shift in the investigation of strongly correlated electronic systems from real materials to the atomic-scale manipulation of artificial lattices and/or superlattices~\cite{Morr2019,Morrpaper1,Morrpaper2,Morrpaper3,Morrpaper4,Morrpaper5,Morrpaper6,Morrpaper7}. 
For example, in the context of the Kondo physics in heavy fermion materials~\cite{Steglich}, the realization of artificial lattices has provided a radically new platform to both explore and manipulate the emergence of strong correlation effects. The resulting many-body phenomena at the nanoscale permit the diverse opportunities for studying the interplay between different degrees of freedom in a controllable manner. In particular, the quantum engineering of nanoscopic Kondo droplets has been demonstrated to be capable to coherently control the droplet's properties such as its Kondo temperature~\cite{Morr2019}. Theoretically, they employed large-$N$ expansion for the treatment of triangular Cu (111) surface lattice, which allows for two types of hexagonal magnetic atom droplets. This study demonstrated the possibility of not only creating coherently coupled Kondo droplets but also modifying the droplet's Kondo temperature via changing a droplet's real space geometry.
As a closely related aspect of the Kondo droplets, the requisite conditions of the coherent Kondo lattice behavior for periodically arranged magnetic moments on a square lattice within the particle-hole symmetric Kondo lattice model (KLM) has been investigated as well recently~\cite{Assaad2019}.

It is well known that another type of models that is believed to qualitatively describe the essential features of the rich physics of heavy fermion systems is the Anderson model, e.g. single-impurity or periodic Anderson model, whose relation to the Kondo models have been extensively explored in the past decades~\cite{SchriefferWolff,mapping}. As effective models in the strong coupling limit, Kondo models describe the $f$-electrons as localized quantum-mechanical spins so that discard the charge degrees of freedom. Hence, given the recent experimental progress of realizing the atomic-scale manipulation of artificial lattices and the theoretical exploration of Kondo droplets~\cite{Morr2019}, it is a good moment to thoroughly investigate the properties of the hexagonal Anderson droplet with the additional involvement of the charge degrees of freedom.

Here we numerically explore the local magnetic properties of a hexagonal Anderson droplet consisting of multiple rings of magnetic atoms on a triangular lattice in the framework of the Anderson model, which has richer physics than its counterpart Kondo model because of the intrinsic charge fluctuations of f-electrons.
Our focus is the evolution of the local properties within the droplet and more importantly their dependence on the spatial structure of the droplet. 

This paper is organized as follows. In Sec.~\ref{model}, we define our Anderson droplet model and the determinant quantum Monte Carlo method employed. Sec.~\ref{results} discusses the local magnetic susceptibilities and closely related density modulations in various droplets. Sec.~\ref{robustness} illustrates more evidence of the dependence of magnetic properties on various parameters. The summary and future issues to be addressed are presented in Sec.~\ref{Conclusion}.

\section{Models and Methods}\label{model}
\begin{figure}
\psfig{figure=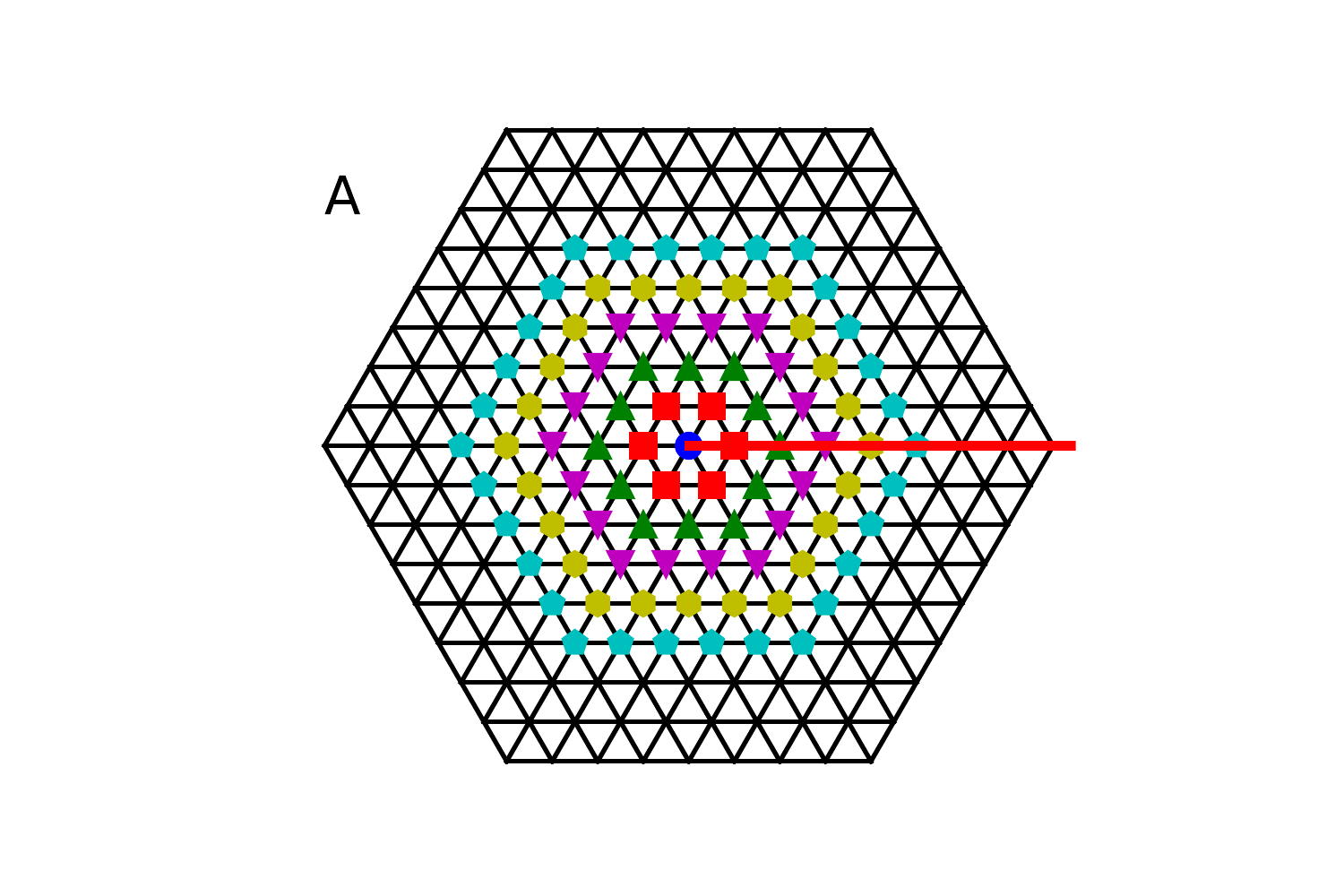,height=3.0cm,width=.23\textwidth, clip=true, trim = 2.5cm 1.2cm 2.5cm 1.2cm} 
\psfig{figure=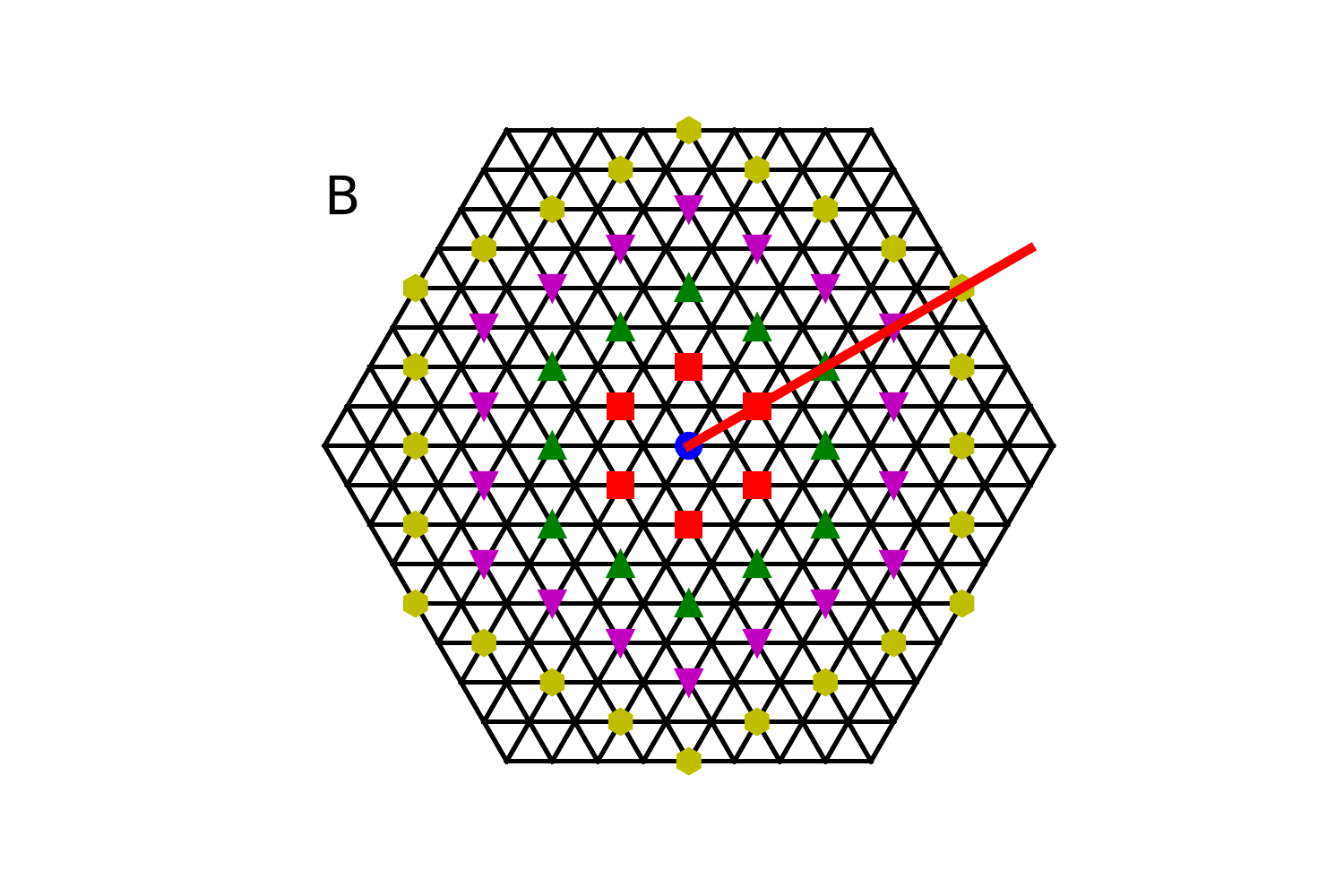,height=3.0cm,width=.23\textwidth, clip=true, trim = 2.5cm 1.2cm 2.5cm 1.2cm} \\
\caption{Lattice structure of two types of Anderson droplet consisting of multiplet rings of Anderson impurites with distance (A) $na_0$ and (B) $\sqrt{3} na_0$ with $a_0\equiv 1$ the lattice constant. $n$ is integer and only $n=1$ case is shown here. The red line denotes the direction adopted for examining the evolution of local properties.}
\label{geom}
\end{figure}

We employ the two-dimensional Anderson droplet model (ADM) in the half-filled form on a triangular lattice 
\begin{eqnarray}
    {\cal H} &=& -t \sum\limits_{\langle ij \rangle,\sigma}
(c^{\dagger}_{i\sigma}c_{j\sigma}^{\vphantom{dagger}}
+c^{\dagger}_{j\sigma}c_{i\sigma}^{\vphantom{dagger}})
        -V \sum\limits_{i\in D, \sigma}
(c^{\dagger}_{i\sigma}f_{i\sigma}^{\vphantom{dagger}}+
f^{\dagger}_{i\sigma}c_{i\sigma}^{\vphantom{dagger}})
\nonumber \\
        &+& U \sum\limits_{i\in D} (n^{f}_{i\uparrow}-\frac{1}{2}) (n^{f}_{i\downarrow}-\frac{1}{2})
- \mu (\sum\limits_{i\sigma} n^{c}_{i\sigma}+ \sum\limits_{i\in D, \sigma} n^{f}_{i\sigma} ) \nonumber
\label{eq:PAM}
\end{eqnarray}
where $c^{\dagger}_{i\sigma}(c_{i\sigma}^{\vphantom{dagger}})$ and $f^{\dagger}_{i\sigma}(f_{i\sigma}^{\vphantom{dagger}})$ are creation(annihilation) operators for the conduction and local $f$-electrons on site $i$ with spin $\sigma$, respectively. $n^{c,f}_{i\sigma}$ are the associated number operators. $t=1$ set to be the energy unit is the hopping amplitude between conduction electrons on nearest-neighbor sites $\langle ij \rangle$ of a triangular lattice. $U$ denotes the local repulsive interaction for $f$-electrons and $V$ is the hybridization between the conduction and $f$-electrons. The chemical potential $\mu$ controls the average density of the system. The set $D$ defines the set of sites where the impurity droplet resides. 
Figure~\ref{geom} illustrates the geometrical structure of our hexagonal lattice with open boundary and two types of droplets, where the distance between the consecutive impurity rings can be $na_0$ and b) $\sqrt{3} na_0$ separately with $a_0 \equiv 1$ the lattice constant and $n$ is integer. This geometry has direct relevance to recent study on the artificially engineered Kondo droplet system~\cite{Morr2019}. From now on, we denote the spatial structure of the droplet by $A/B \{n:N_r\}$ with $n$ the distance between consecutive rings and $N_r$ the number of rings such that Fig.~\ref{geom} shows $A\{1:5\}$ and $B\{1:4\}$ droplets. With this notation, for example, the first ring of $A\{3:2\}$ droplet has the same location as the third ring of $A\{1:4\}$ droplet.

Although the ADM model breaks the translational symmetry of the lattice, it has close relation to the well-known periodic Anderson model (PAM), which is conventionally believed to capture the essential physics of heavy-fermion materials~\cite{Steglich}.
Accordingly, PAM has been extensively explored numerically in various contexts, for example, of the phase diagram~\cite{Kotliar2005,Kotliar2008}, universal Knight shift anomaly~\cite{MJ}, d-wave superconductivity~\cite{TremblayPRX}, Mott metal-insulator transition~\cite{NFL2008,Mott2016,Capone2016} etc. ADM can also be viewed as a special form of depleted PAM~\cite{depletedPAM} and has relevance to the PAM with impurities~\cite{impPAM1,impPAM2}.

The phase diagram of the PAM on triangular lattice has been explored extensively in the past decades, which hosts richer phases than its counterpart on square lattice~\cite{triangularAssaad,triangularMFT}. Following the phase diagram reported in ~\cite{triangularAssaad}, we focus on the characteristic intermediate coupling strength $U/t=4.0$ and $c-f$ hybridization strength $V/t=1.0$ such that they are of the same order of magnitude. More detailed dependence on the parameters such as $U$ and $V$ will be addressed in Sec.~\ref{robustness}. To treat with these energy scales on the equal footing, we solve the ADM by means of the conventional finite temperature determinant Quantum Monte Carlo (DQMC)~\cite{DQMC}. 
Note that our hexagonal shape triangular lattice with open boundary is highly inhomogeneous and non-periodic so that both the conduction and $f$ electron density distributions will affect the local properties throughout the lattice. As discussed below, this inhomogeneity of density fluctuations is believed to be decisive in the determination of the local magnetic properties.
In this work, we have treated with two characteristic types of systems: (1) the whole system is half-filled $\rho= [ \sum \limits_{i\sigma}\langle n^{c}_{i\sigma} \rangle + \sum \limits_{i\in D,\sigma}\langle n^{f}_{i\sigma} \rangle ] /N =1$ with $N$ the total number of lattice sites (including the droplet) by tuning the chemical potential $\mu$ and (2) the droplet impurities are almost half-filled by setting fixed $\mu=0$ for all droplet geometries. We emphasize that these two cases, which is equivalent in the conventional PAM on square lattice, differ in our lattice geometry. In the former system, both the conduction and $f$ electron densities are away from and can exceed half-filling. In contrast, the latter system partially removes the charge fluctuations of $f$-electron whose density is almost half-filled while the conduction electron below half-filling presents mostly the spatial density modulation. 
Because of the geometric frustration, the infamous Fermionic minus sign problem prevents us from the arbitrary choice of parameters for various droplet geometries. Therefore, to study a large enough lattice at low enough temperature with manageable sign problem and computational cost, most results presented are for lattices with boundary length $L=10$ sites such that the total number of sites is $3L^2+3L+1=331$, which has constrains on the maximal possible number of impurity rings for a particular droplet ring distance $n$. 

\section{Main Results}\label{results}
\subsection{Local susceptibility}
\begin{figure*}
\psfig{figure=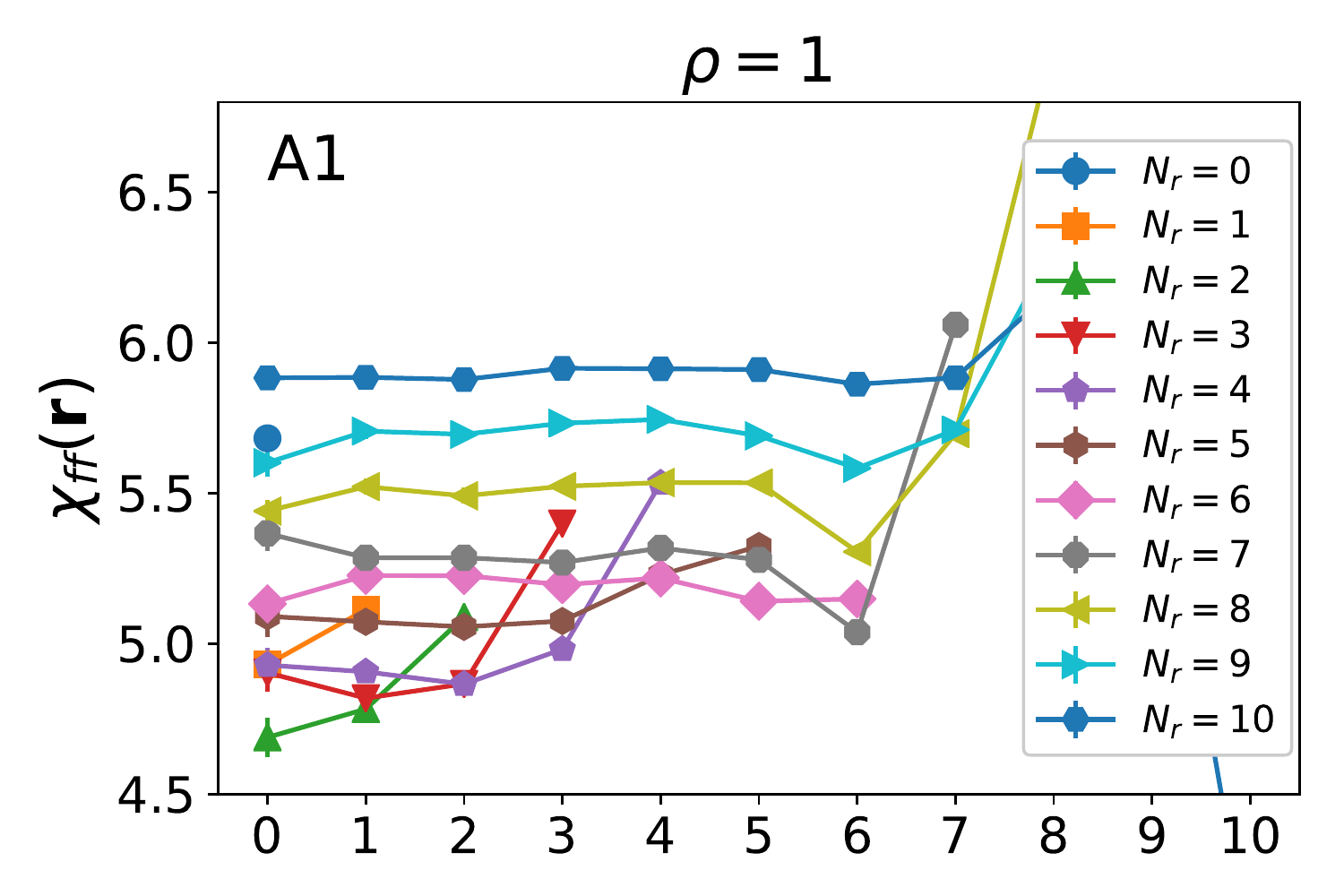}.pdf,height=3.0cm,width=.25\textwidth, clip=true, trim = 0.0cm 0.0cm 0.0cm 0.0cm} 
\psfig{figure=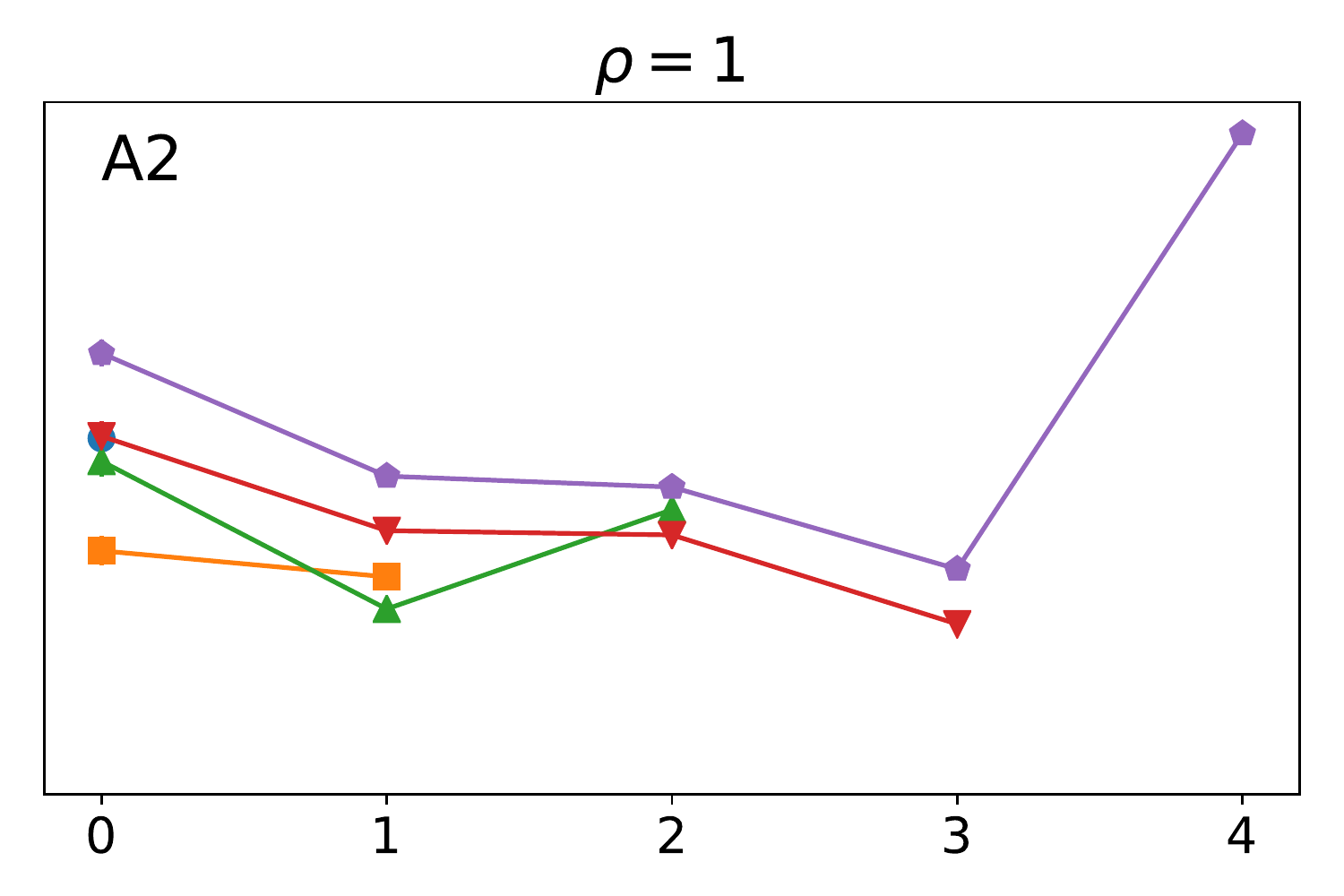}.pdf,height=3.0cm,width=.22\textwidth, clip=true, trim = 0.0cm 0.0cm 0.0cm 0.0cm}
\psfig{figure=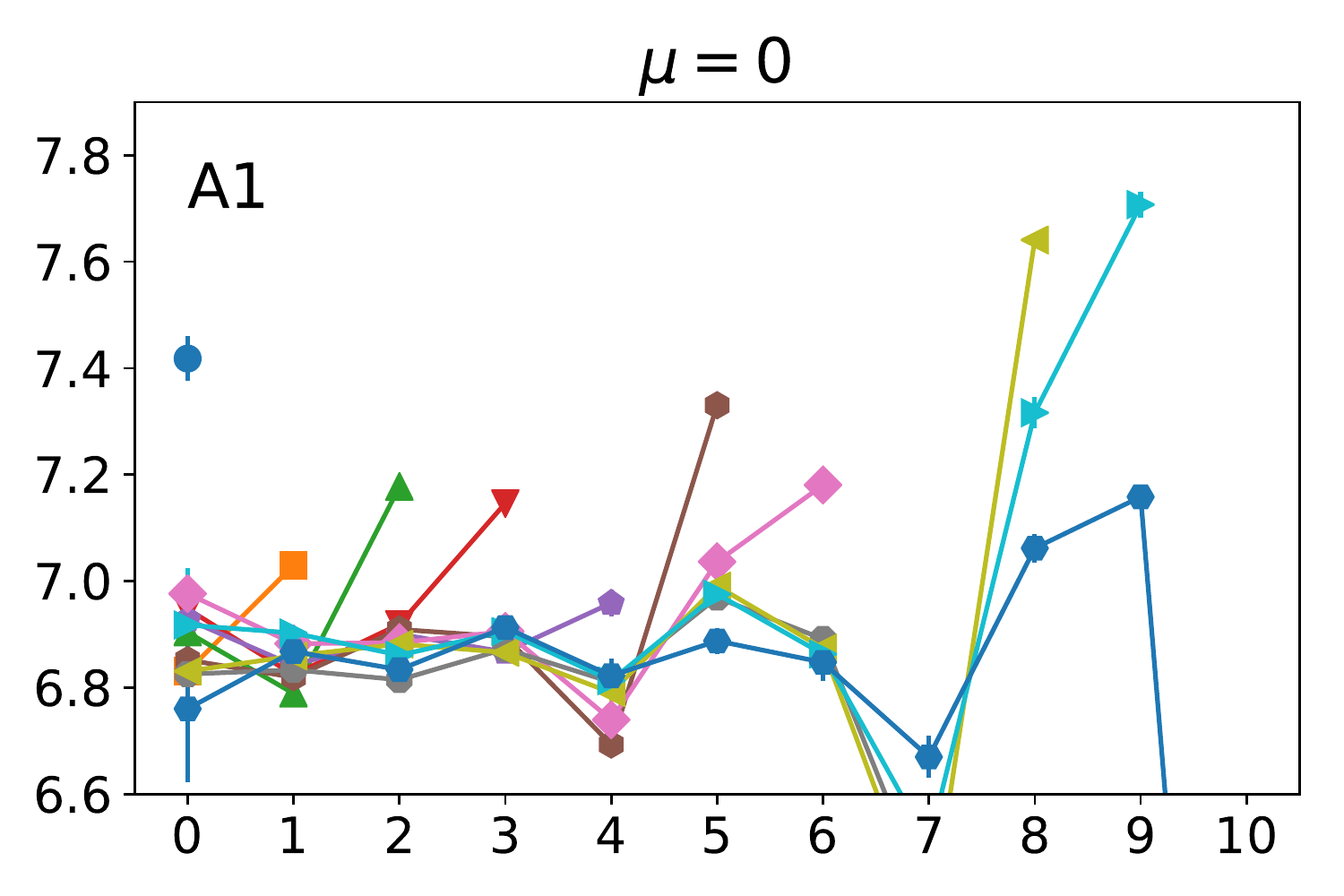}.pdf,height=3.0cm,width=.25\textwidth, clip=true, trim = 0.0cm 0.0cm 0.0cm 0.0cm} 
\psfig{figure=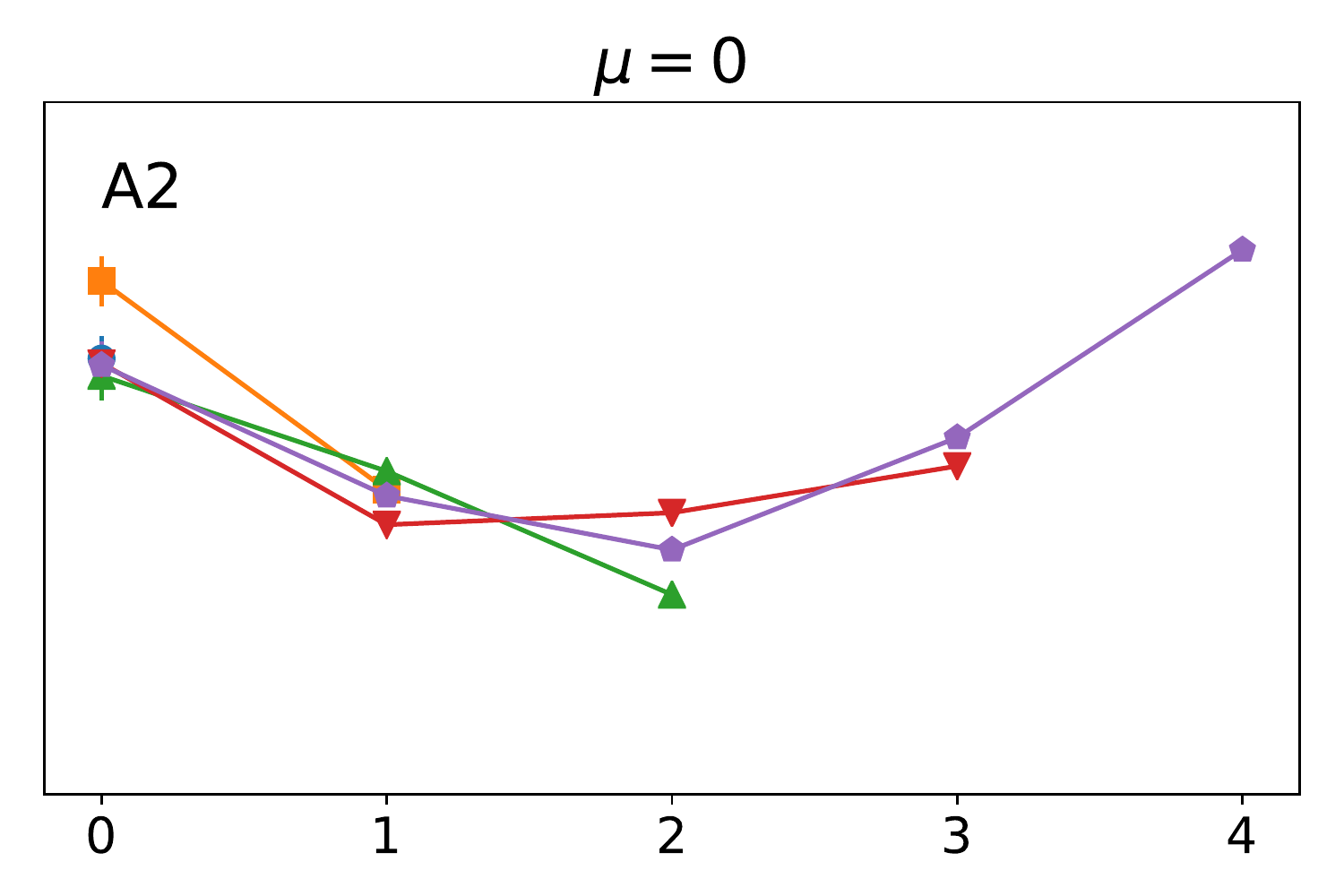}.pdf,height=3.0cm,width=.22\textwidth, clip=true, trim = 0.0cm 0.0cm 0.0cm 0.0cm} \\
\psfig{figure=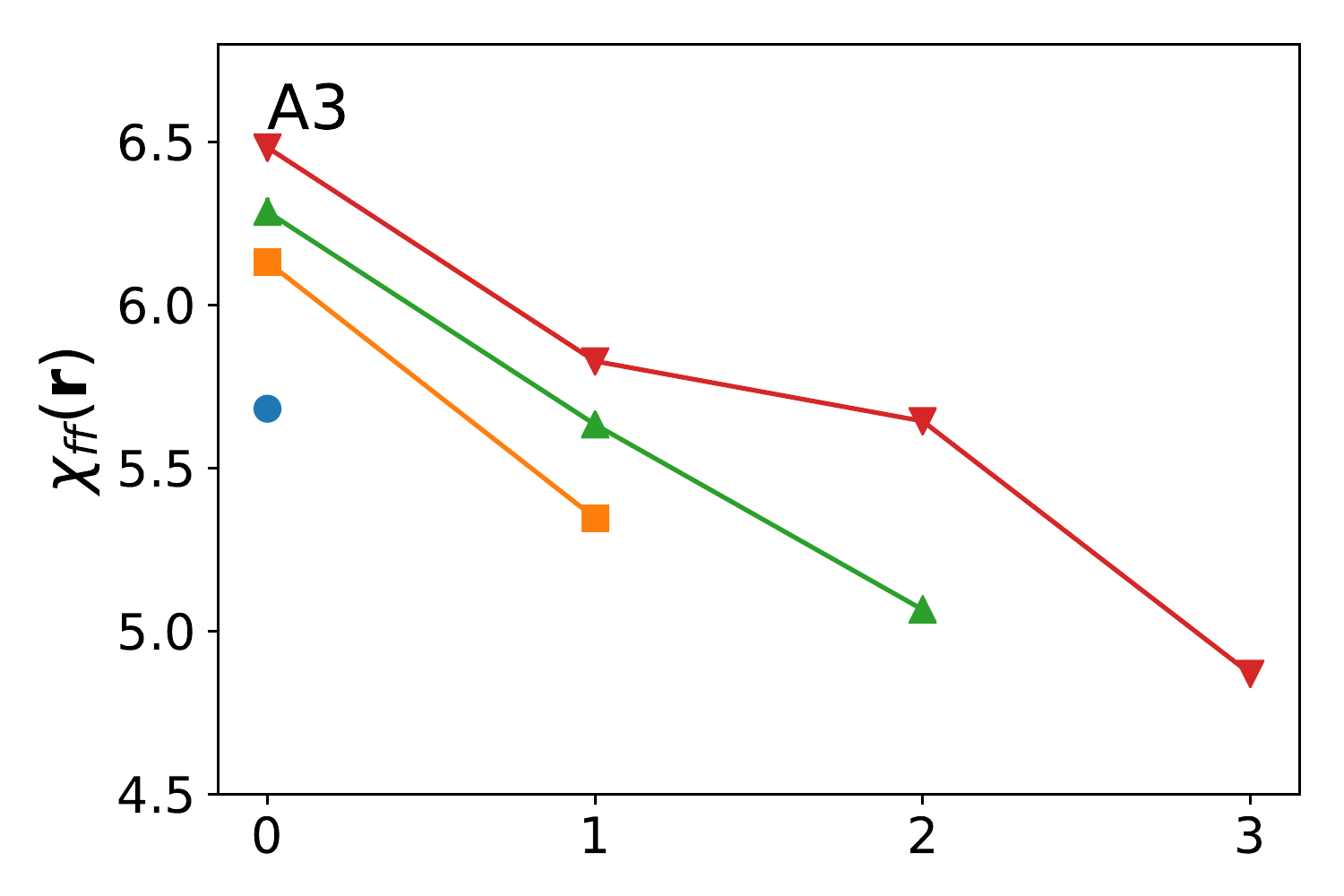}.pdf,height=2.8cm,width=.25\textwidth, clip=true, trim = 0.0cm 0.0cm 0.0cm 0.0cm}
\psfig{figure=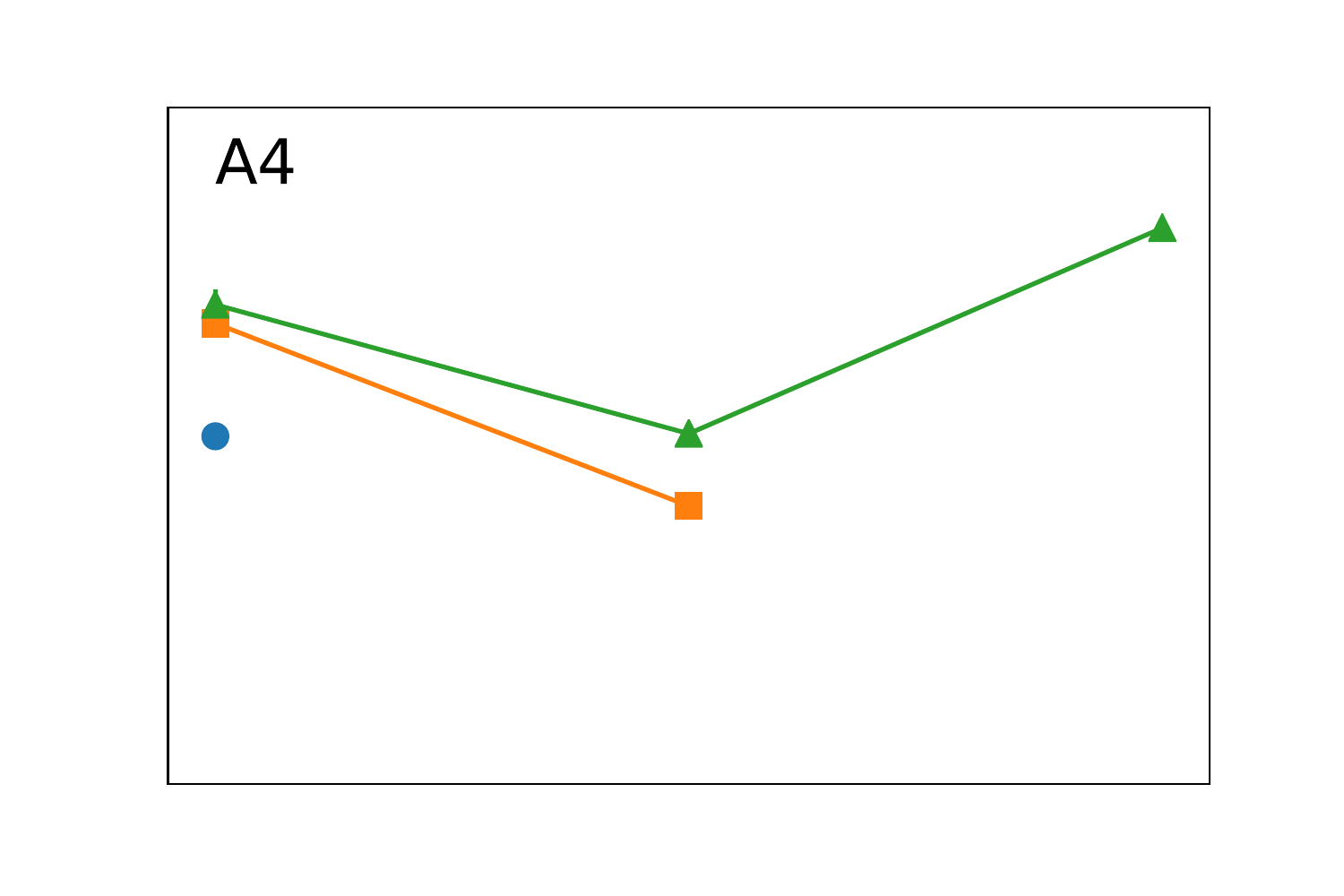}.pdf,height=2.8cm,width=.22\textwidth, clip=true, trim = 0.0cm 0.0cm 0.0cm 0.0cm} 
\psfig{figure=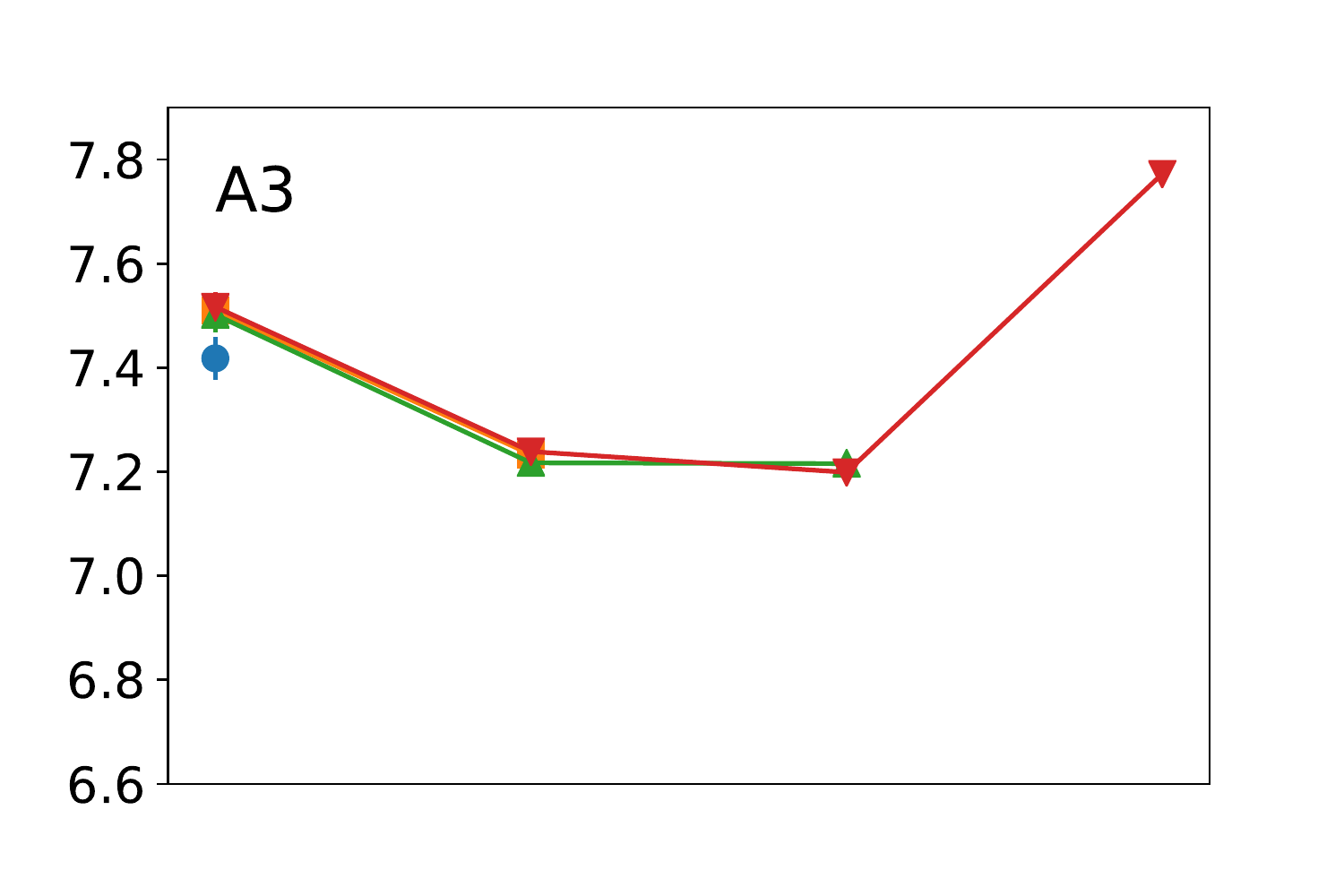}.pdf,height=2.8cm,width=.25\textwidth, clip=true, trim = 0.0cm 0.0cm 0.0cm 0.0cm} 
\psfig{figure=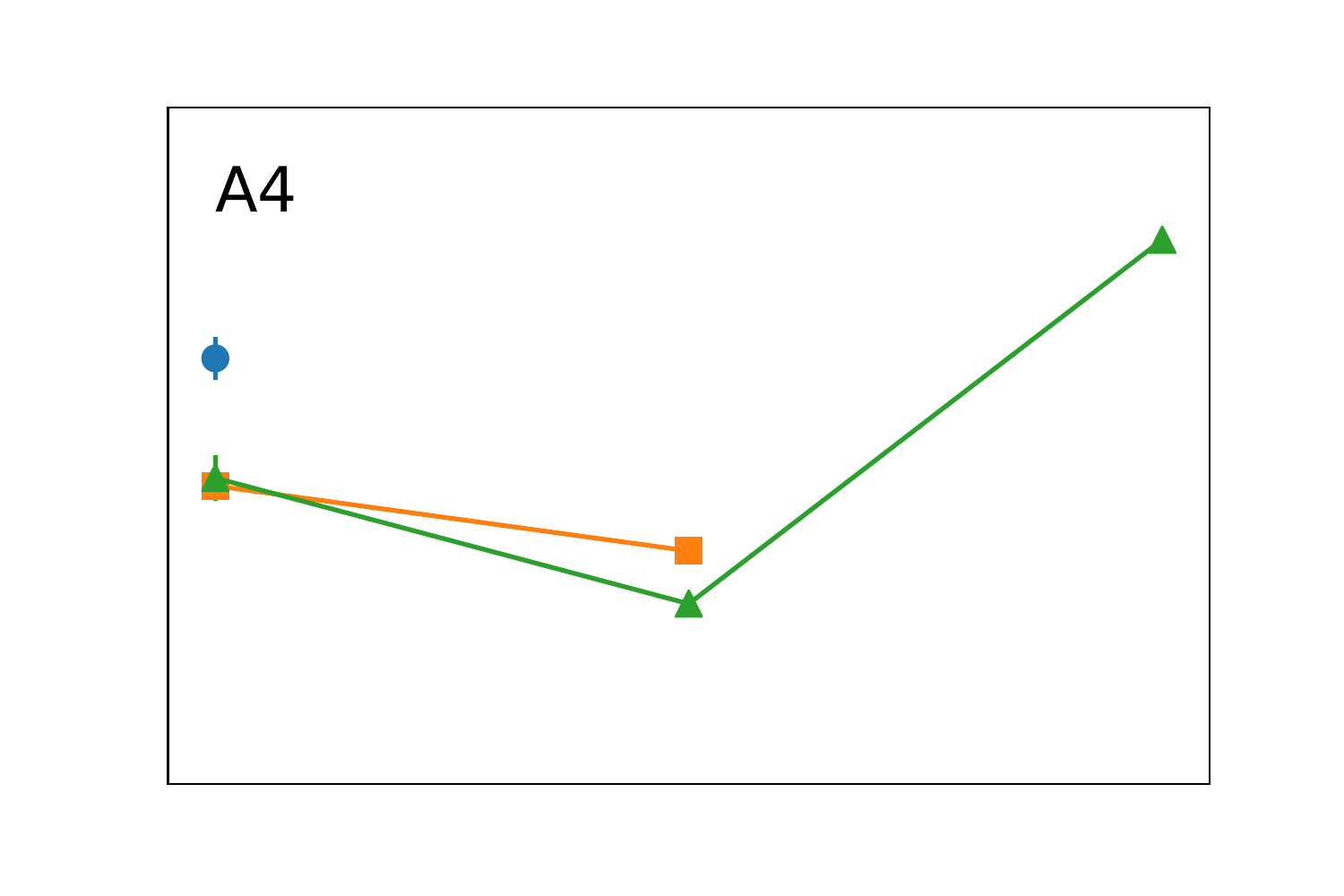}.pdf,height=2.8cm,width=.22\textwidth, clip=true, trim = 0.0cm 0.0cm 0.0cm 0.0cm}  \\
\psfig{figure=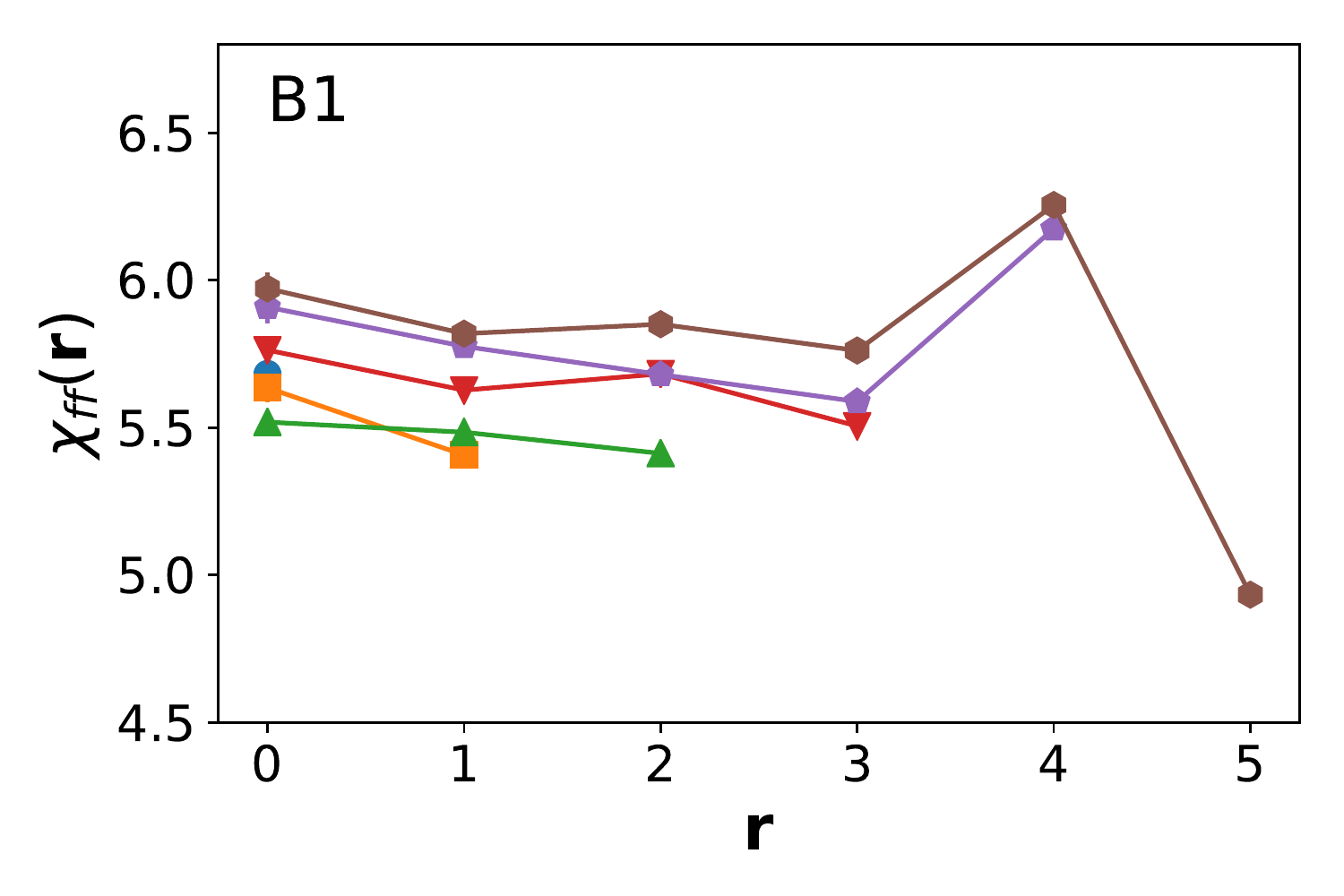}.pdf,height=3.0cm,width=.25\textwidth, clip=true, trim = 0.0cm 0.0cm 0.0cm 0.0cm} 
\psfig{figure=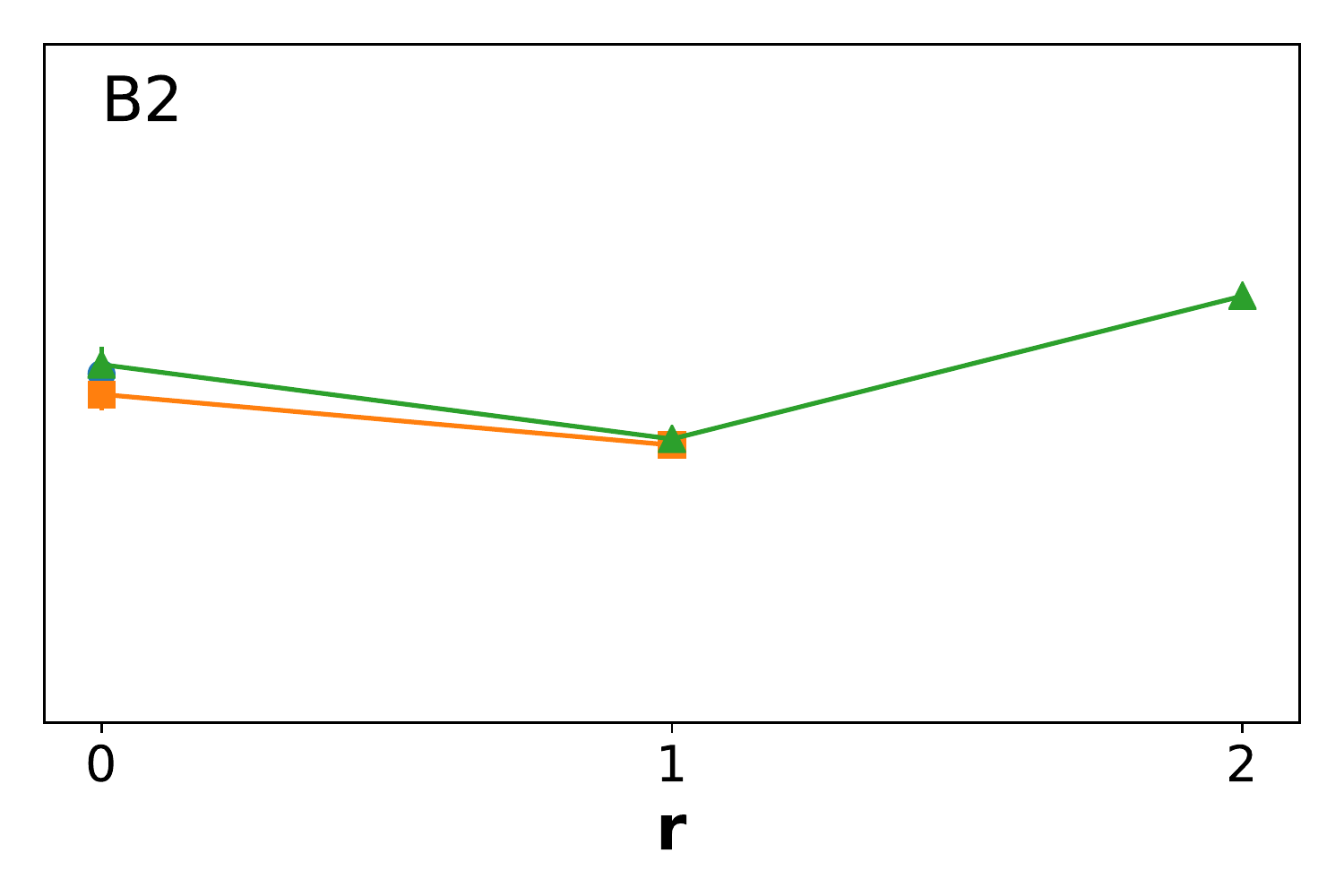}.pdf,height=3.0cm,width=.22\textwidth, clip=true, trim = 0.0cm 0.0cm 0.0cm 0.0cm} 
\psfig{figure=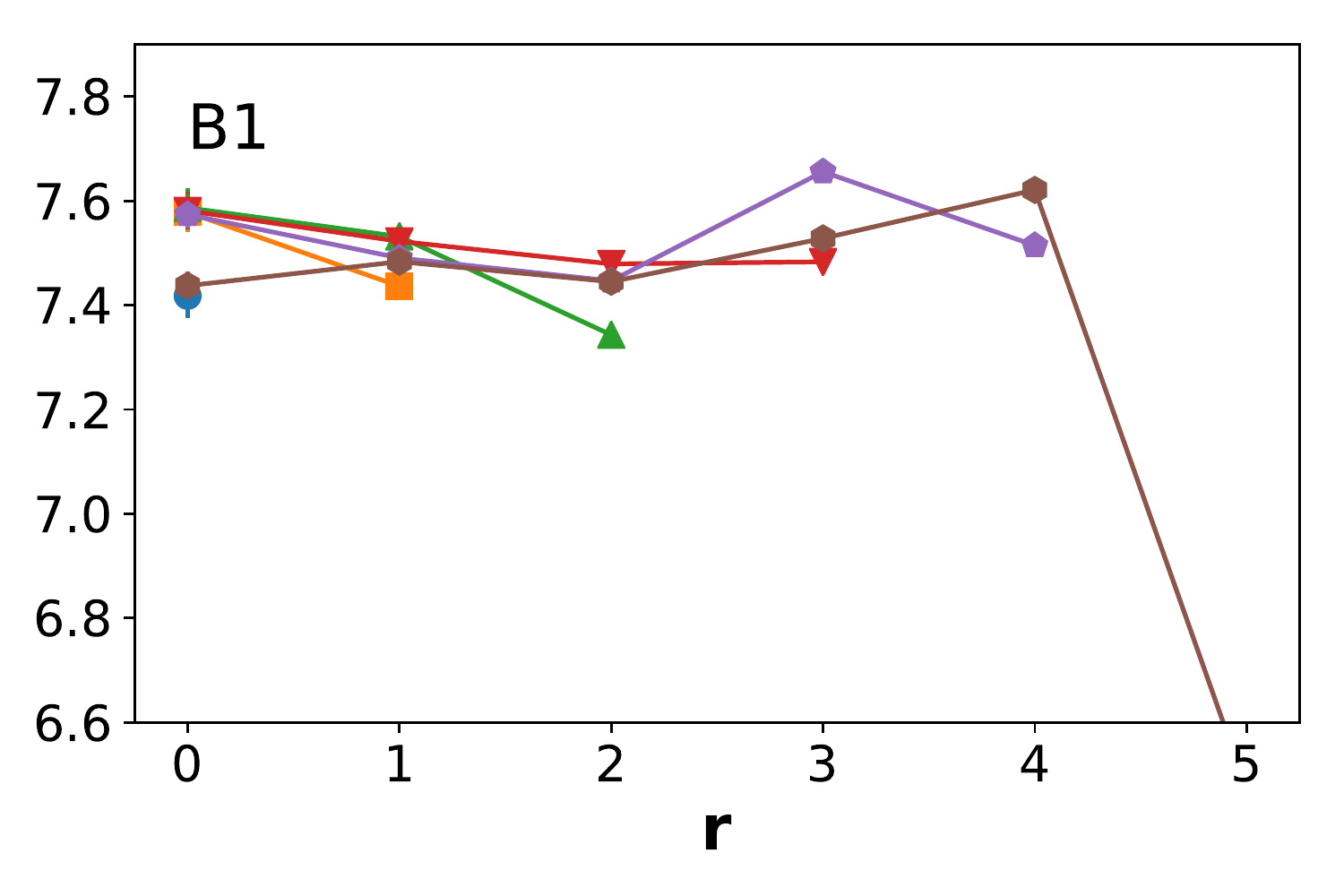}.pdf,height=3.0cm,width=.25\textwidth, clip=true, trim = 0.0cm 0.0cm 0.0cm 0.0cm} 
\psfig{figure=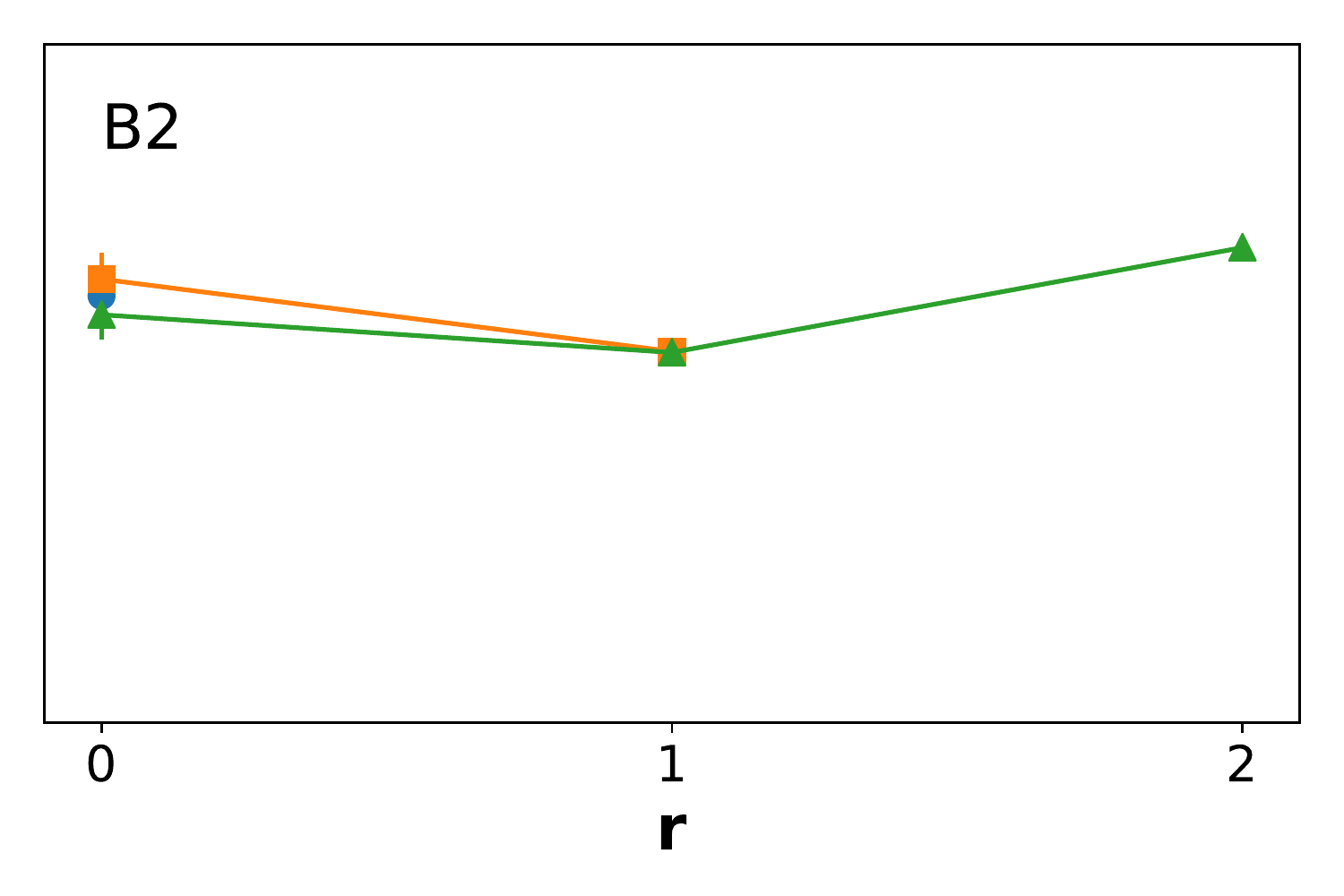}.pdf,height=3.0cm,width=.22\textwidth, clip=true, trim = 0.0cm 0.0cm 0.0cm 0.0cm} 
\caption{Local $f$-electron susceptibility $\chi_{ff}(\mathbf{r})$ for various A and B droplet geometries in both systems of $\rho=1$ and $\mu=0$, where we have simplified the label as $A/B n \equiv A/B \{n:N_r\}$ for various cases of $N_r$. $T=t/10$ is chosen to access large enough droplets (especially for $A1$ droplets).}
\label{chiff_B10}
\end{figure*}

The key quantity throughout the paper is the local magnetic susceptibility defined as
\begin{equation}\label{}
  \chi_{ab}(\mathbf{r}) = \int_{0}^{\beta} d\tau \langle [n^{a}_{\mathbf{r}\uparrow}(\tau)-n^{a}_{\mathbf{r}\downarrow}(\tau)] 
  [n^{b}_{\mathbf{r}\uparrow}(0)-n^{b}_{\mathbf{r}\downarrow}(0)] \rangle
\end{equation}
with $a, b$ denoting the conduction and $f$ electron respectively and $\mathbf{r}$ is the location of the characteristic impurity along the red examination line in Fig.~\ref{geom} on the $\mathbf{r}$-th impurity ring so that the central impurity is at $\mathbf{r}=0$. Note that this red examination line passes the lattice corner and the middle of the lattice boundary for A and B type of droplets respectively. We mention that the reduced number of nearest impurities so that differing spin correlation via RKKY interaction and also different neighboring conduction electron sea so that the Kondo screening will largely affect the the properties near the lattice boundary~\cite{Assaad2019}. In some cases, the following results will cover the data near or at the boundary although that is not our focus apparently.

Figure~\ref{chiff_B10} illustrates the evolution of the local $f$-electron susceptibility $\chi_{ff}(\mathbf{r})$ versus $\mathbf{r}$ for various A and B droplet geometries in both systems of $\rho=1$ and $\mu=0$, where we have simplified the label as $A/B n \equiv A/B \{n:N_r\}$ for various cases of $N_r$. To access the large enough droplets (especially for $A1$ droplet), the temperature is chosen to be $T=t/10$, which is low enough to identify the essential properties presented here. The major features persist at a lower temperature $T=t/20$ with a smaller largest droplet size ($N_r=5$ is practically the largest accessible $A1$ droplet due to the sign problem).

Firstly, $\chi_{ff}(\mathbf{r})$ in both systems of $\rho=1$ and $\mu=0$ oscillates with increasing $\mathbf{r}$ for $A/B \{n:N_r>2\}$ droplets (most clearly for $A1$ droplet) before reaching the outermost ring, where this `regular' oscillation breaks down due to the lattice boundary effect mentioned before or a similar effect occurred at the droplet boundary~\cite{Assaad2019}. Neglecting the droplets, e.g. $A\{1:9\}$ and $A\{1:10\}$, whose outermost ring approaches to or locates at the lattice boundary, we note that $A1$'s outermost ring has an upturn of $\chi_{ff}$ while other droplets, e.g. $A\{2:3\}$ ($\rho=1$ system) and $A\{3:2\}$ ($\mu=0$ system), can have the opposite trends. This complication stems from the density fluctuation (dominantly at the spatial region of the droplet boundary) of the system, which intertwines with the lattice and/or droplet boundary effects. 
In fact, as discussed in detail in Section III.B, the local density fluctuation, which mostly comes from the conduction electrons especially in $\mu=0$ systems anti-correlates with the oscillations of $\chi_{ff}(\mathbf{r})$, which signifies the important role played by the charge degrees of freedom imposed by the inhomegeneous lattice.

Secondly, the oscillation of $\chi_{ff}(\mathbf{r})$ gradually diminishes with increasing $N_r$, namely the droplet size, as most clearly shown for $A1$ droplet of $\rho=1$ system. In other words, the droplet's central region becomes more and more homogeneous and coherent~\cite{comment}. For other droplets with larger distance $n$ between consecutive rings, we are limited by the lattice size to identify the diminishment of $\chi_{ff}(\mathbf{r})$.

Thirdly, the weaker dependence of $\chi_{ff}(\mathbf{r})$ on $N_r$ for $B$-type droplets compared with their $A$-type counterparts signifies the impact of the differing geometric arrangement of impurities, in particular, the distance between consecutive rings and the absence of nearest-neighbor impurities in $B$-droplets. In the sense of the minimal hopping distance between the consecutive rings, $B1$ ($B2$) droplet is more similar to $A2$ ($A4$) droplet.

The major difference between systems of $\rho=1$ and $\mu=0$ lies that $\chi_{ff}(\mathbf{r})$ in the latter system gradually saturates with $N_r$ while in the former system it remains growing even for large $N_r$. This originates from the stronger charge effect in the globally half-filled system $\rho=1$. Taking $A1$ droplet for example, the site-dependent density keeps decreasing and approaches to half-filling upon increasing $N_r$ so that $\chi_{ff}(\mathbf{r})$ keeps growing in the former system; while the density saturates for large enough $N_r$ droplet in $\mu=0$ systems (See Sec. III.B). The strong charge effects can be suppressed to some extent by lowering the temperature, which has been verified at $T=t/20$ despite of the limitation of accessing smaller droplet size. At this point, we emphasize that the `full' suppression of charge effects requires pushing to much lower temperature, which is difficult, if not possible, in our DQMC simulations. Kondo lattice model is more appropriate in this regard. All in all, the rich behavior of $\chi_{ff}(\mathbf{r})$ that is dependent on the droplet geometry indicates the possibility of artificial manipulation of the magnetic properties in a controllable manner.

\begin{figure}
\psfig{figure=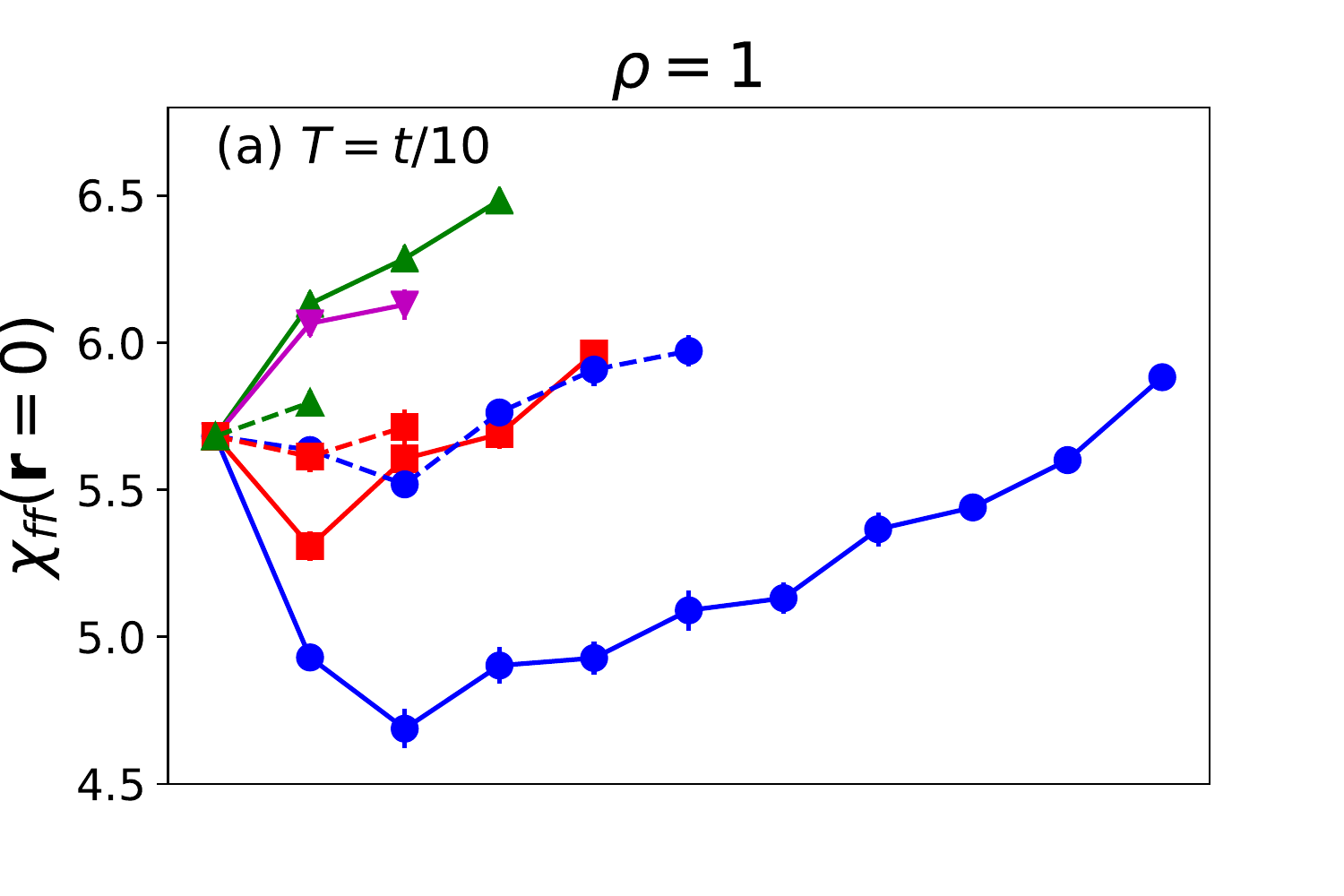}.pdf,height=2.5cm,width=.23\textwidth, clip=true, trim = 0.0cm 1.0cm 1.0cm 0.5cm} 
\psfig{figure=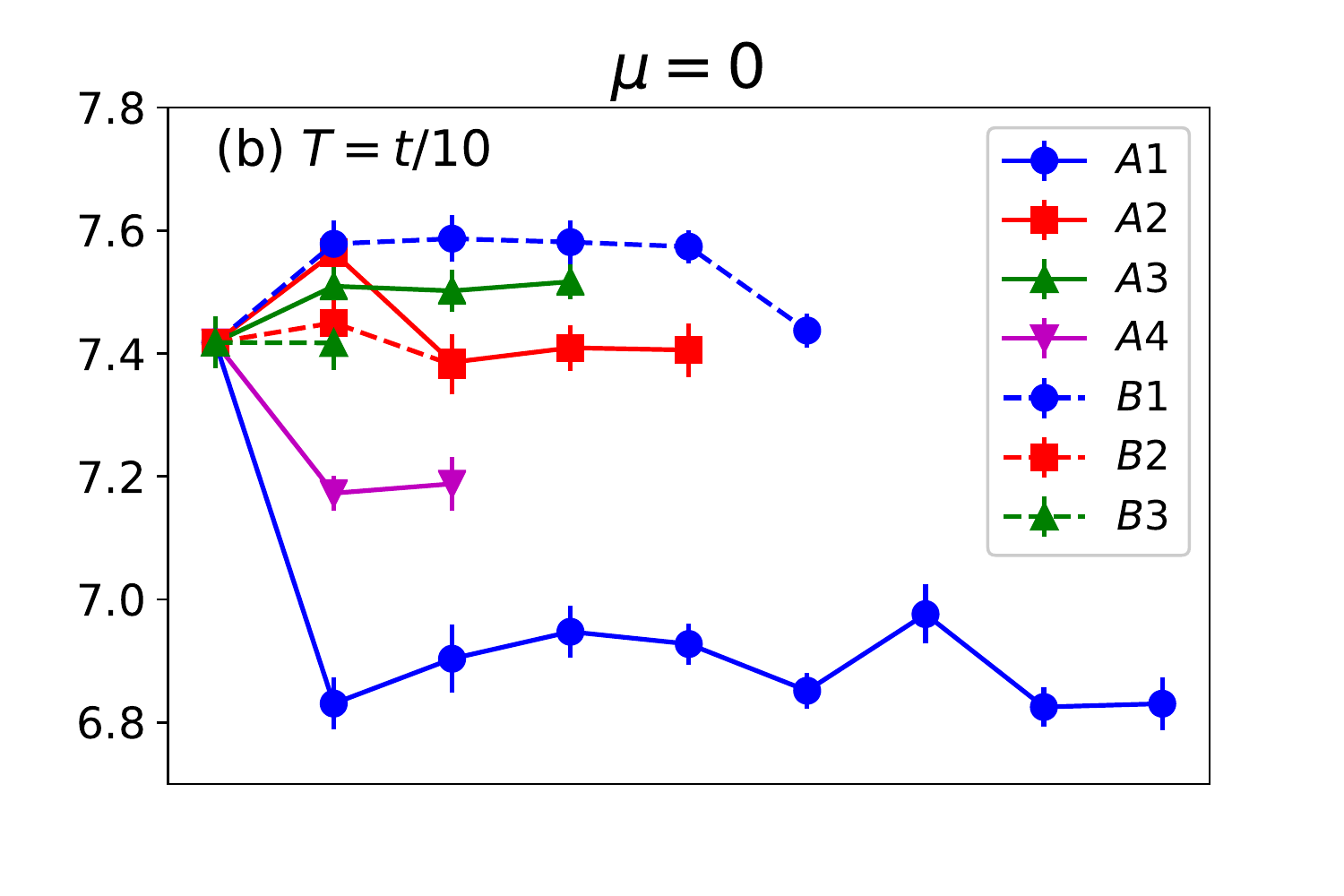}.pdf,height=2.5cm,width=.23\textwidth, clip=true, trim = 0.9cm 1.0cm 1.0cm 0.5cm} \\
\psfig{figure=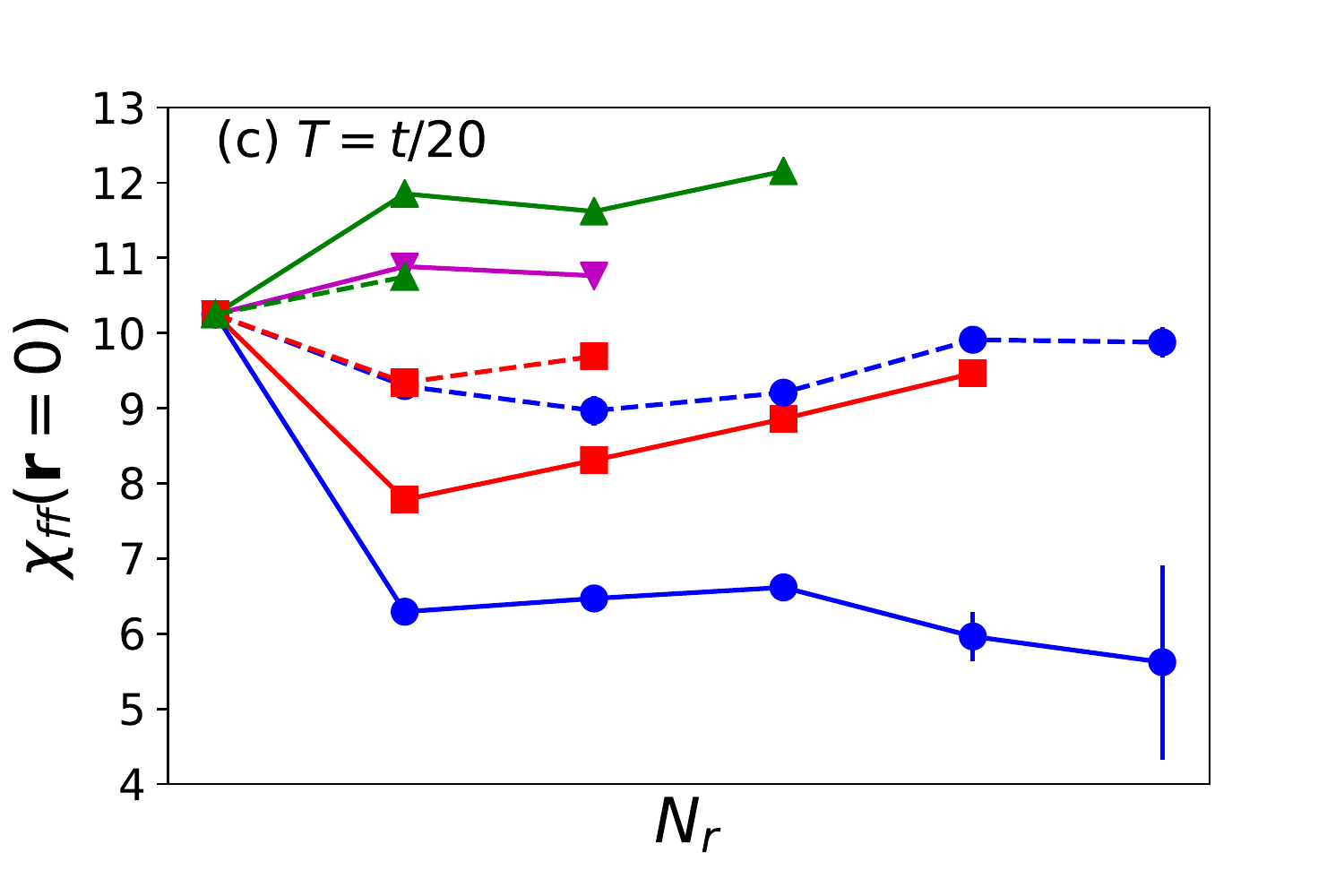}.pdf,height=2.7cm,width=.23\textwidth, clip=true, trim = 0.0cm 0.0cm 1.0cm 0.9cm} 
\psfig{figure=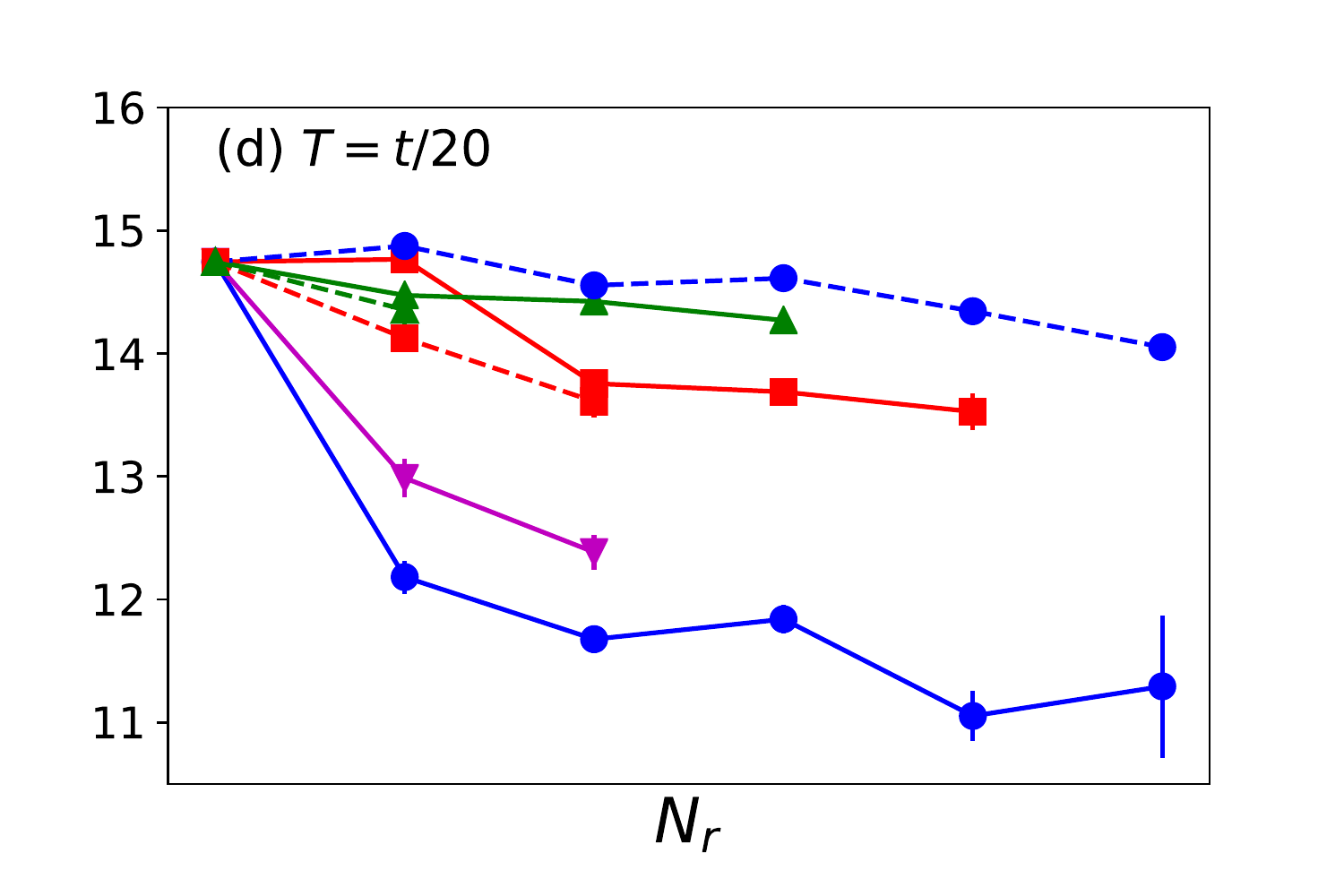}.pdf,height=2.7cm,width=.23\textwidth, clip=true, trim = 0.9cm 0.0cm 1.0cm 0.9cm} 
\caption{Comparison between $\chi_{ff}(\mathbf{r}=0)$ at the central impurity as a function of the number of impurity rings $N_r$ for various A and B droplets.}
\label{center_chiff}
\end{figure}

One fascinating feature of $\chi_{ff}(\mathbf{r})$ is the strong dependence of $\chi_{ff}(\mathbf{r}=0)$ at the central impurity on the droplet geometry. Figure~\ref{center_chiff} illustrates its evolution upon increasing $N_r$ for various types of droplets. 
$\chi_{ff}(\mathbf{r}=0)$ reflects the competition between inter-impurity antiferromagnetic spin correlation with the surrounding impurity rings mediated via RKKY interaction and Kondo screening of the conduction electrons. Moreover, the charge fluctuations in our systems interplay with these two factors to complicate the whole picture.
Fig.~\ref{center_chiff}(a-b) compare the behavior of $\chi_{ff}(\mathbf{r}=0)$ between two types of systems at $T=t/10$ and (c-d) present the comparison at lower temperature $T=t/20$.
For $A$-type droplet with smallest distance between rings e.g. $n=1$, $\chi_{ff}(\mathbf{r}=0)$ decreases rapidly from its single-impurity value and then (a) keeps growing in $\rho=1$ systems while (b) gradually saturates after an oscillating behavior in $\mu=0$ systems with increasing $N_r$. As discussed before, this difference stems from the stronger charge effects in $\rho=1$ systems, which can be partially suppressed at lower temperature $T=t/20$ shown in (c). This saturation similar to that reported for the KLM on square lattice~\cite{Assaad2019} can be attributed to the buildup of spin correlations induced from the neighboring droplet rings, which gives rise to the collective-like screening of the central impurity.
As pointed out in Fig.~\ref{chiff_B10}, $\chi_{ff}(\mathbf{r}=0)$ in $B$-type droplets has more moderate dependence on $N_r$. The moderate deviation from the single impurity case implies the compensation between local Kondo, inter-impurity spin correlation, and charge fluctuation effects. Unfortunately, it is impossible to access much larger lattice and/or lower temperature to identify the saturation of all droplet geometries, especially the $B$-type with larger distance between rings.

The most unusual feature in Fig.~\ref{center_chiff} due to the interplay between various intertwined effects manifests when increasing the distance $n$ between impurity rings, where the fate of the decrease of $\chi_{ff}(\mathbf{r}=0)$ upon $N_r$ is significantly modified. In fact, $\chi_{ff}(\mathbf{r}=0)$ can even be enhanced in $A3, A4, B3$ droplets in $\rho=1$ and $A2, B1$ droplets in $\mu=0$ systems (a-b), which persists at lower temperature $T=t/20$ (c-d). As verified via the density profile in Sec. III.B, this remarkable enhancement implies the significant charge redistribution due to the lattice inhomogeneity, which in turn results in the tunable magnetic susceptibility for a particular impurity embedded in a system. 

\begin{figure}
\psfig{figure=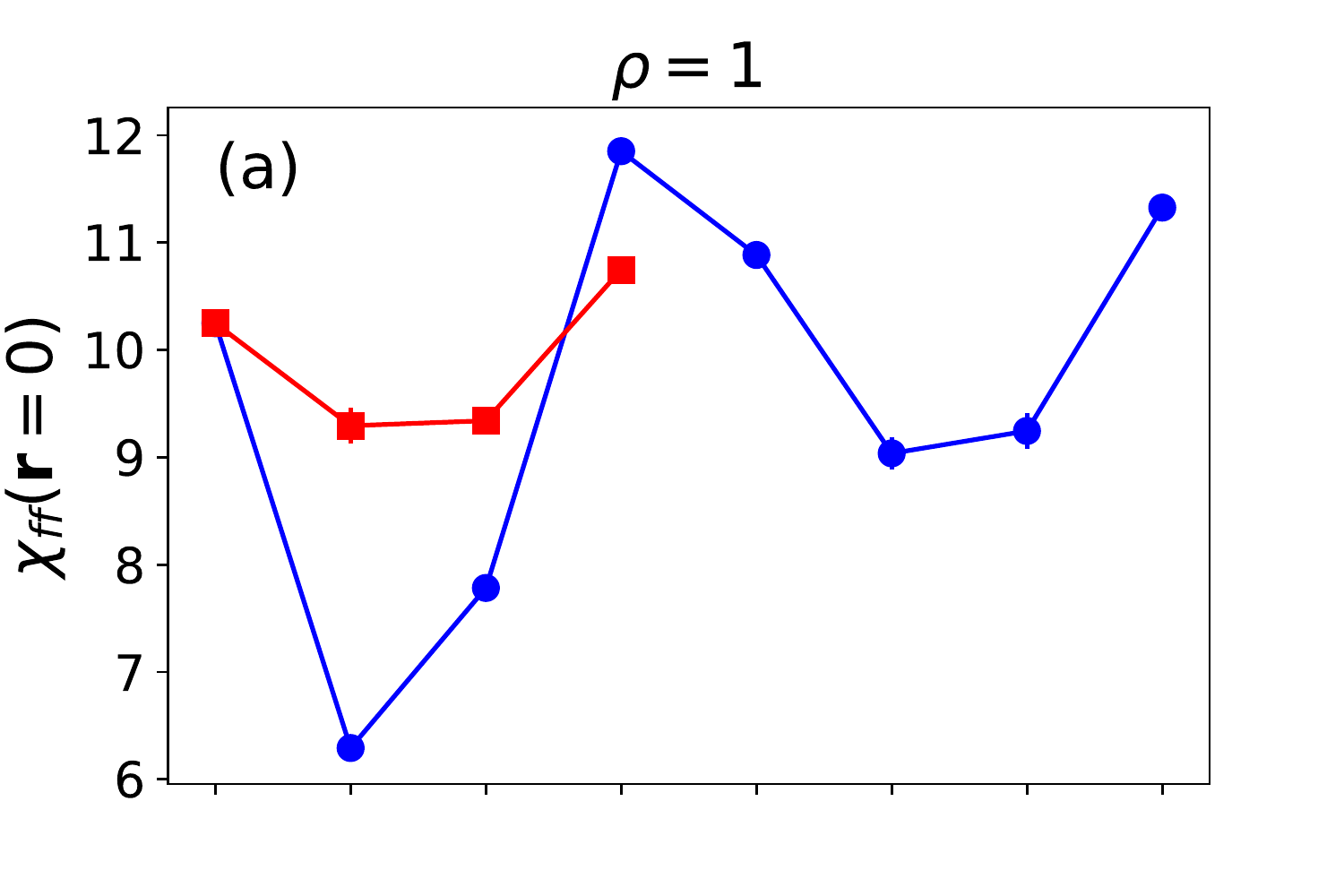}.pdf,height=2.5cm,width=.23\textwidth, clip=true, trim = 0.0cm 1.0cm 1.2cm 0.3cm}
\psfig{figure=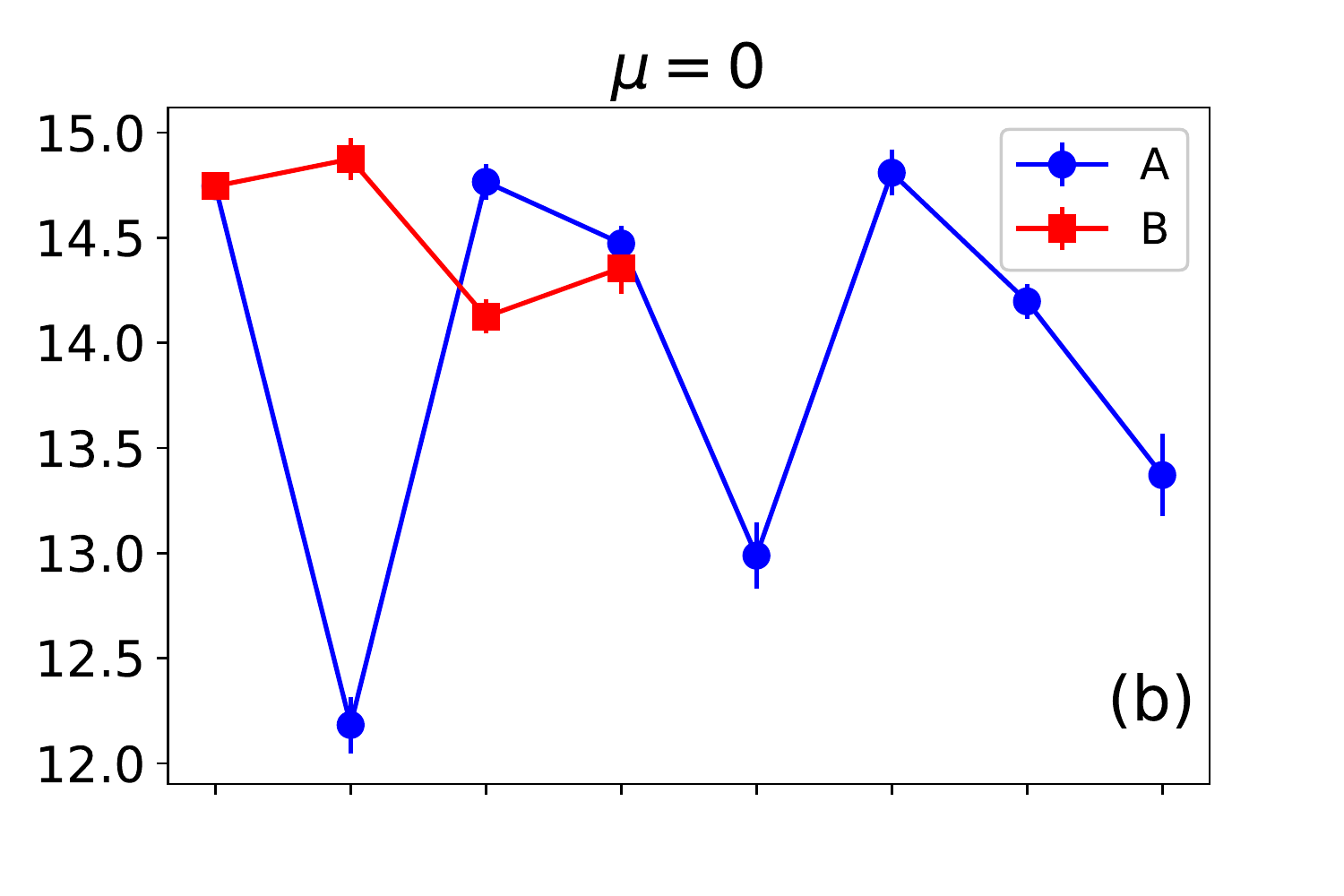}.pdf,height=2.5cm,width=.23\textwidth, clip=true, trim = 0.0cm 1.0cm 1.2cm 0.5cm} \\
\psfig{figure=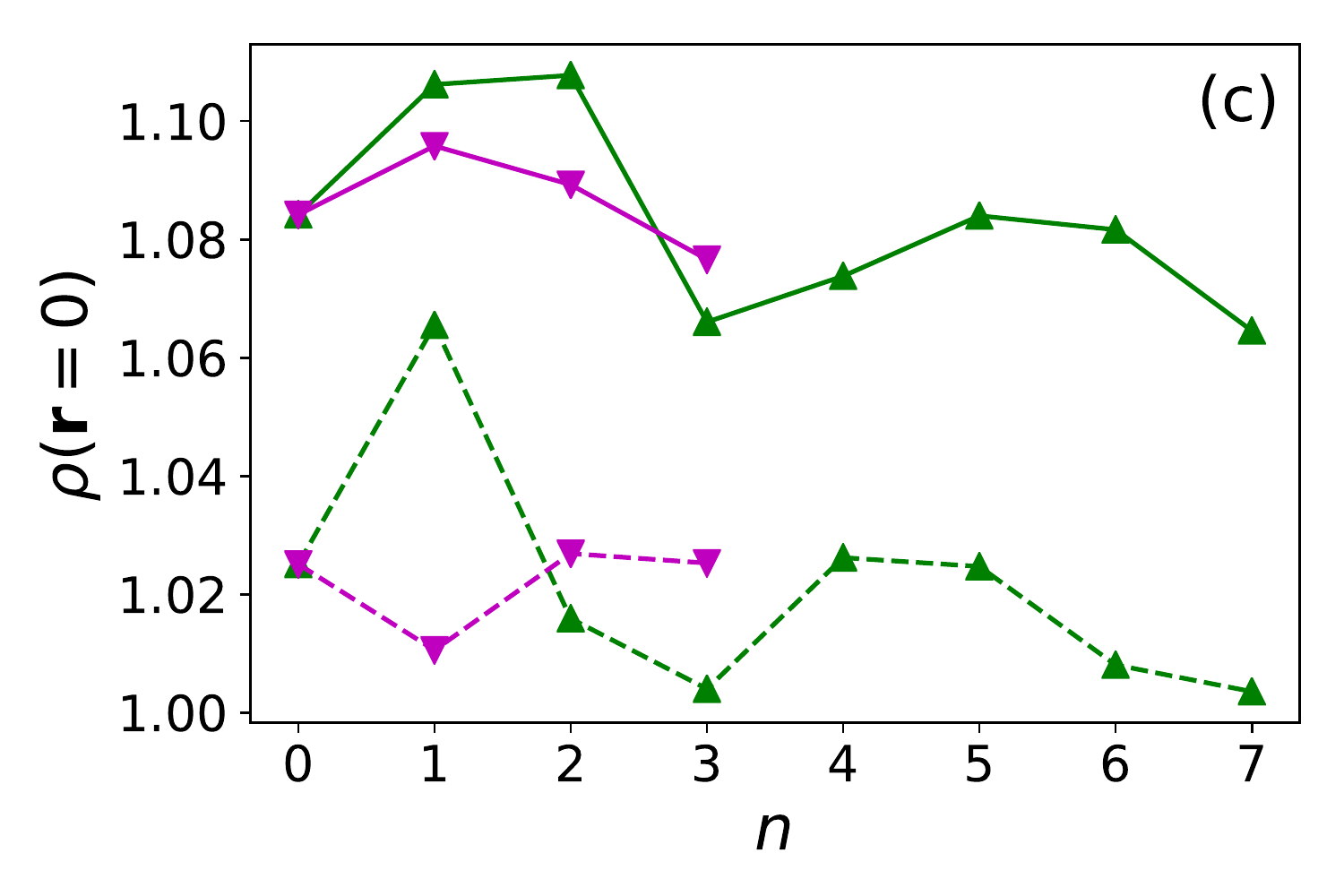}.pdf,height=2.8cm,width=.23\textwidth, clip=true, trim = 0.5cm 0.1cm 0.2cm 0.3cm}
\psfig{figure=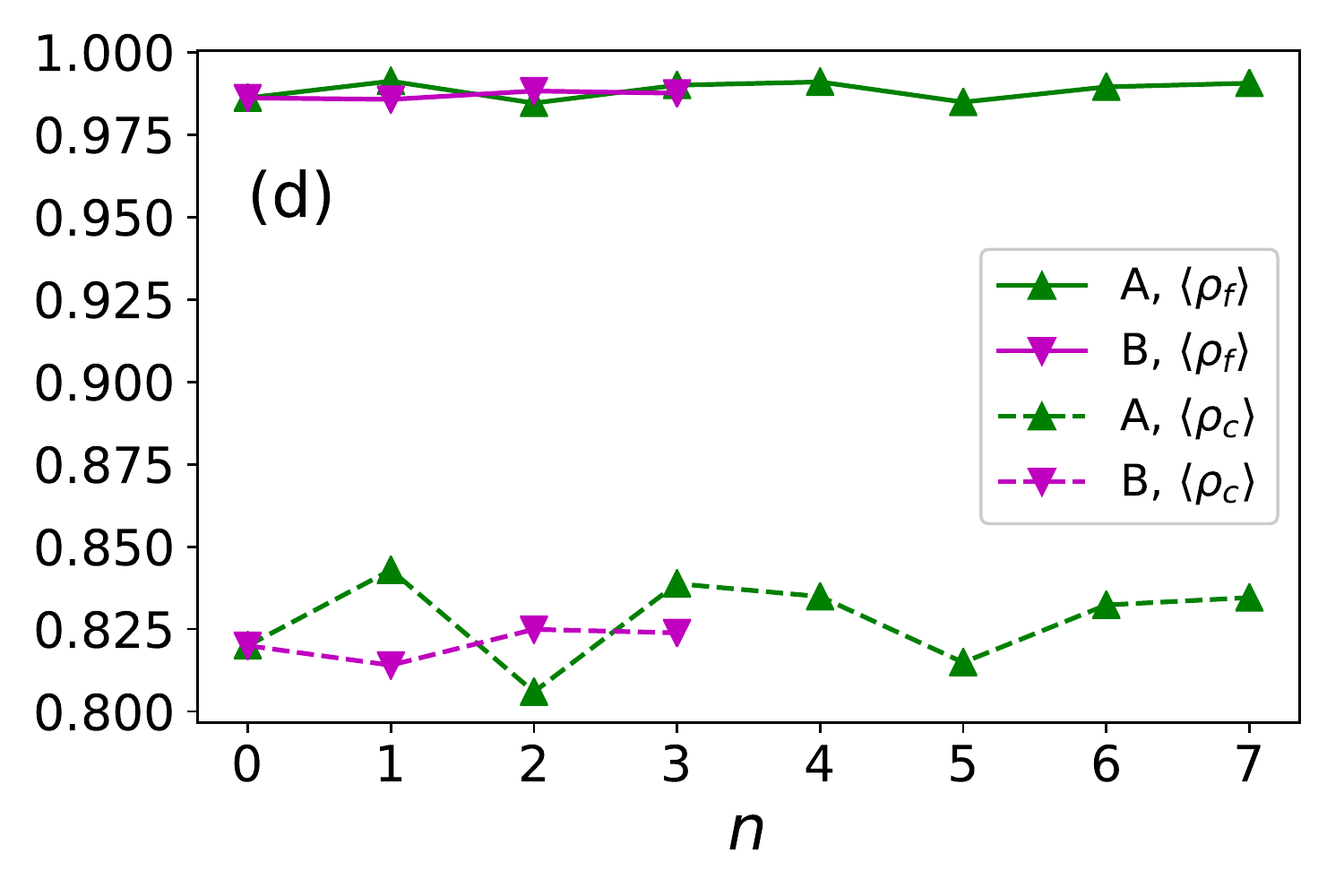}.pdf,height=2.8cm,width=.23\textwidth, clip=true, trim = 0.2cm 0.1cm 0.2cm 0.3cm}
\caption{(a-b) Comparison between $\chi_{ff}(\mathbf{r}=0)$ for droplets with a single impurity ring $N_r=1$ but varied distance $n$ to the central impurity at $T=t/20$; (c-d) The local orbital dependent occupancy at the central impurity anticorrelates with $\chi_{ff}(\mathbf{r}=0)$.}
\label{center_chiff_Nr1}
\end{figure}

To further understand the nontrivial dependence of the local magnetism upon the droplet geometry, we also explored $\chi_{ff}(\mathbf{r}=0)$ for droplets with a single impurity ring $N_r=1$ but varied distance $n$ to the central impurity. As can be seen in Figure~\ref{center_chiff_Nr1}(a-b), it oscillates periodically with $n$ in both $\rho=1$ and $\mu=0$ systems although the lattice size forbids accessing more oscillations for $B$-type droplets, which vividly depicts the tunability of the magnetic properties via modulation of the droplet geometry. The strong charge effect is illustrated in Fig.~\ref{center_chiff_Nr1}(c-d), which displays the anticorrelation between the local density profile at the lattice center and $\chi_{ff}(\mathbf{r}=0)$. Although the charge fluctuation for $f$-electron can be partially suppressed by enforcing its nearly half-filled occupancy via $\mu=0$, the density profile of the conduction electrons also plays an important role in determining the magnetic properties of the droplet.

\subsection{Density fluctuation}
\begin{figure*}
\psfig{figure=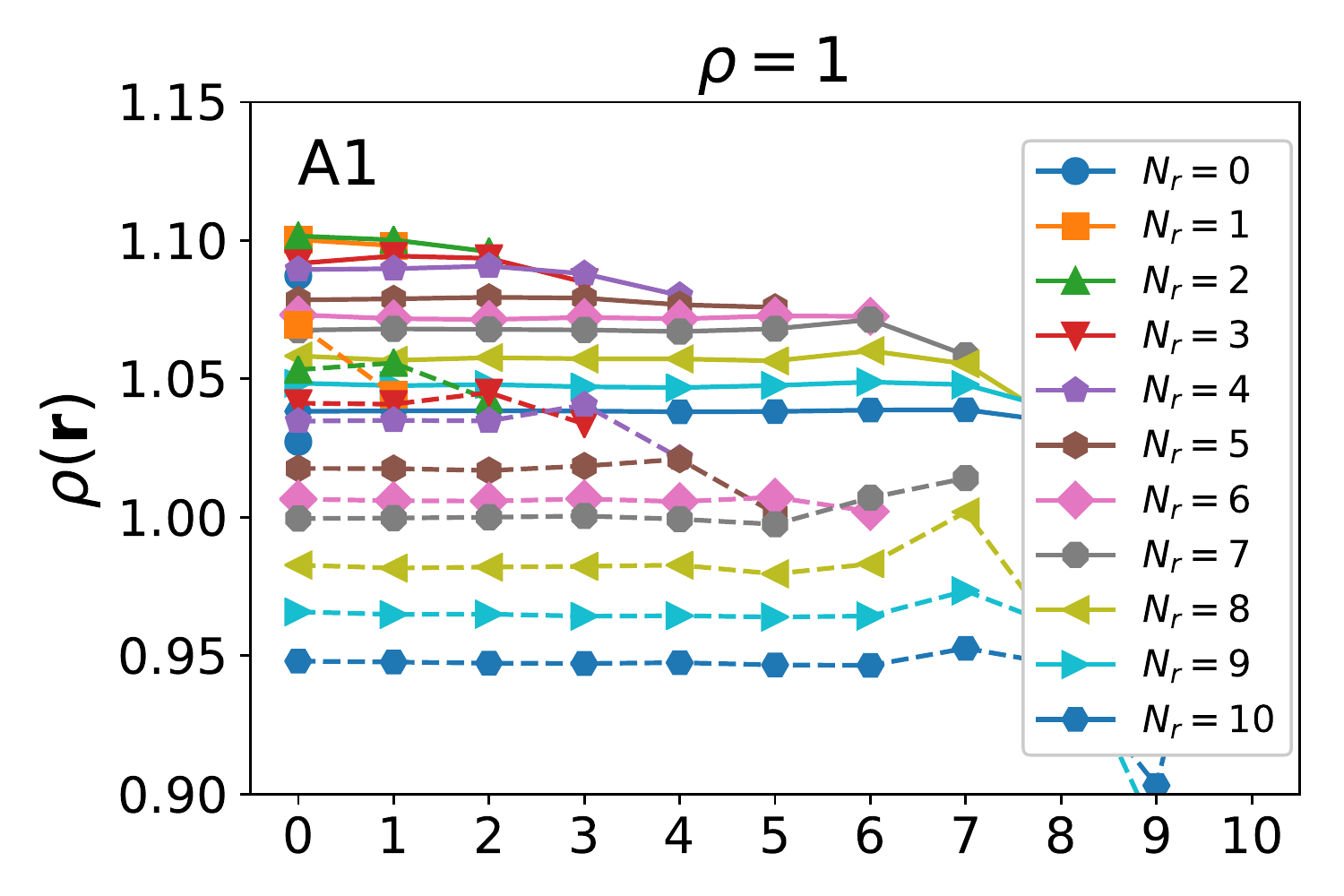}.pdf,height=3.0cm,width=.25\textwidth, clip=true, trim = 0.0cm 0.0cm 0.0cm 0.0cm} 
\psfig{figure=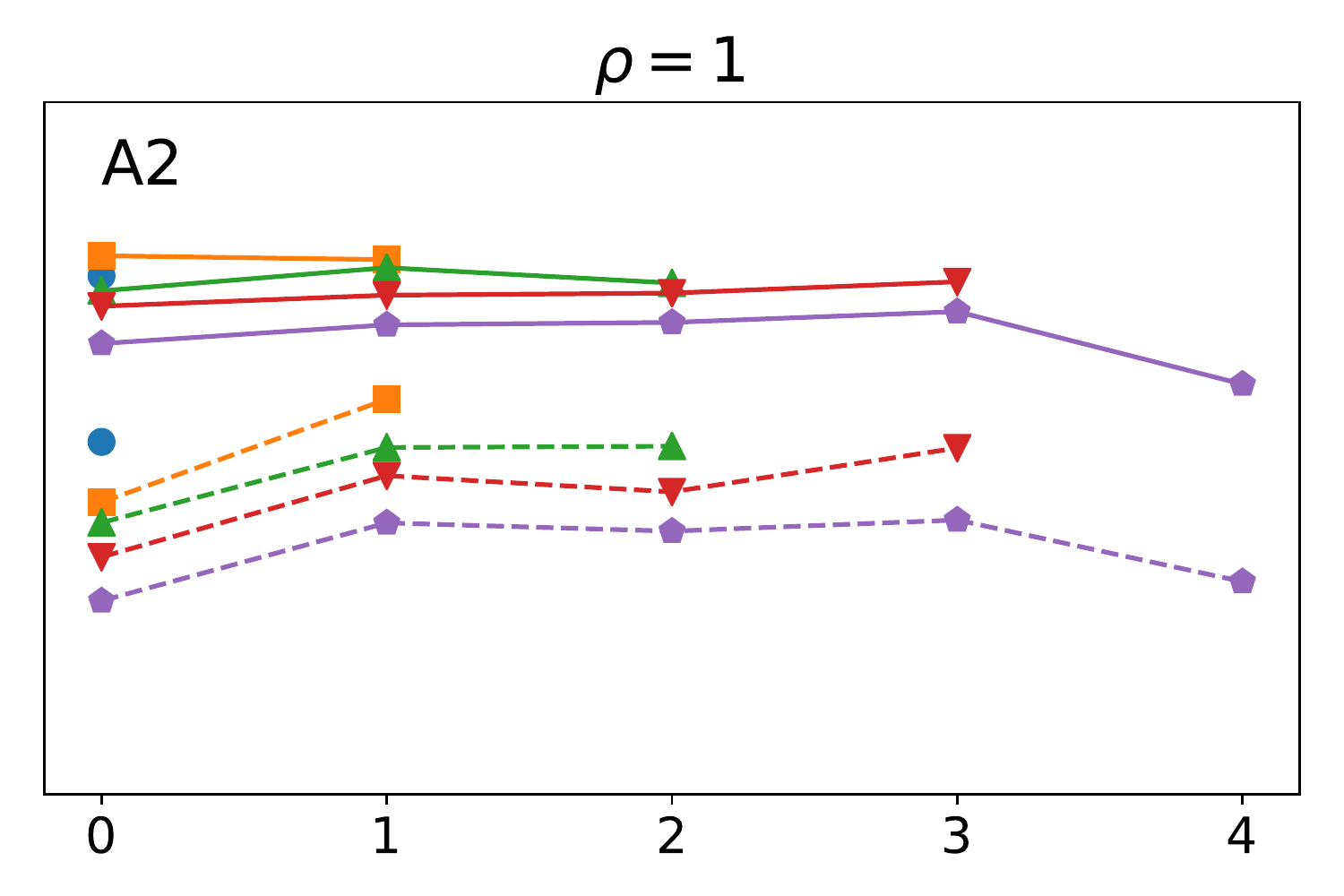}.pdf,height=3.0cm,width=.22\textwidth, clip=true, trim = 0.0cm 0.0cm 0.0cm 0.0cm}
\psfig{figure=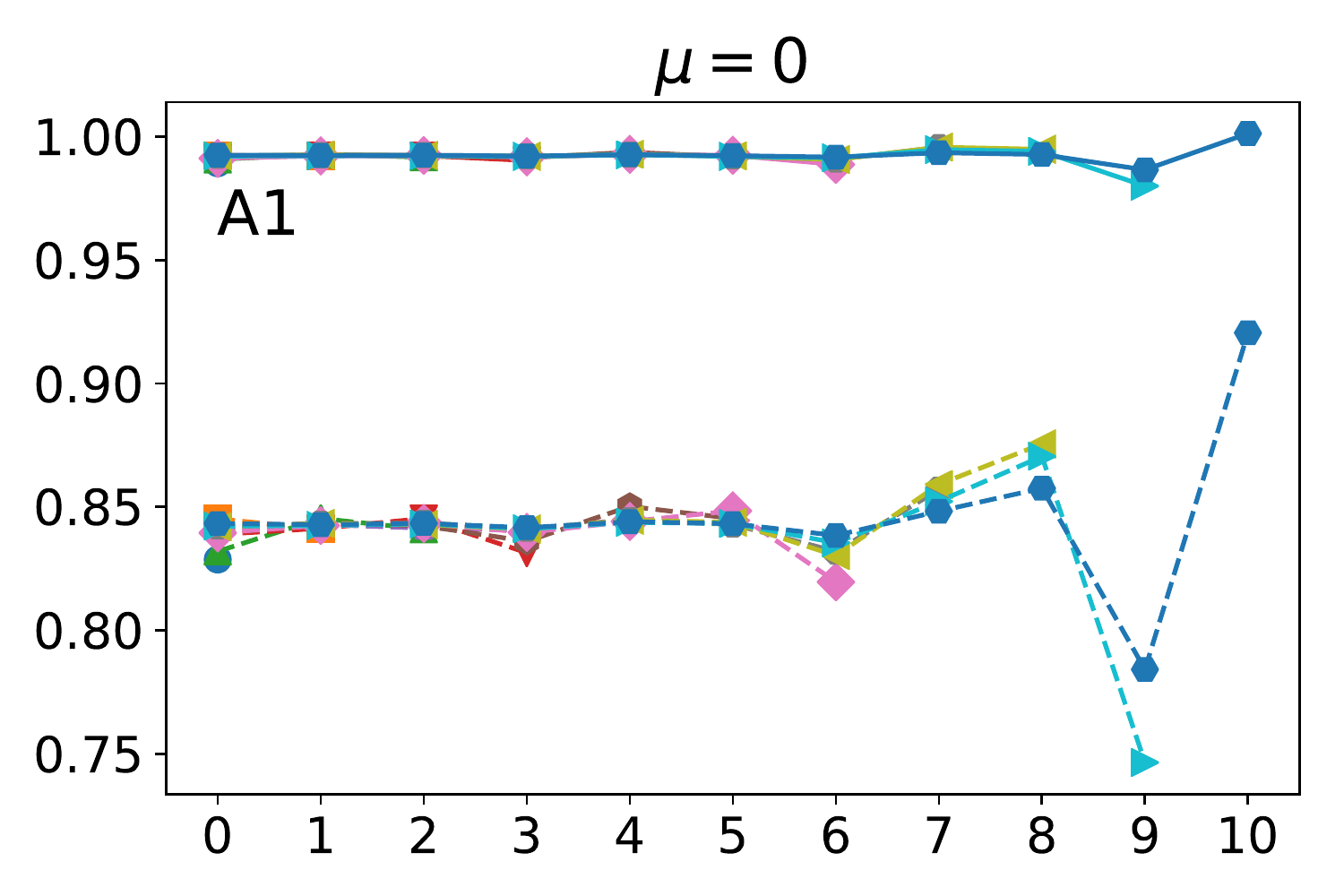}.pdf,height=3.0cm,width=.25\textwidth, clip=true, trim = 0.0cm 0.0cm 0.0cm 0.0cm} 
\psfig{figure=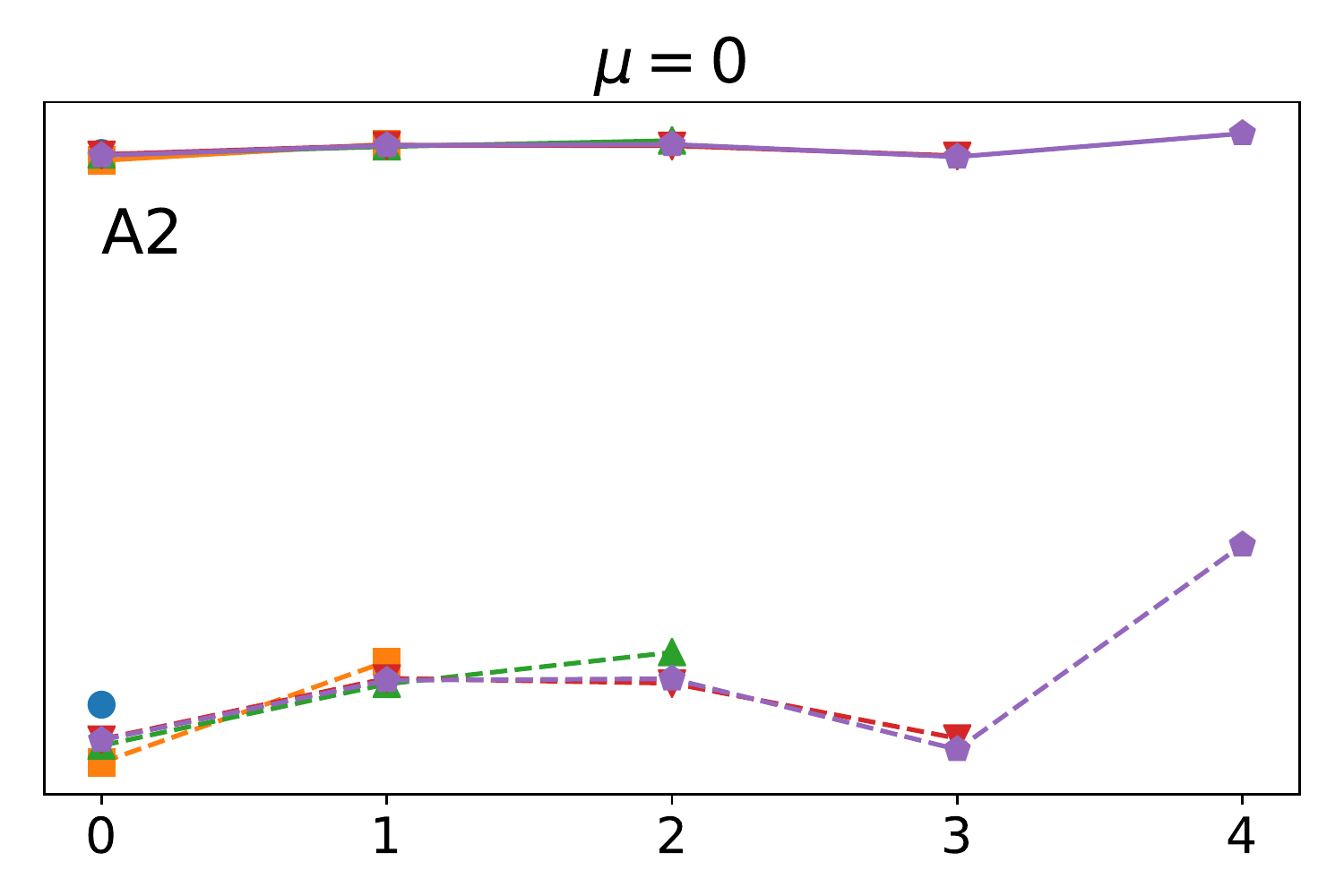}.pdf,height=3.0cm,width=.22\textwidth, clip=true, trim = 0.0cm 0.0cm 0.0cm 0.0cm} \\
\psfig{figure=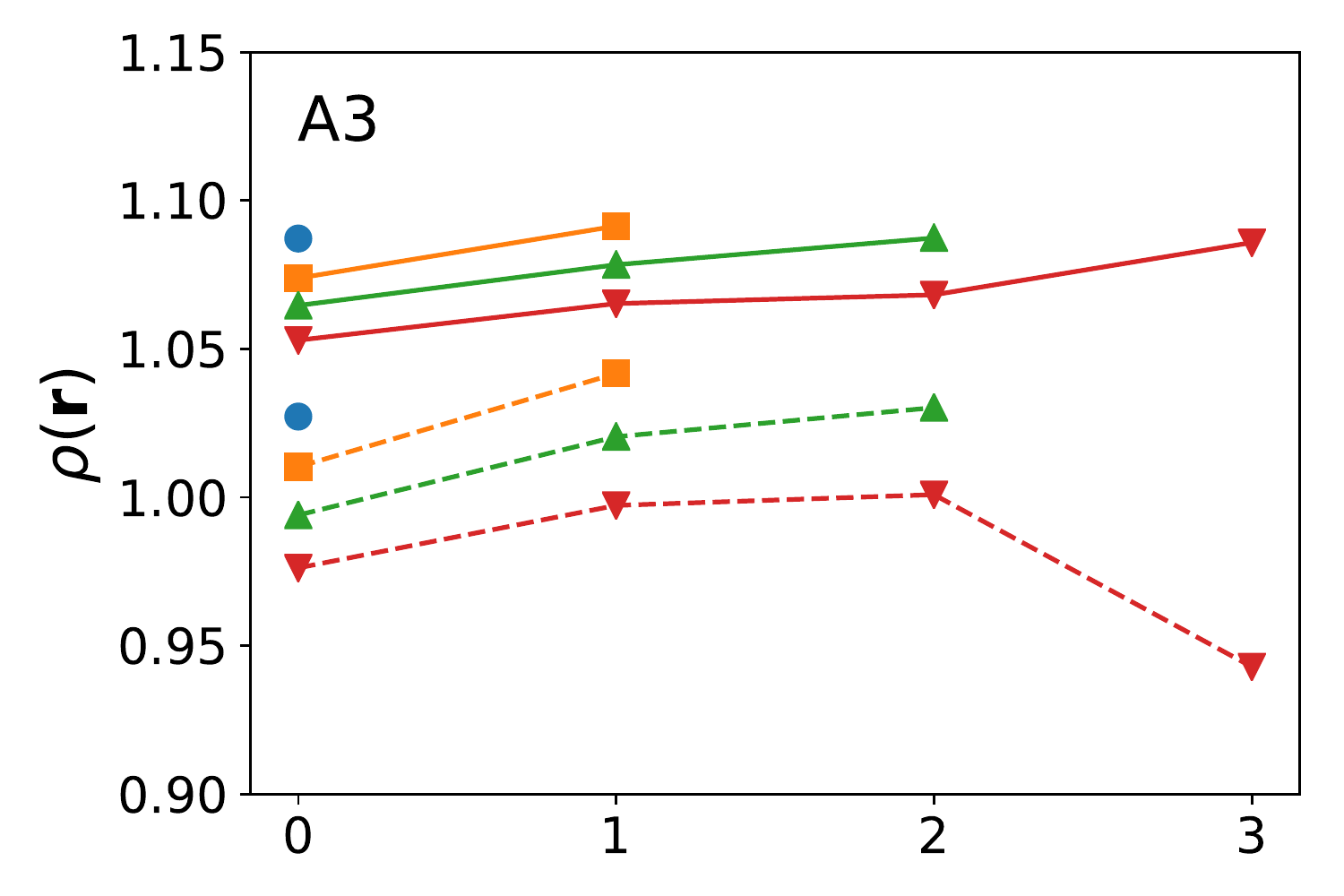}.pdf,height=2.8cm,width=.25\textwidth, clip=true, trim = 0.0cm 0.0cm 0.0cm 0.0cm}
\psfig{figure=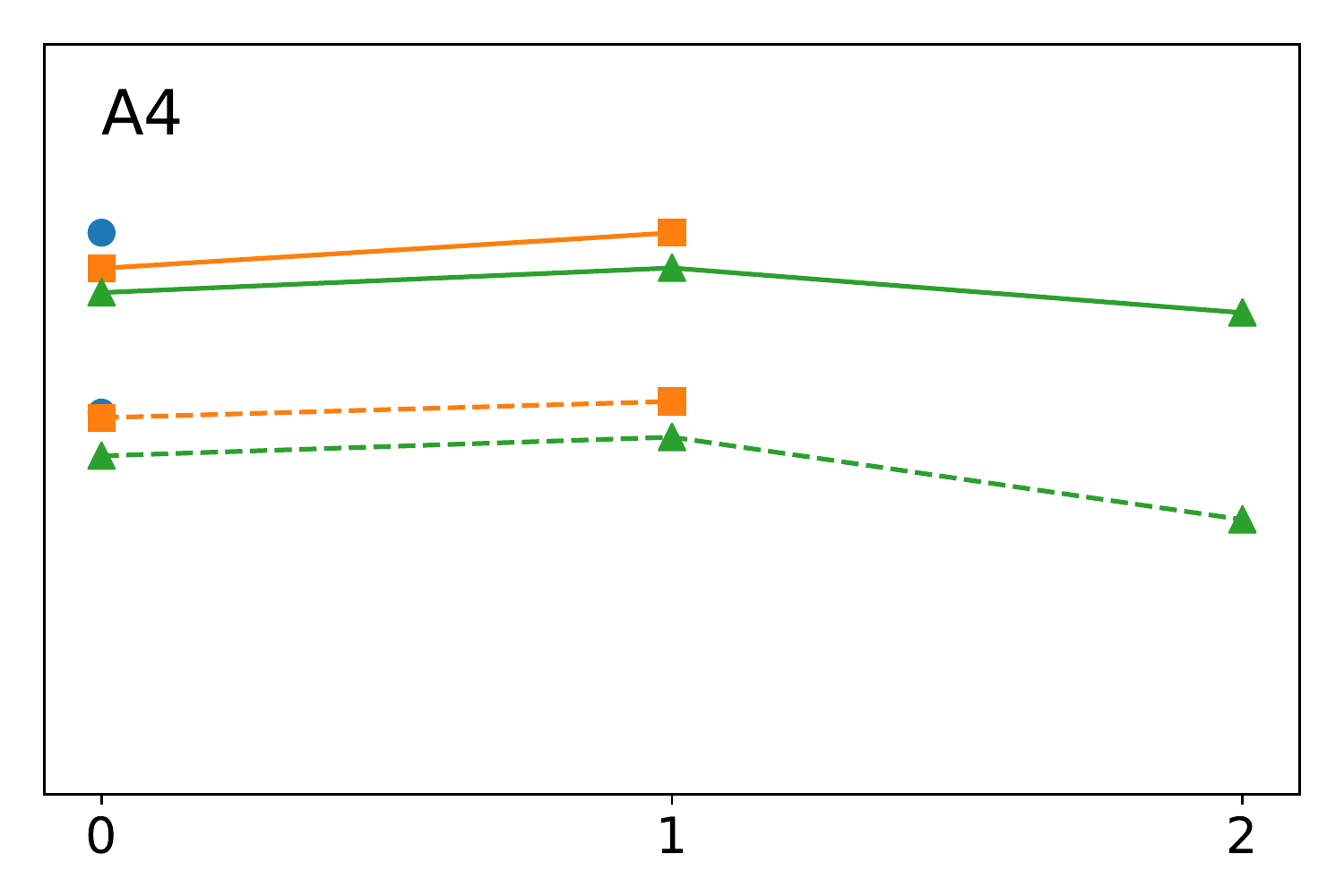}.pdf,height=2.8cm,width=.22\textwidth, clip=true, trim = 0.0cm 0.0cm 0.0cm 0.0cm} 
\psfig{figure=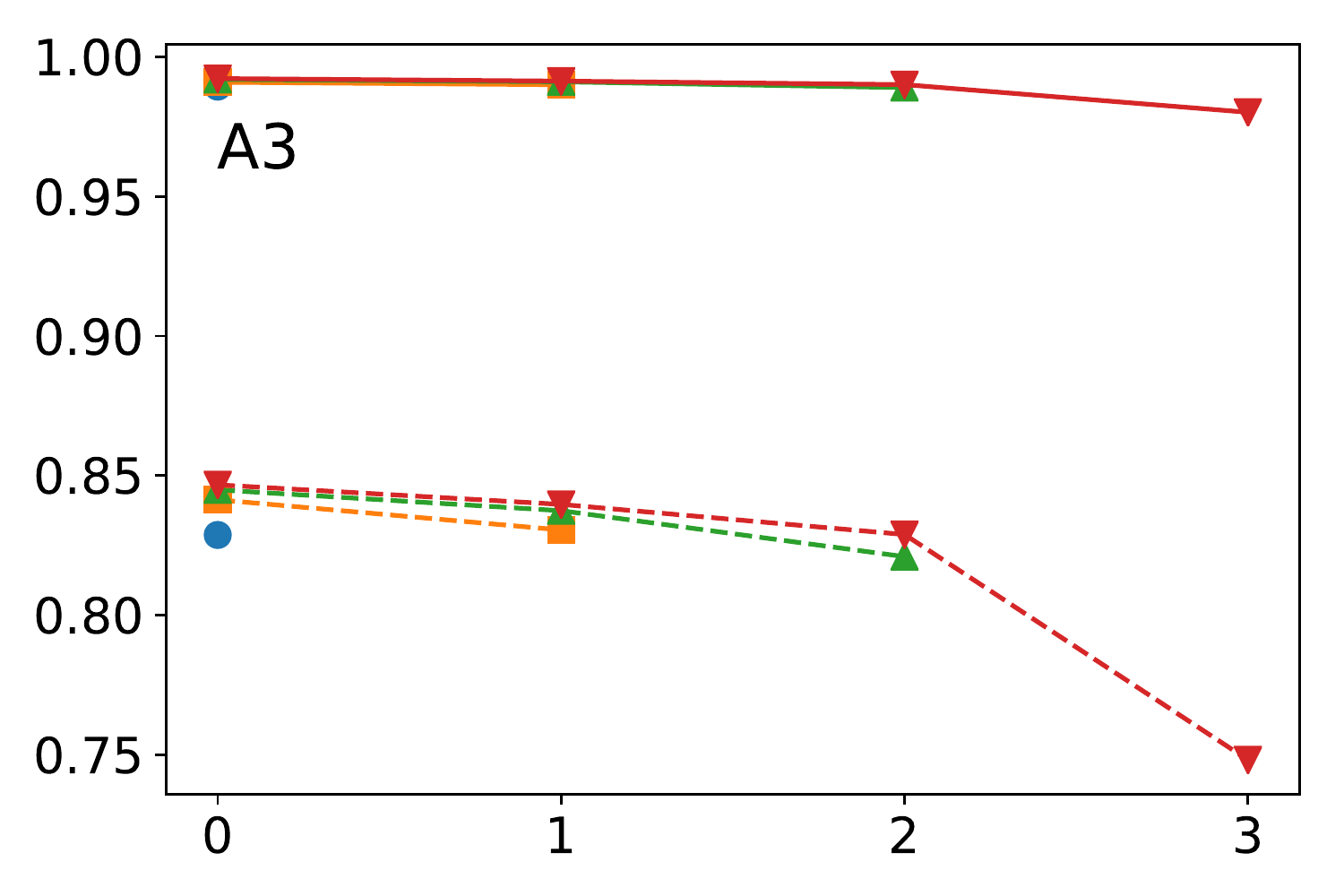}.pdf,height=2.8cm,width=.25\textwidth, clip=true, trim = 0.0cm 0.0cm 0.0cm 0.0cm} 
\psfig{figure=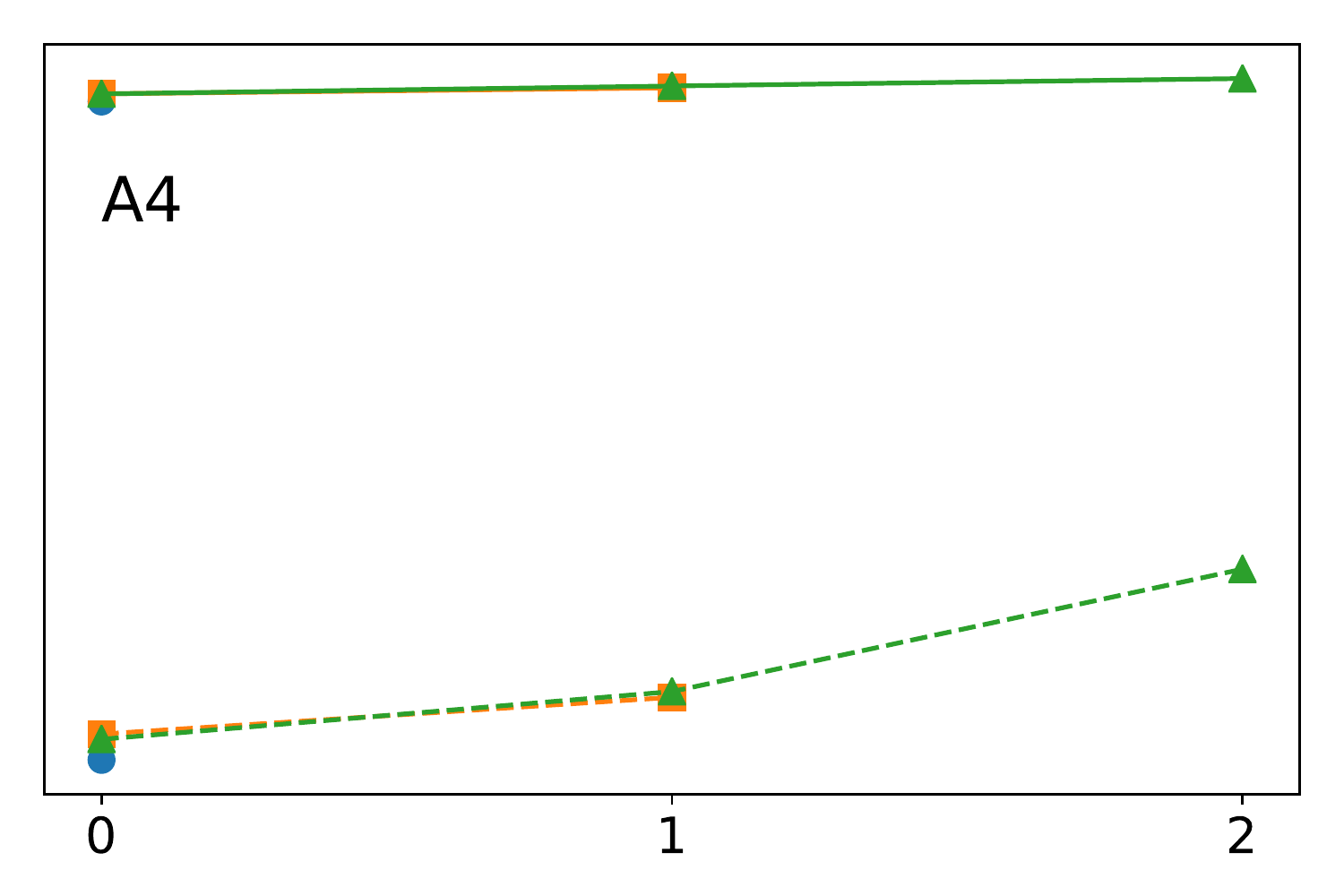}.pdf,height=2.8cm,width=.22\textwidth, clip=true, trim = 0.0cm 0.0cm 0.0cm 0.0cm}  \\
\psfig{figure=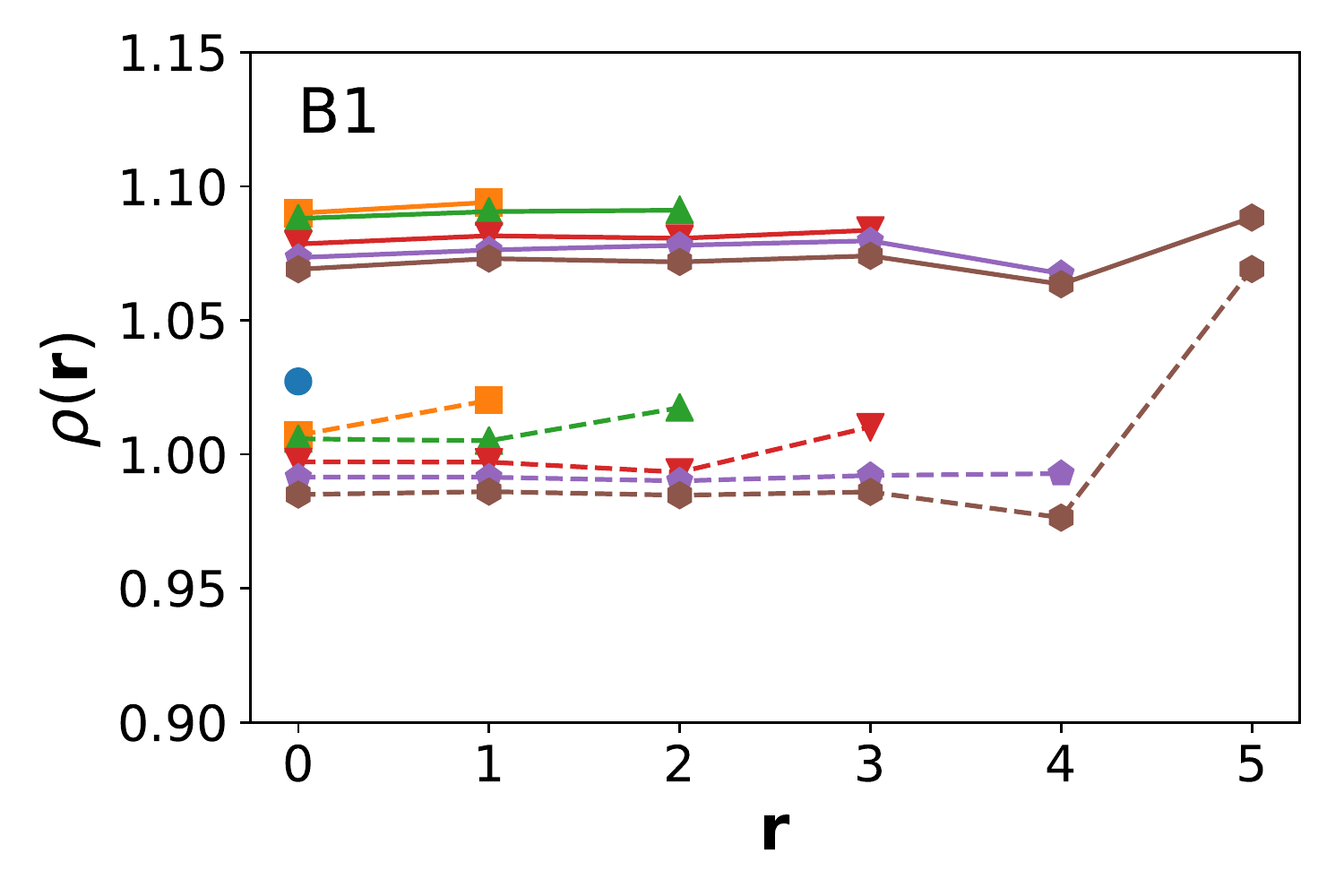}.pdf,height=3.0cm,width=.25\textwidth, clip=true, trim = 0.0cm 0.0cm 0.0cm 0.0cm} 
\psfig{figure=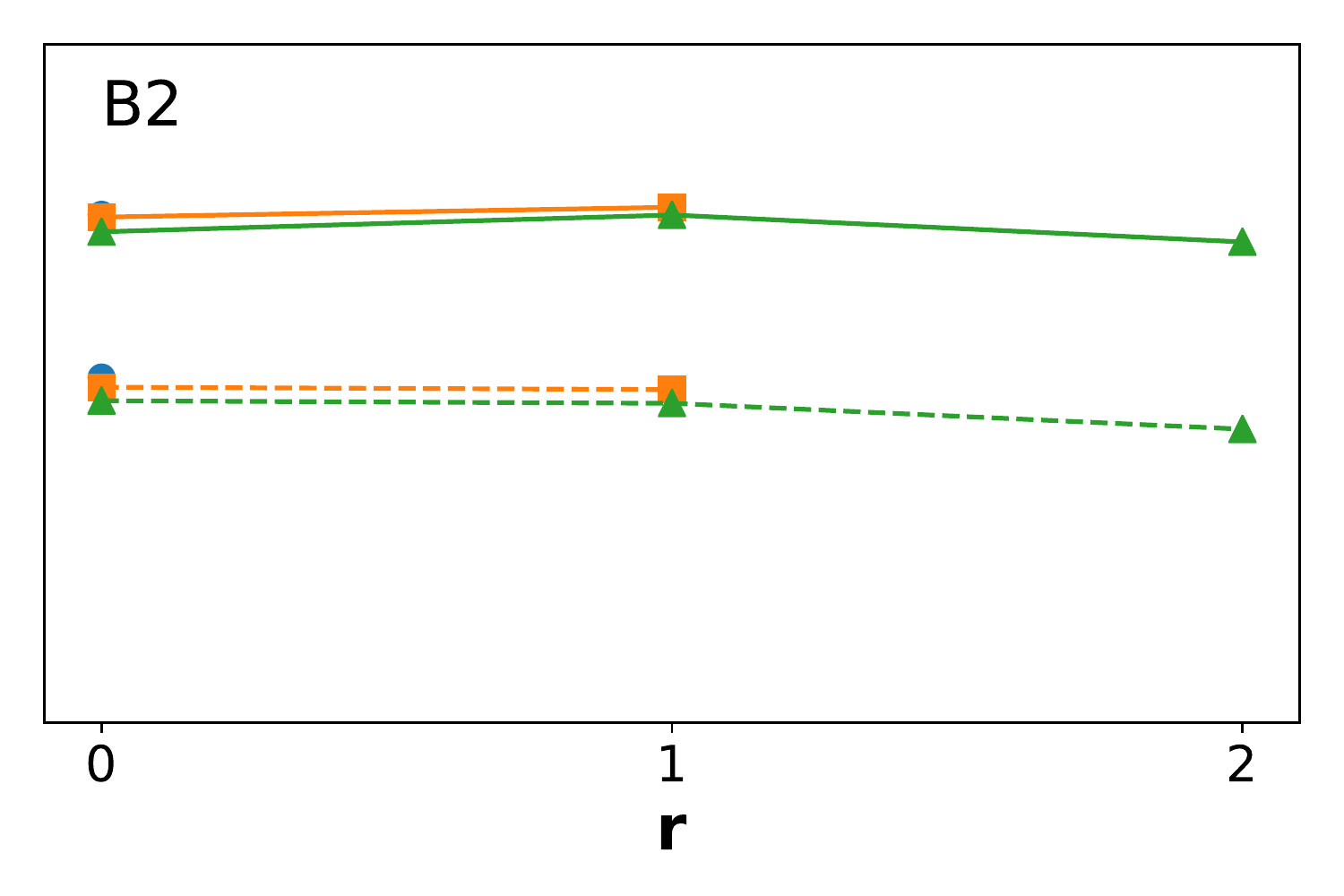}.pdf,height=3.0cm,width=.22\textwidth, clip=true, trim = 0.0cm 0.0cm 0.0cm 0.0cm} 
\psfig{figure=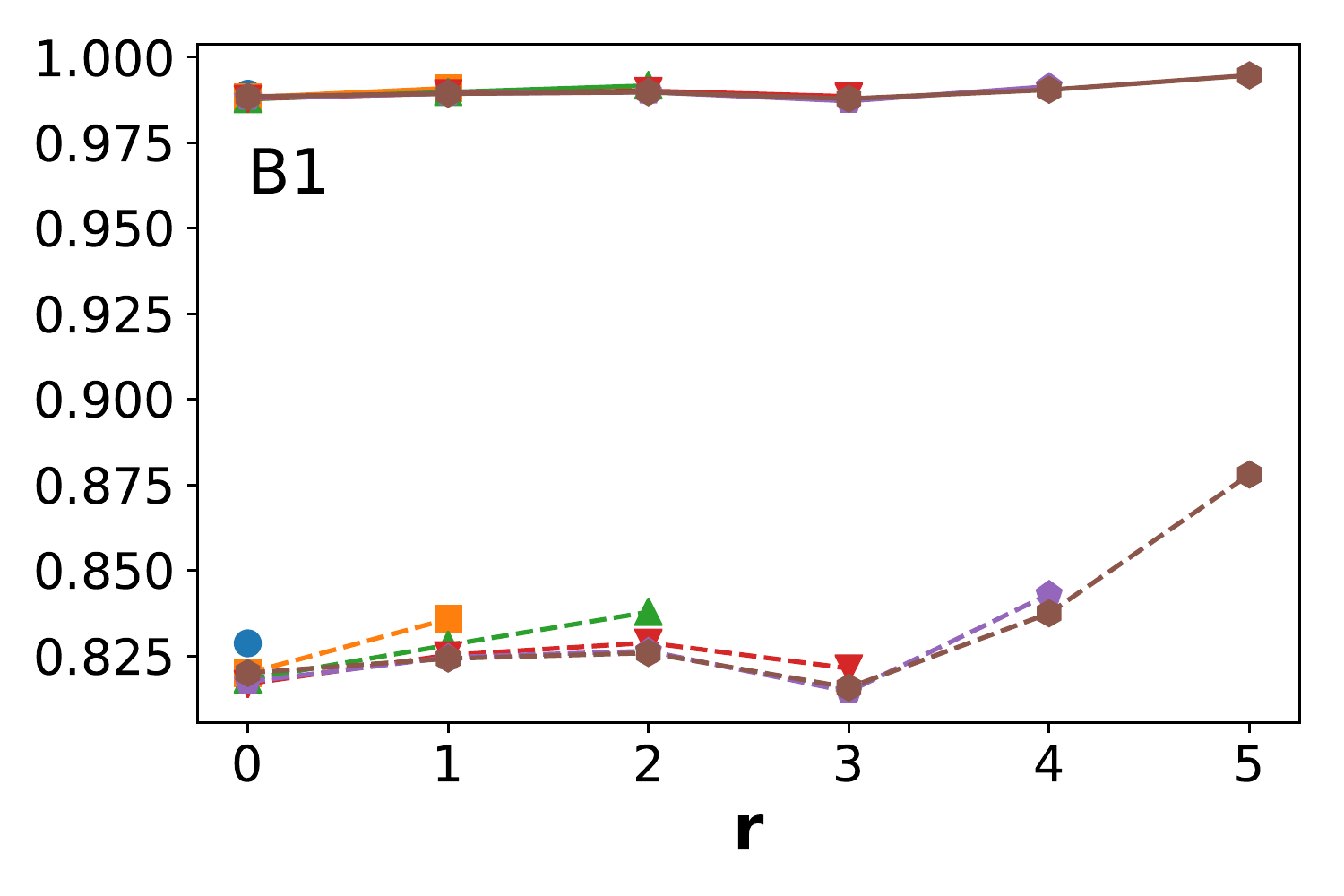}.pdf,height=3.0cm,width=.25\textwidth, clip=true, trim = 0.0cm 0.0cm 0.0cm 0.0cm} 
\psfig{figure=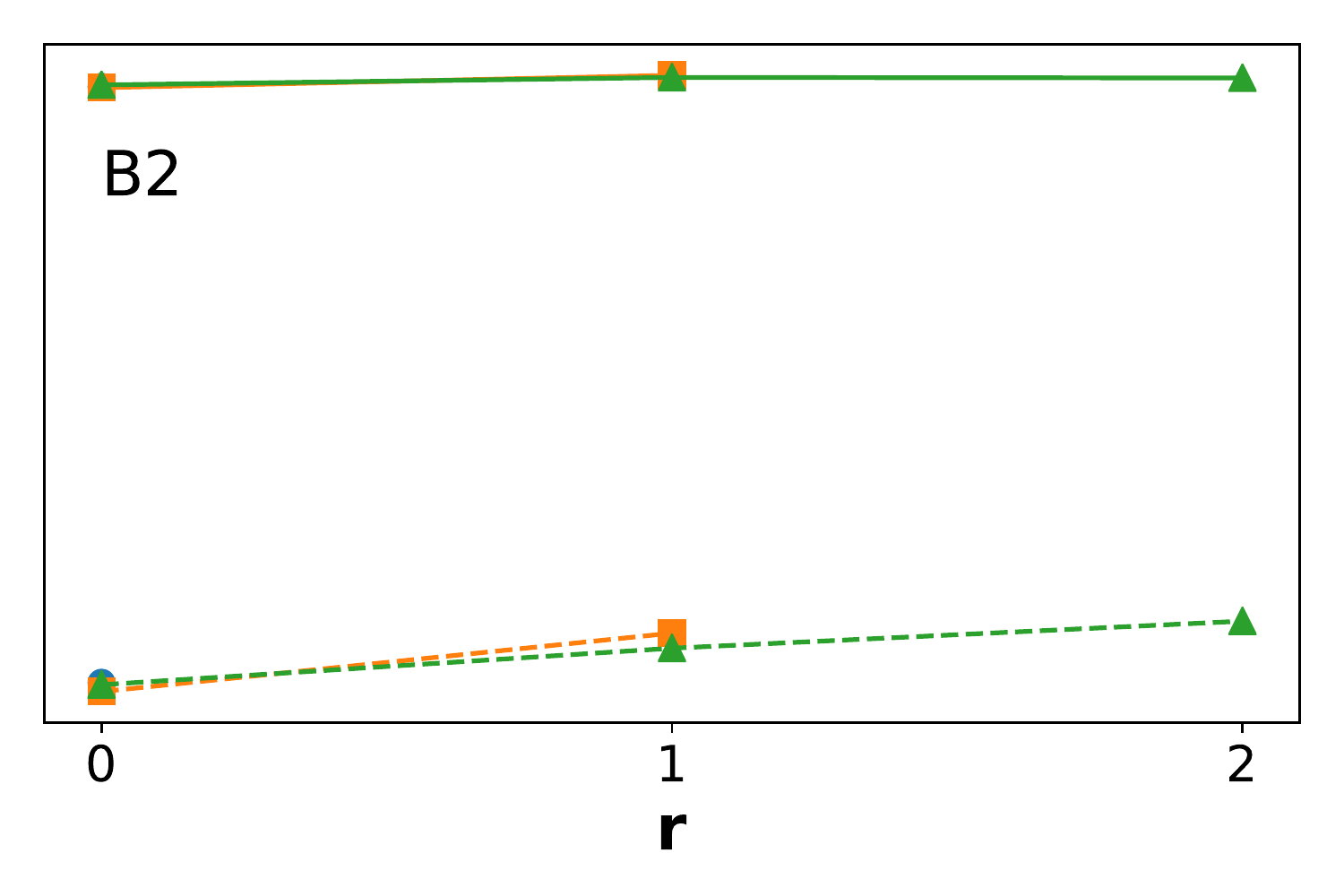}.pdf,height=3.0cm,width=.22\textwidth, clip=true, trim = 0.0cm 0.0cm 0.0cm 0.0cm}  
\caption{Local orbital-resolved occupancy of droplet site $\rho_f(\mathbf{r})$ (solid lines) and conduction electrons $\rho_c(\mathbf{r})$ (dashed lines) in various systems at $T=t/10$. The general anticorrelation to $\chi_{ff}(\mathbf{r})$ shown in Fig.~2 is visible.}
\label{den_B10}
\end{figure*}

Previously we briefly mentioned that the local density fluctuation anticorrelates with the oscillations of $\chi_{ff}(\mathbf{r})$. To provide more insights on the origin of the evolution of the magnetic properties with the droplet geometry, Figure~\ref{den_B10} illustrates the local occupancy of droplet site $\rho_f(\mathbf{r})$ (solid lines) and conduction electrons $\rho_c(\mathbf{r})$ (dashed lines) in various systems, which shows this general anticorrelation vividly compared with Fig.~\ref{chiff_B10}. Specifically, taking $A1$ droplet of $\rho=1$ system for example, the peak of $\rho_f(\mathbf{r}=0)$ versus $N_r$ occurs at $N_r=2$, which coincides with the occurence of the valley of $\chi_{ff}(\mathbf{r}=0)$ at the same $N_r$ in Fig.~\ref{center_chiff}(a).

The most important feature is the coincidence between $\chi_{ff}(\mathbf{r})$ and the closeness of $\rho_f(\mathbf{r})$ to unity. As $\rho_f(\mathbf{r})$ approaches to unity with $\mathbf{r}$ and/or $N_r$, $\chi_{ff}(\mathbf{r})$ increases accordingly, which is a well-known consequence of the local magnetic moment associated with the forbidden double occupancy. This is clearly evidenced by the generally larger $\chi_{ff}(\mathbf{r})$ in $\mu=0$ systems because of the almost half-filled droplet sites $\rho_f(\mathbf{r}) \sim 1$.
In other words, the evolution of $\chi_{ff}(\mathbf{r})$, especially the unusual enhancement of $\chi_{ff}(\mathbf{r}=0)$ with $N_r$ (Fig.~\ref{center_chiff}) presented in Sec. III.A is strongly tied to the density redistribution via varying the droplet geometry. Apparently, this density fluctuation is closely related to the non-periodicity of the lattice due to both the droplet geometry and the open boundary employed. Therefore, in essence, the possibility of artificial manipulation of the magnetic properties of the Anderson droplet in our current confined lattice system is realized via the potentially controllable density variation enforced by a finite boundary.

The flatness of the density profile for systems with large $N_r$ implies that the inner region of the droplet becomes more and more homogeneous and coherent upon increasing $N_r$. For other droplets with larger distance $n$ between consecutive rings, we are limited by the lattice size to identify the ultimate flatness of $\rho(\mathbf{r})$. The anomalous oscillations at the outermost ring are due to the lattice and/or droplet boundary effects similar to $\chi_{ff}(\mathbf{r})$. 
Besides, the general weaker dependence of $\rho(\mathbf{r})$ for $B$-type droplets on the droplet geometry matches with the trend of $\chi_{ff}(\mathbf{r})$.
In $\mu=0$ systems, the density fluctuation mostly comes from the conduction electrons because the droplet sites are enforced to be nearly half-filled. 
The major difference between systems of $\rho=1$ and $\mu=0$ lies in the saturation or not of $\rho(\mathbf{r})$ with increasing $N_r$. Apparently, the local density saturation of the latter system mirrors the saturation of $\chi_{ff}(\mathbf{r})$.

\subsection{Interorbital local susceptibility}
\begin{figure*}
\psfig{figure=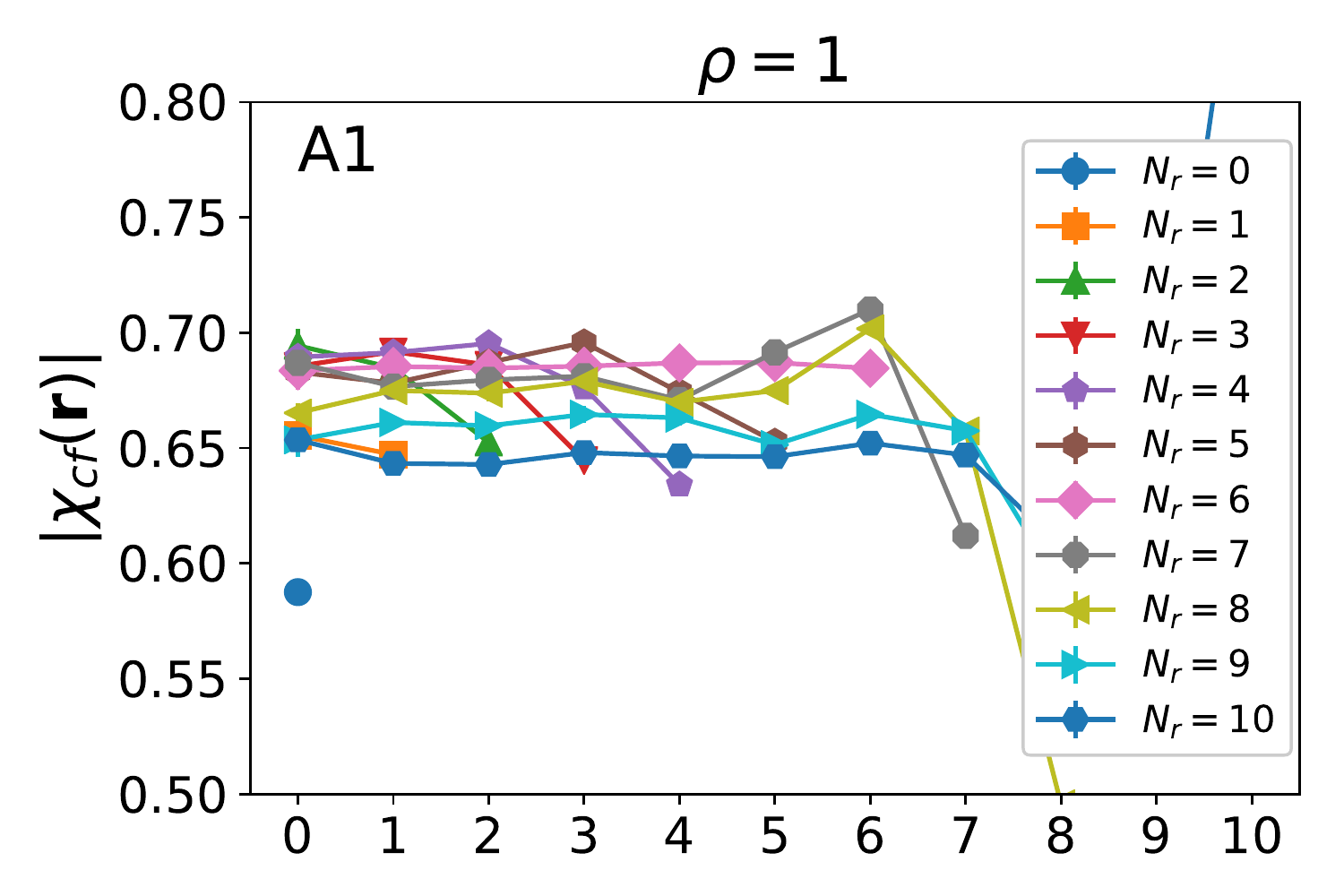}.pdf,height=3.0cm,width=.25\textwidth, clip=true, trim = 0.0cm 0.0cm 0.0cm 0.0cm} 
\psfig{figure=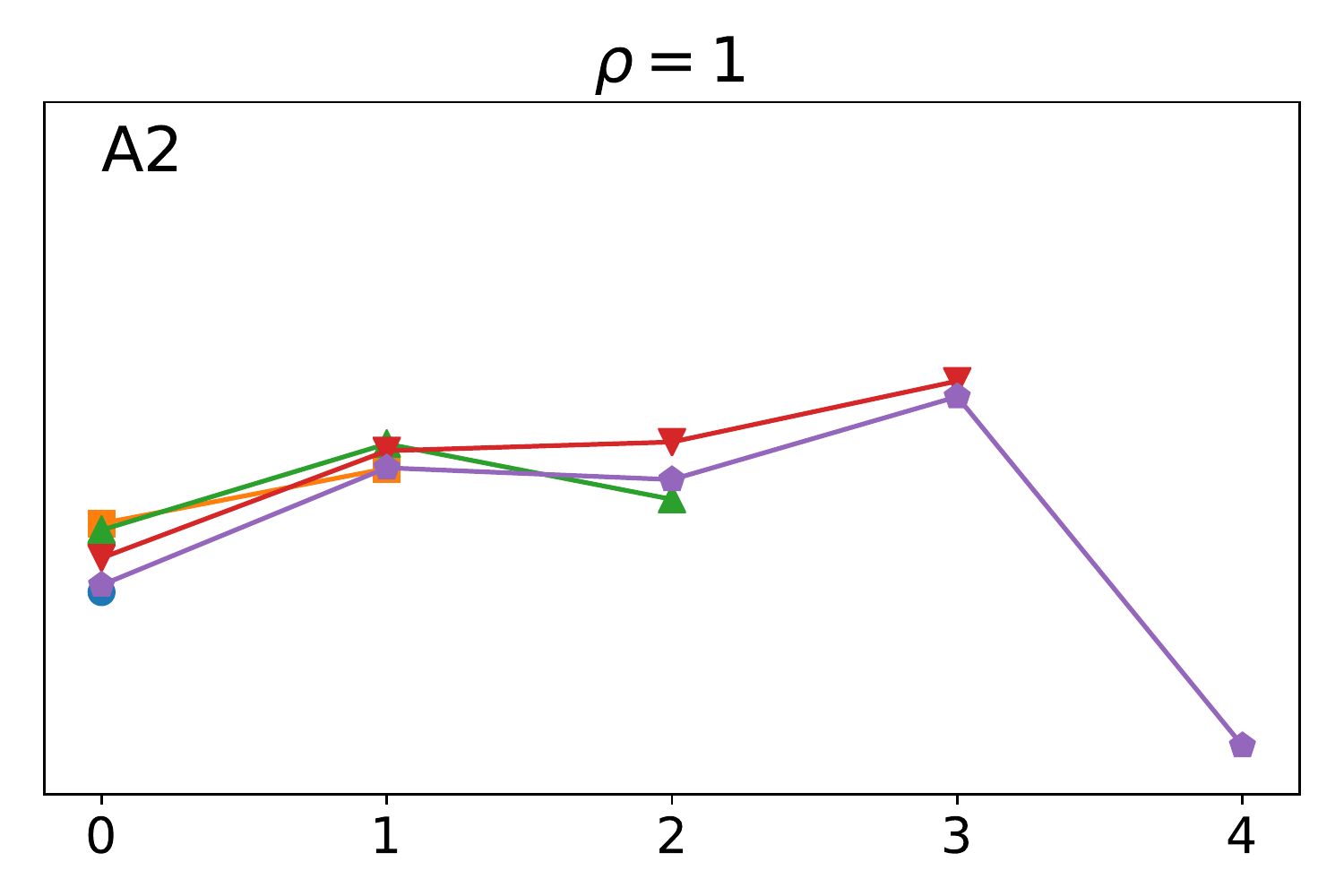}.pdf,height=3.0cm,width=.22\textwidth, clip=true, trim = 0.0cm 0.0cm 0.0cm 0.0cm}
\psfig{figure=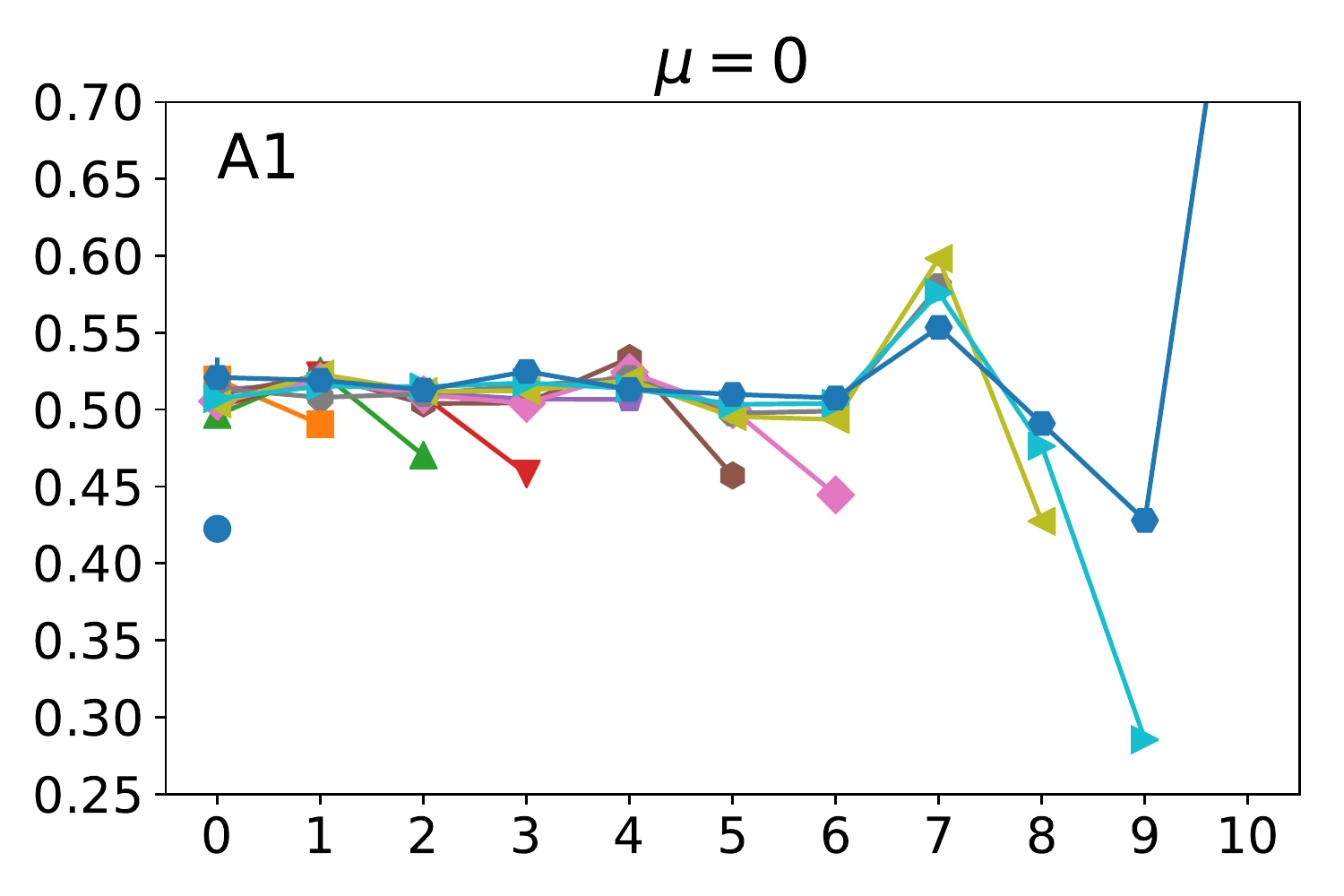}.pdf,height=3.0cm,width=.25\textwidth, clip=true, trim = 0.0cm 0.0cm 0.0cm 0.0cm} 
\psfig{figure=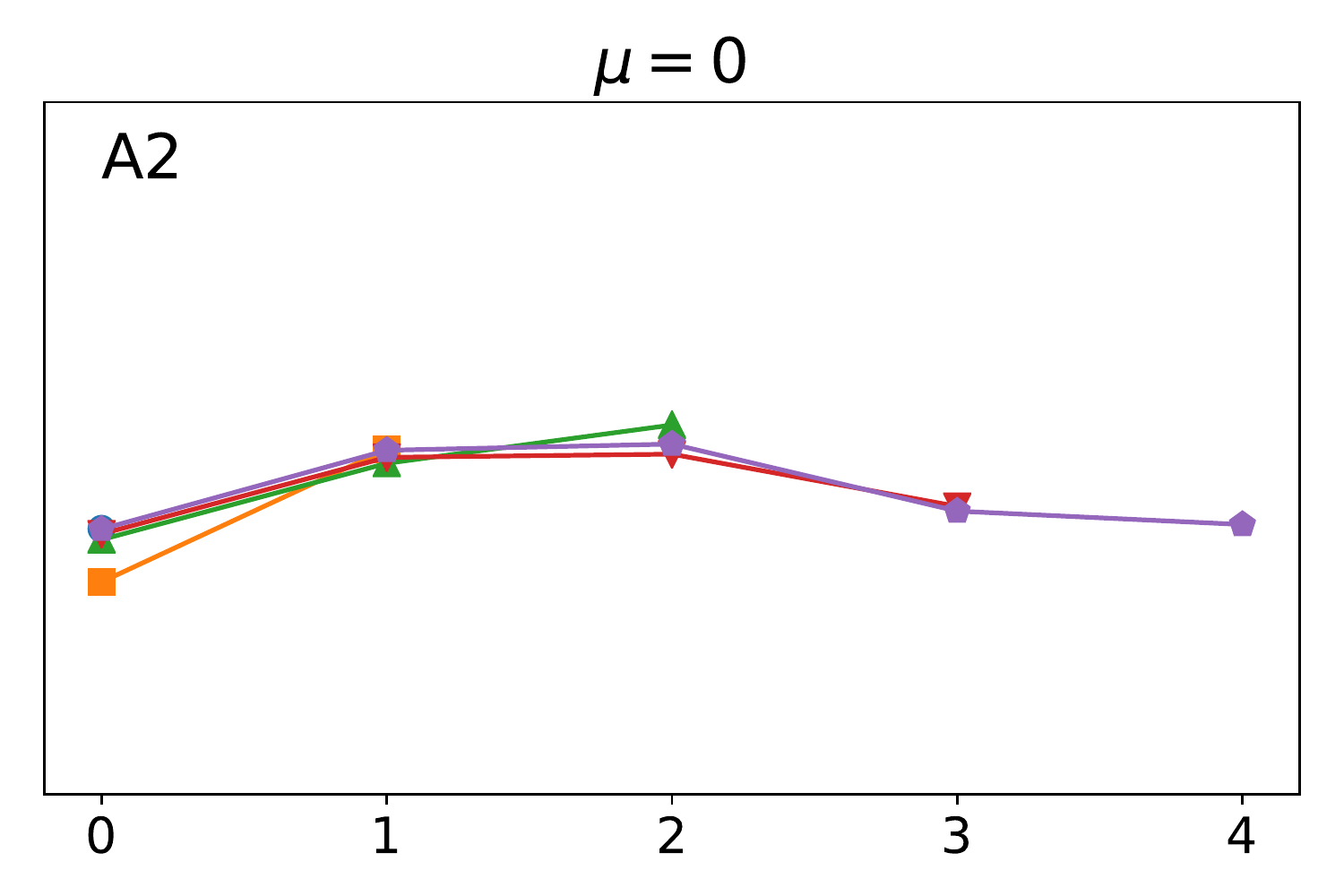}.pdf,height=3.0cm,width=.22\textwidth, clip=true, trim = 0.0cm 0.0cm 0.0cm 0.0cm} \\
\psfig{figure=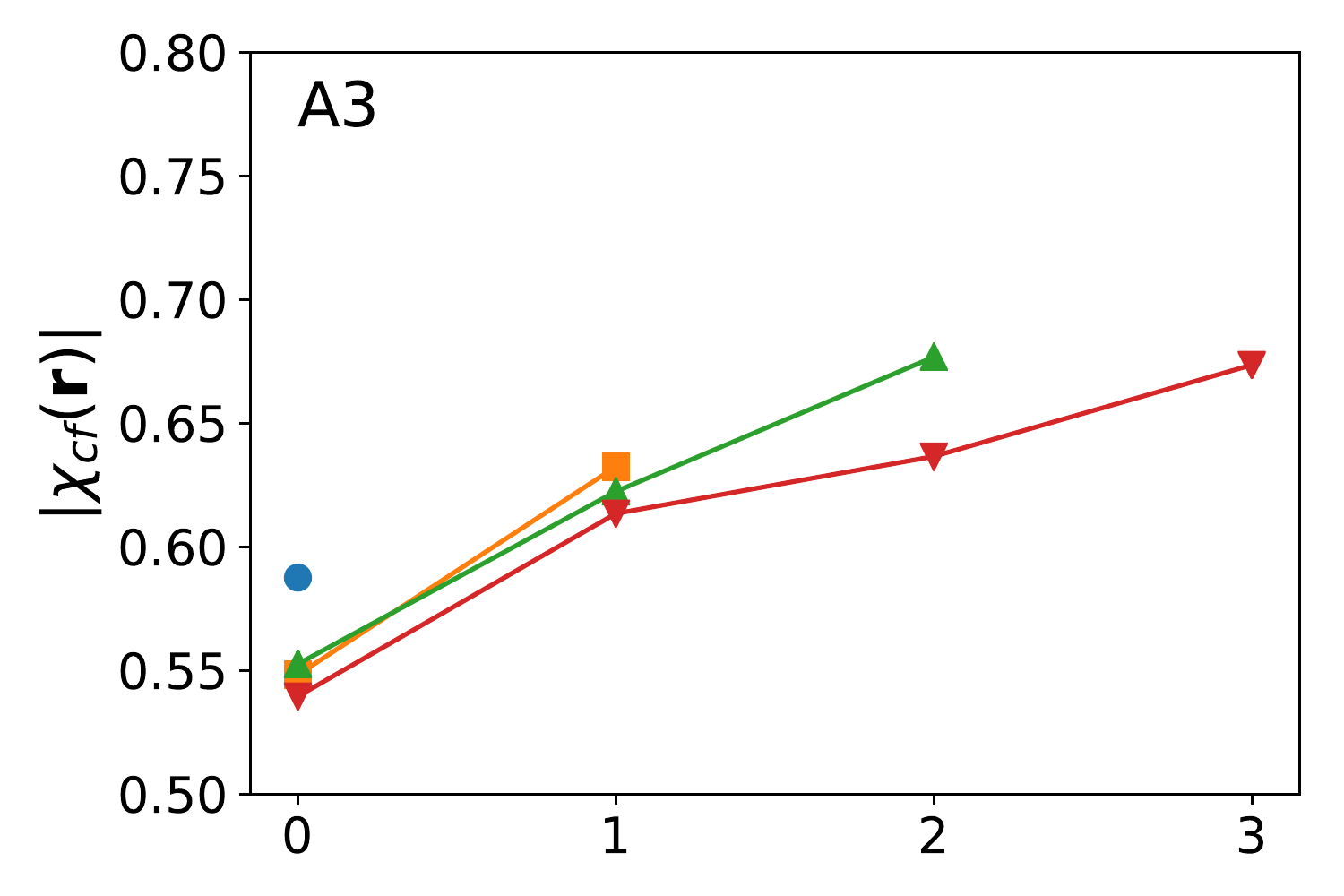}.pdf,height=2.8cm,width=.25\textwidth, clip=true, trim = 0.0cm 0.0cm 0.0cm 0.0cm}
\psfig{figure=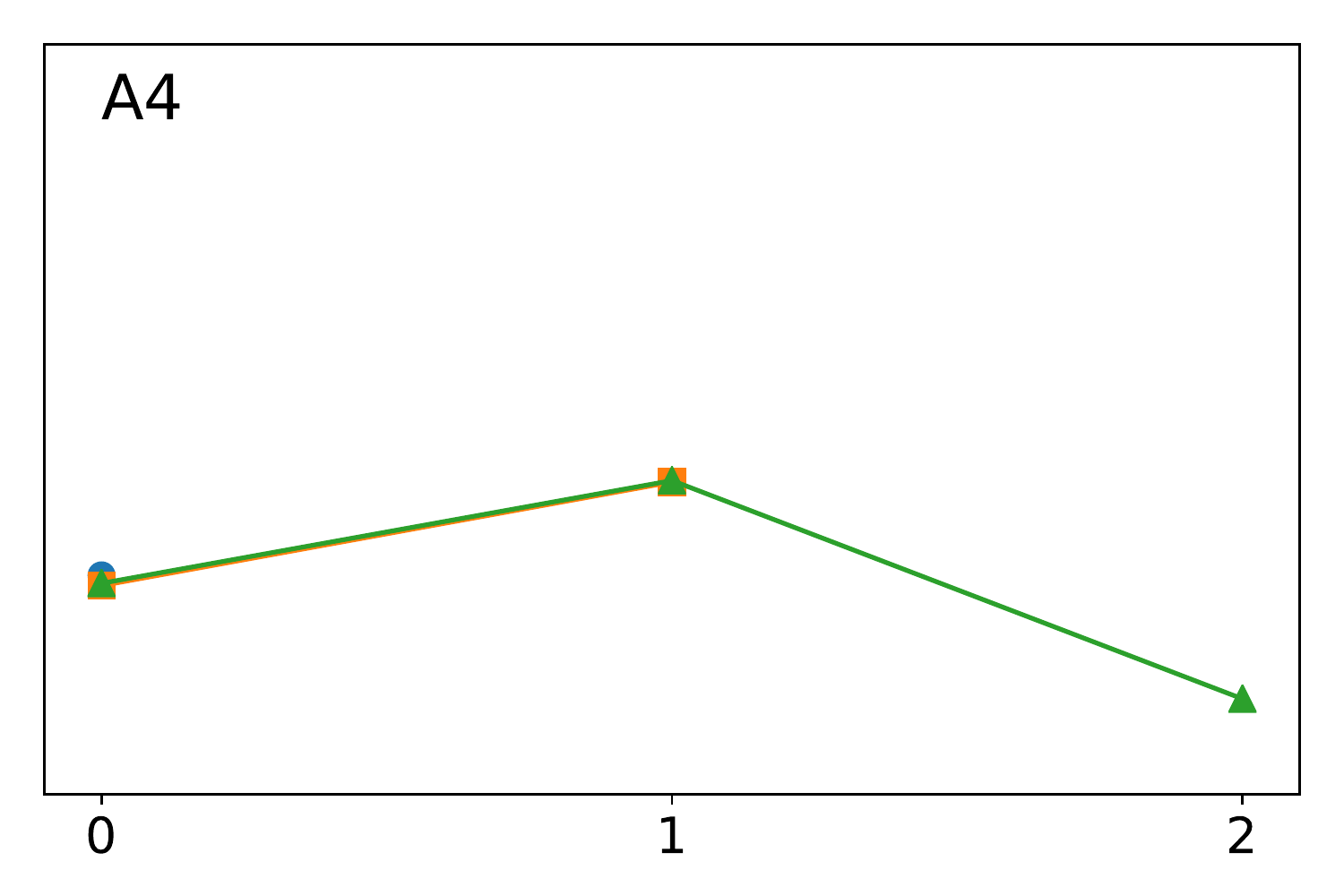}.pdf,height=2.8cm,width=.22\textwidth, clip=true, trim = 0.0cm 0.0cm 0.0cm 0.0cm} 
\psfig{figure=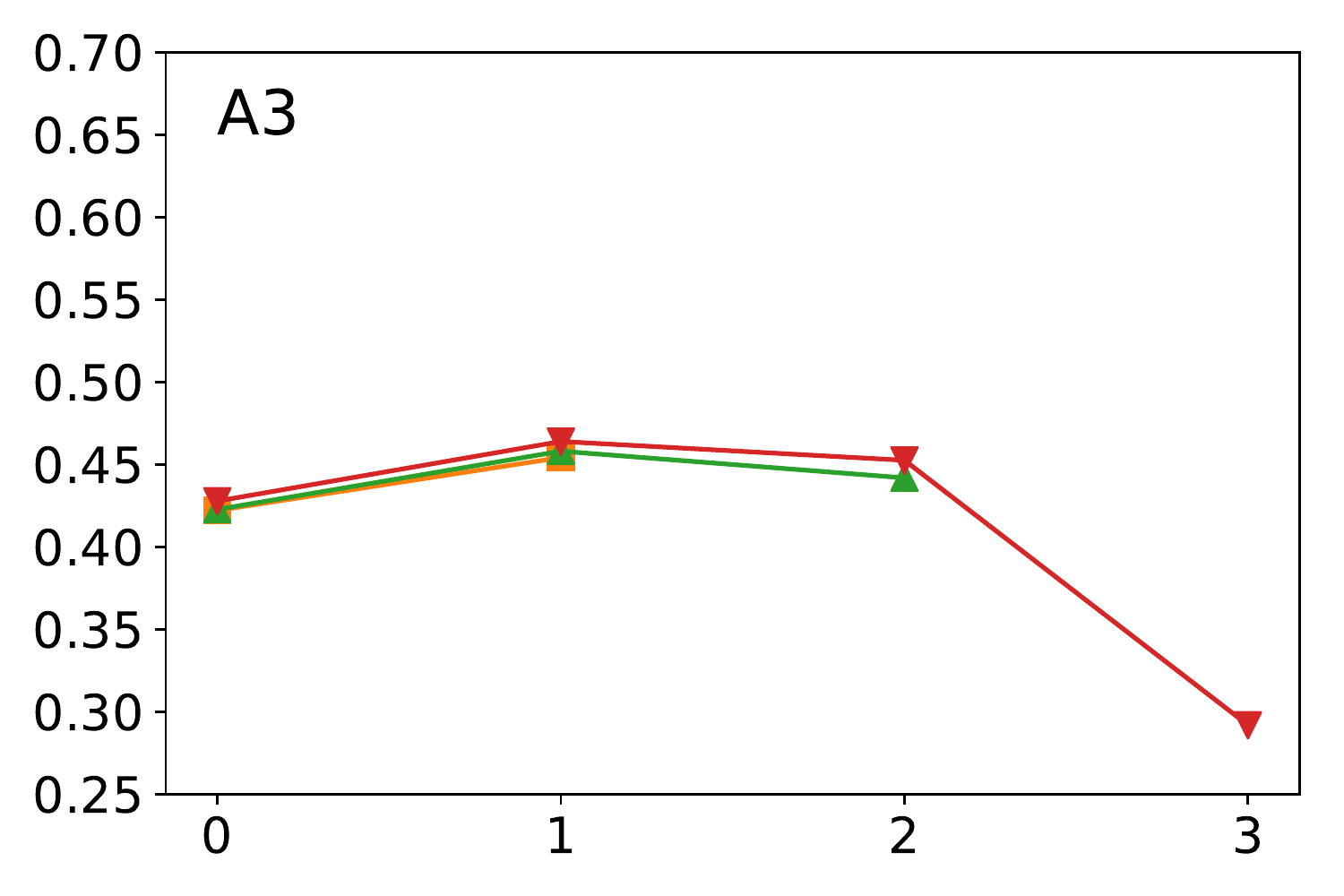}.pdf,height=2.8cm,width=.25\textwidth, clip=true, trim = 0.0cm 0.0cm 0.0cm 0.0cm} 
\psfig{figure=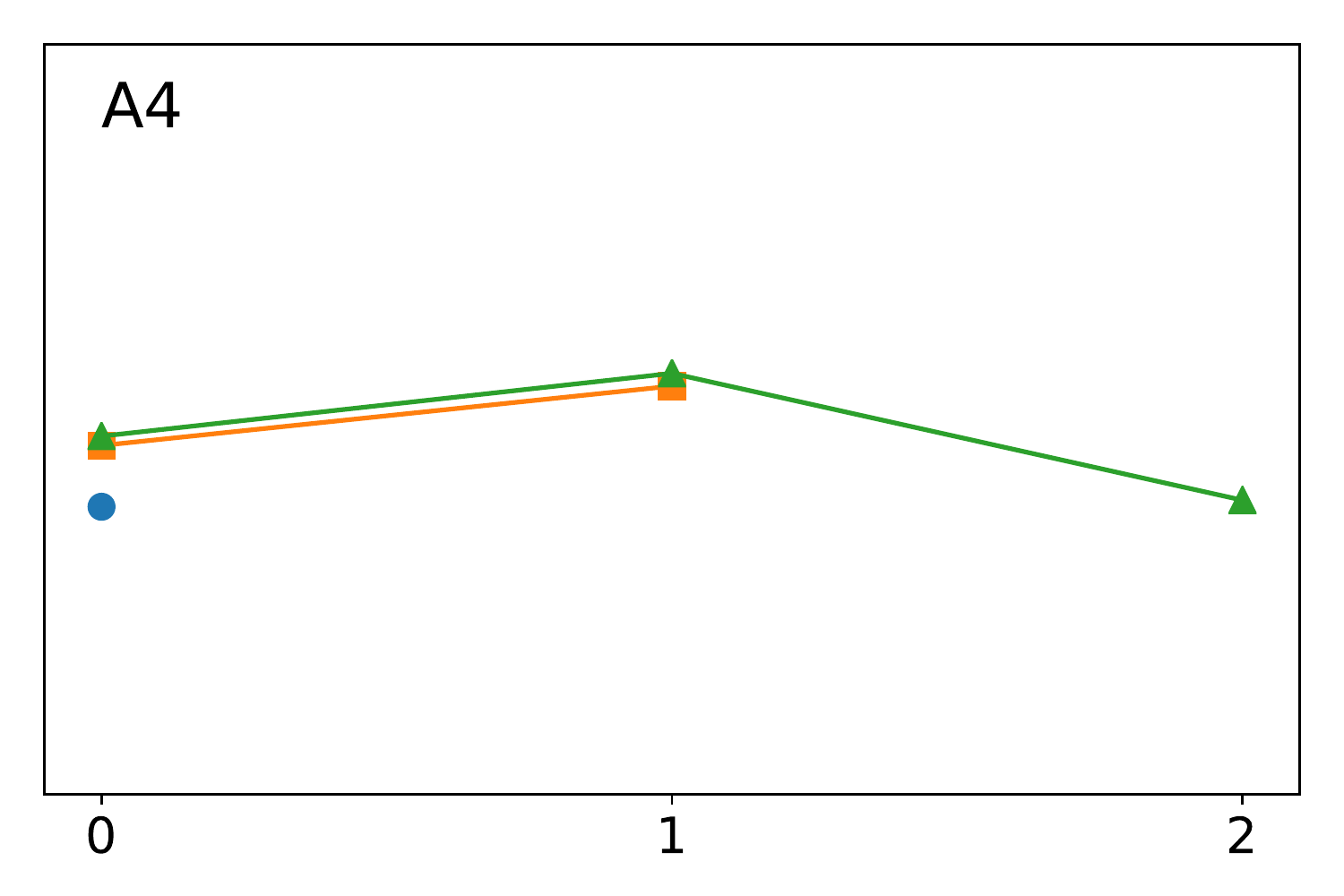}.pdf,height=2.8cm,width=.22\textwidth, clip=true, trim = 0.0cm 0.0cm 0.0cm 0.0cm}  \\
\psfig{figure=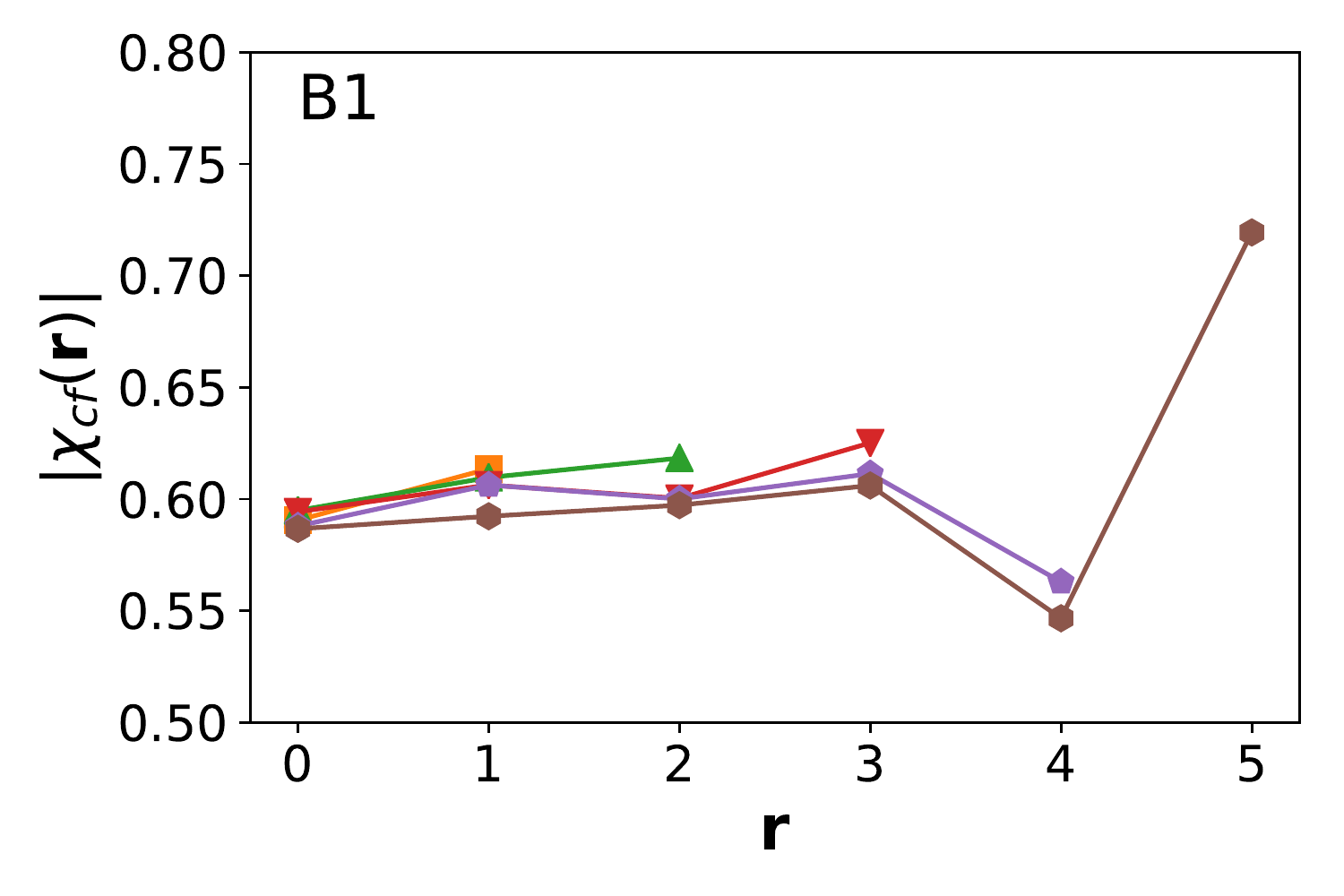}.pdf,height=3.0cm,width=.25\textwidth, clip=true, trim = 0.0cm 0.0cm 0.0cm 0.0cm} 
\psfig{figure=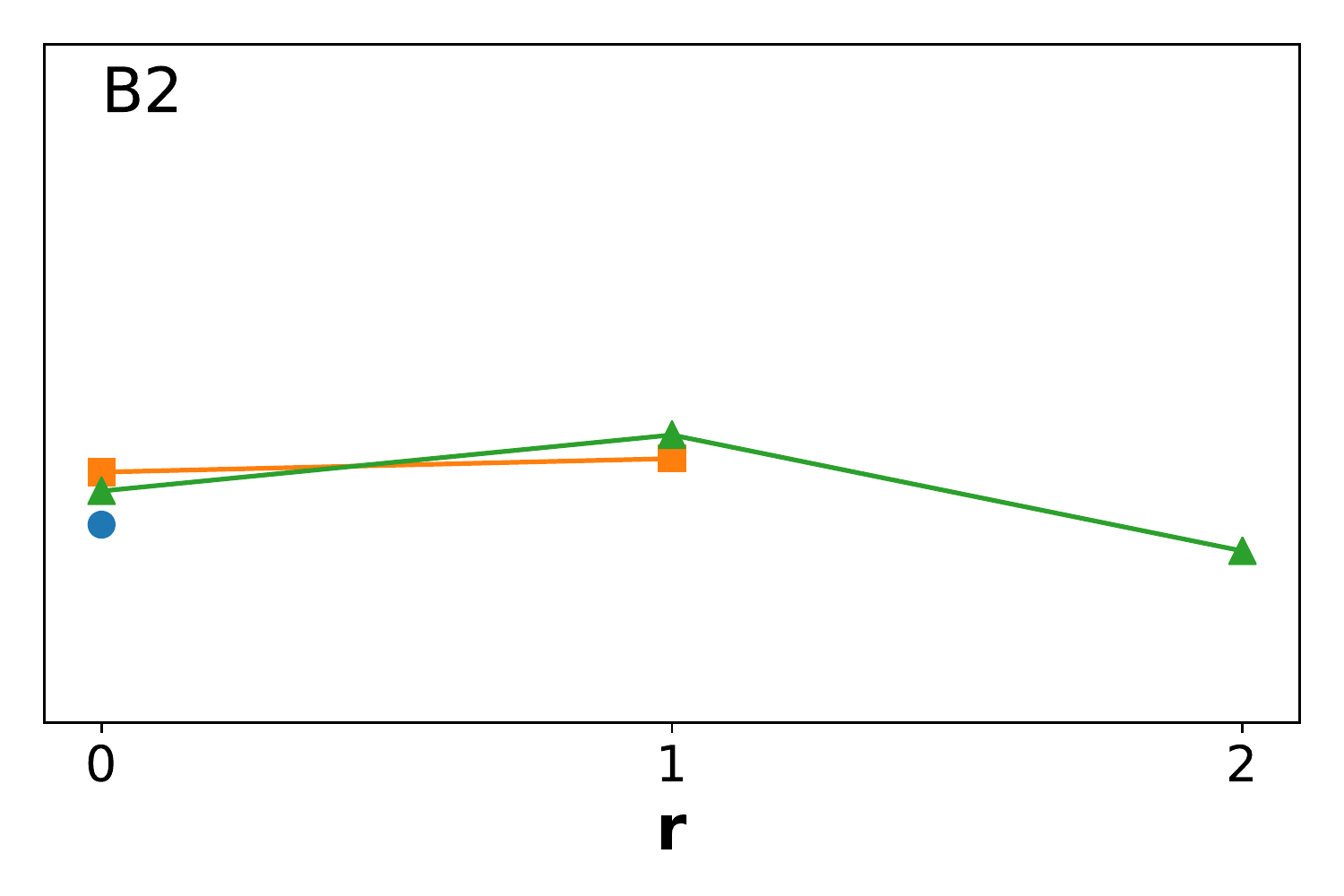}.pdf,height=3.0cm,width=.22\textwidth, clip=true, trim = 0.0cm 0.0cm 0.0cm 0.0cm} 
\psfig{figure=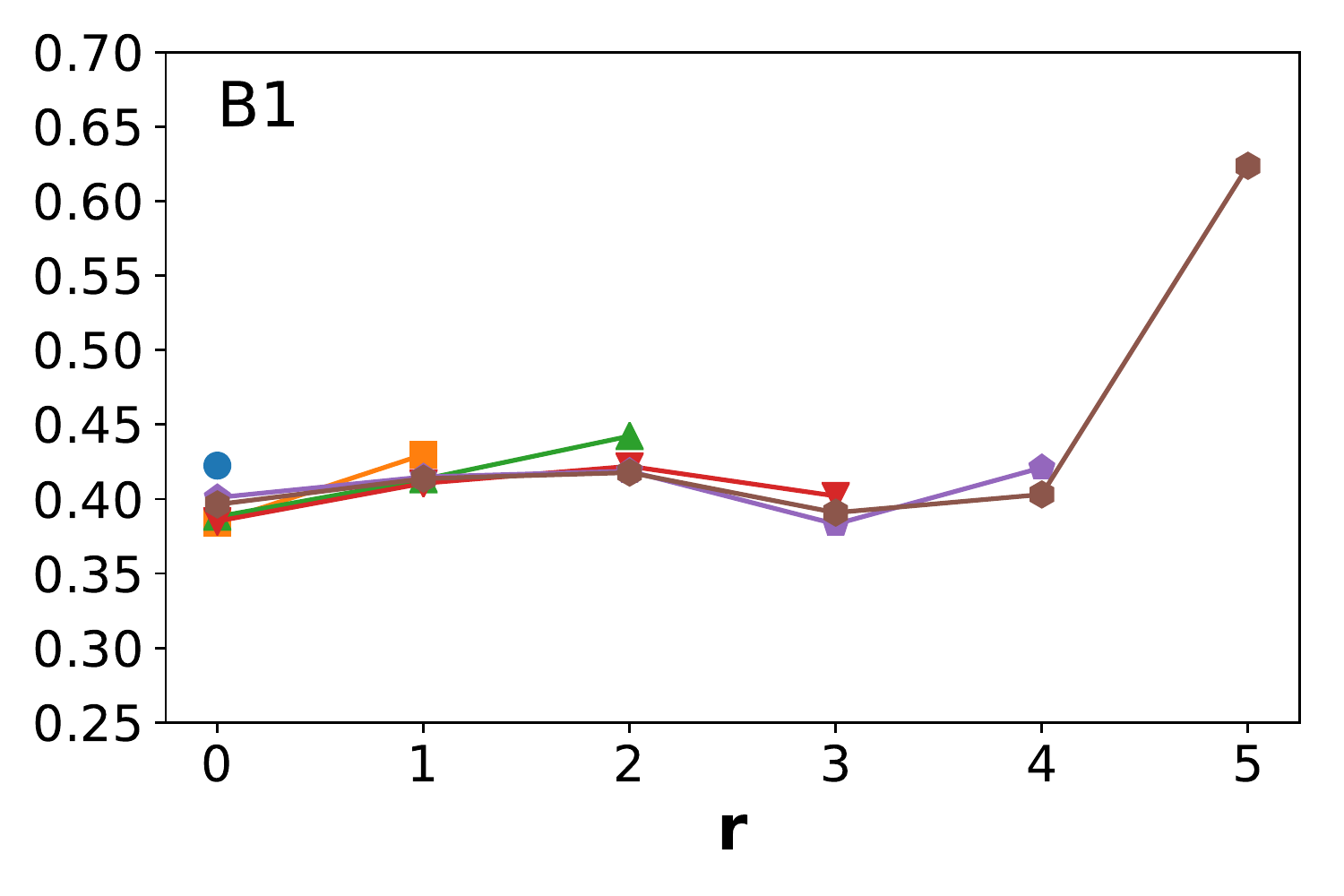}.pdf,height=3.0cm,width=.25\textwidth, clip=true, trim = 0.0cm 0.0cm 0.0cm 0.0cm} 
\psfig{figure=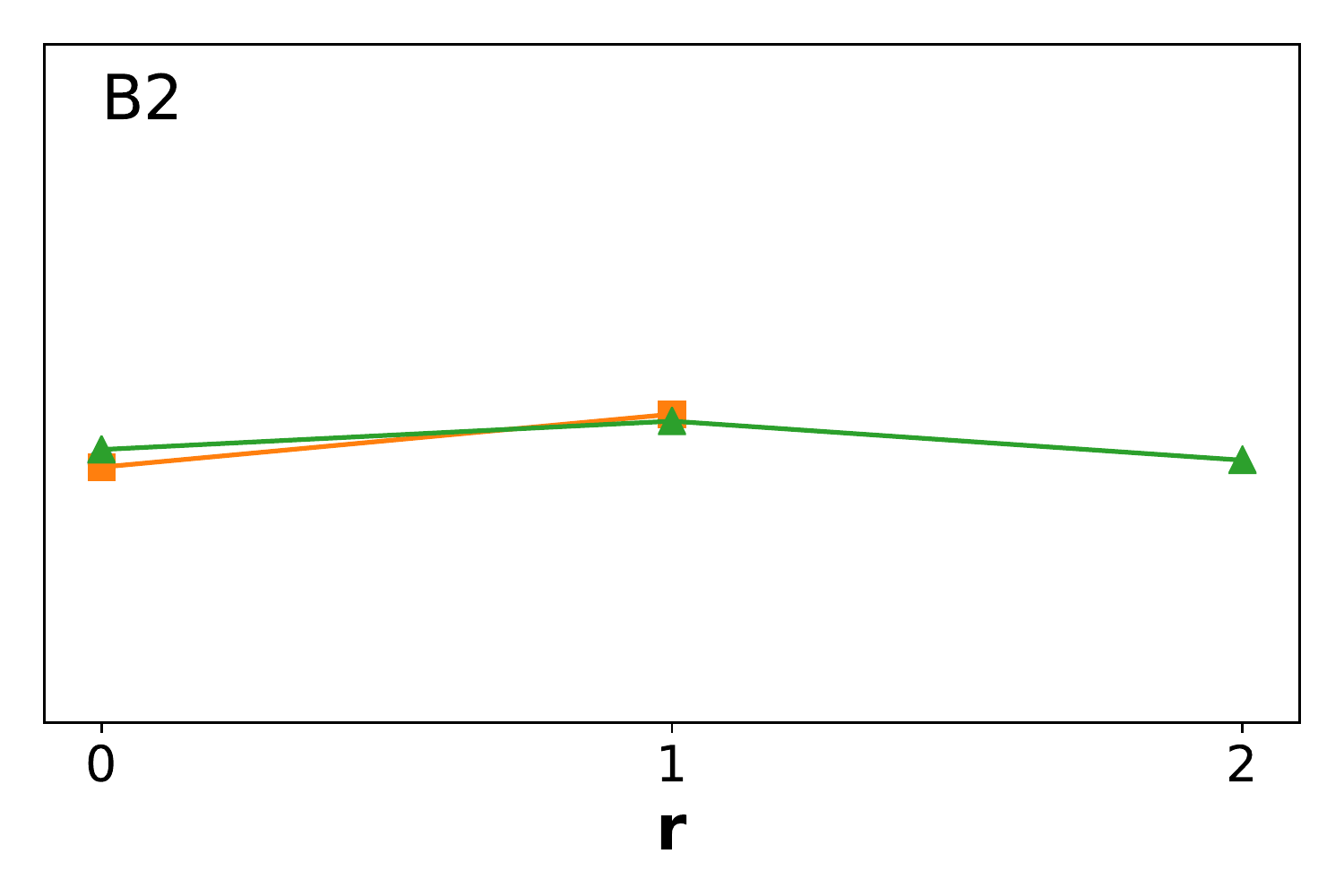}.pdf,height=3.0cm,width=.22\textwidth, clip=true, trim = 0.0cm 0.0cm 0.0cm 0.0cm} 
\caption{Local inter-orbital susceptibility $|\chi_{cf}(\mathbf{r})|$ for various systems at $T=t/10$. The general weaker dependence on $N_r$ and its anticorrelation to $\chi_{ff}(\mathbf{r})$ shown in Fig.~2 are visible.}
\label{chicf_B10}
\end{figure*}

It is not straightforward to compare our results to the lastest exploration of Co adatoms on Cu(111) surface~\cite{Morr2019} due to the intrinsic difficulty of extracting the local hybridization strength via the analytical continuation of local interorbital Green's function. 
Instead, we illustrate the interorbital magnetic susceptibility $|\chi_{cf}(\mathbf{r})|$ that is found to largely anticorrelate with $\chi_{ff}(\mathbf{r})$, which naturally reflects the local competition between the Kondo screening and inter-impurity spin correlation. Figure~\ref{chicf_B10} displays the evolution of $|\chi_{cf}(\mathbf{r})|$ versus the location $\mathbf{r}$ for various droplets. Clearly, when $|\chi_{cf}(\mathbf{r})|$ increases (decreases) with $\mathbf{r}$ or $N_r$, the opposite trend occurs for $\chi_{ff}(\mathbf{r})$. The dominant feature is the generally weaker dependence of $|\chi_{cf}(\mathbf{r})|$ on $N_r$ compared with $\chi_{ff}(\mathbf{r})$, which reflects the locality of $c-f$ hybridization in contrary to the spatial character of inter-impurity spin correlation manifested in $\chi_{ff}(\mathbf{r})$.

\section{Parameter dependence of magnetic properties}\label{robustness}
To provide more evidence on the robustness of our main results illustrated in Sec. III, in the following we discuss the effects of various parameters in more general settings. In addition, we will elaborate more upon the close relation between the local charge density and magnetic properties.

\subsection{Finite Size Effects}
\begin{figure}
\psfig{figure=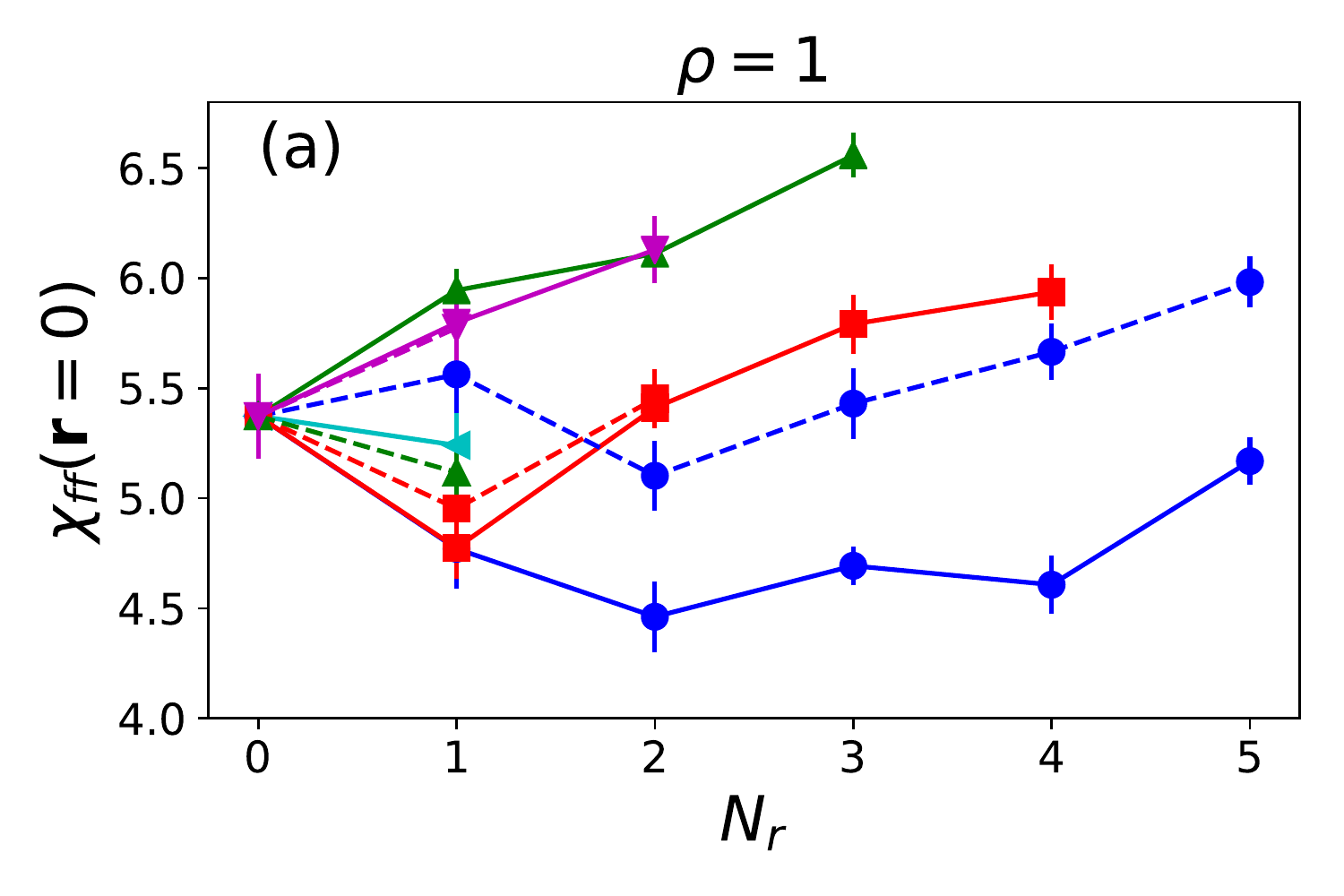}.pdf, height=3.0cm,width=.23\textwidth, clip} 
\psfig{figure=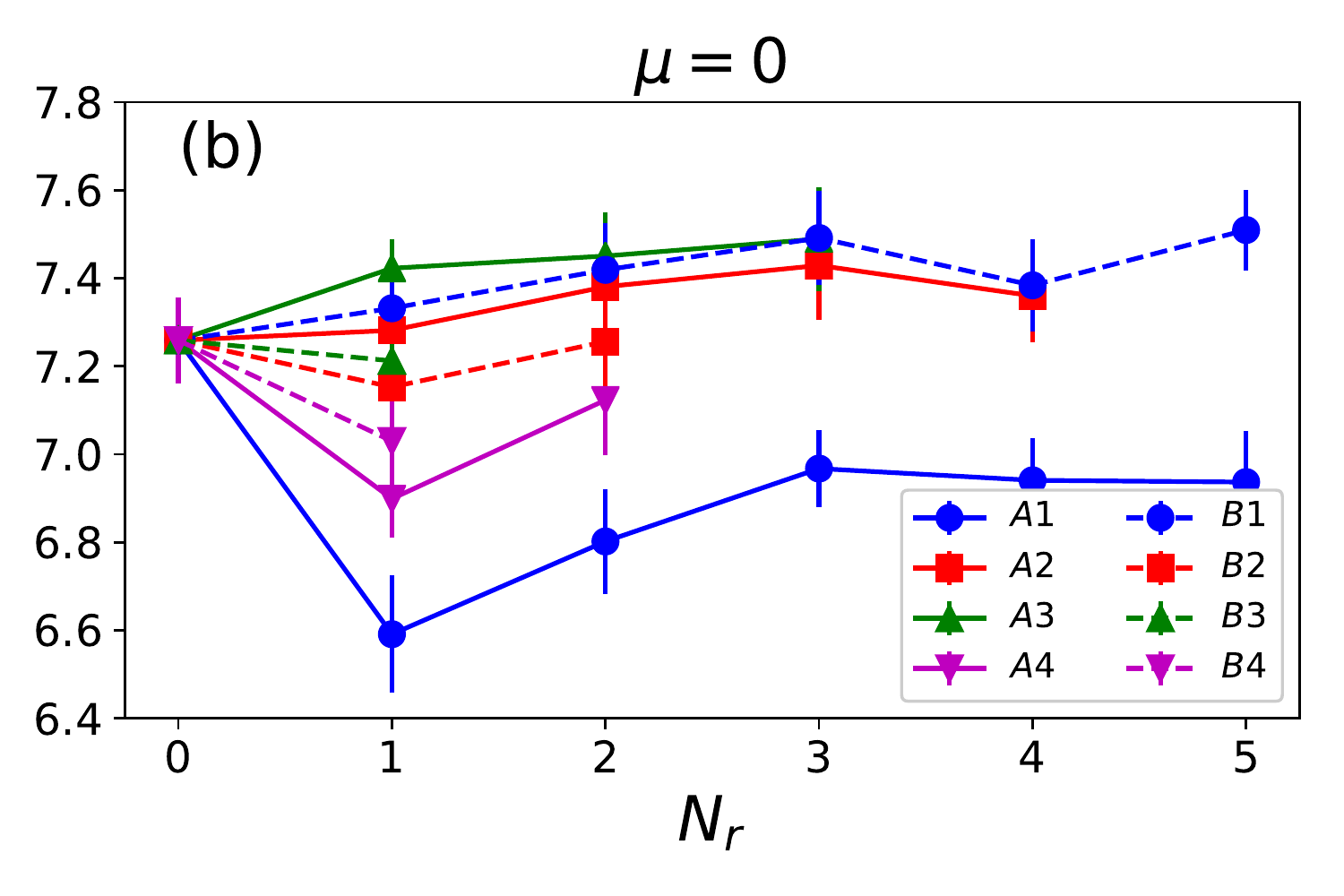}.pdf, height=3.0cm,width=.23\textwidth, clip} \\
\psfig{figure=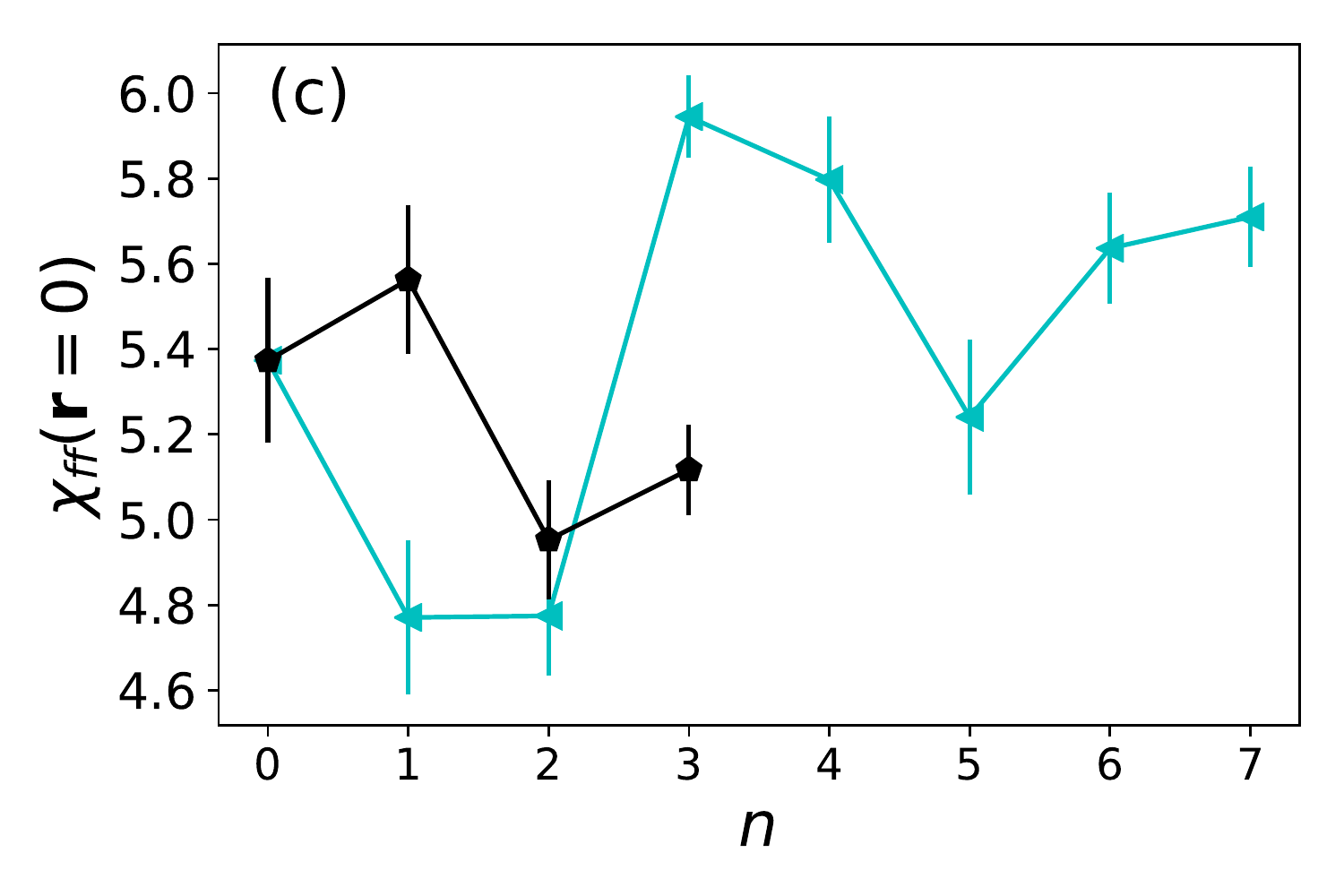}.pdf, height=3.0cm,width=.23\textwidth, clip} 
\psfig{figure=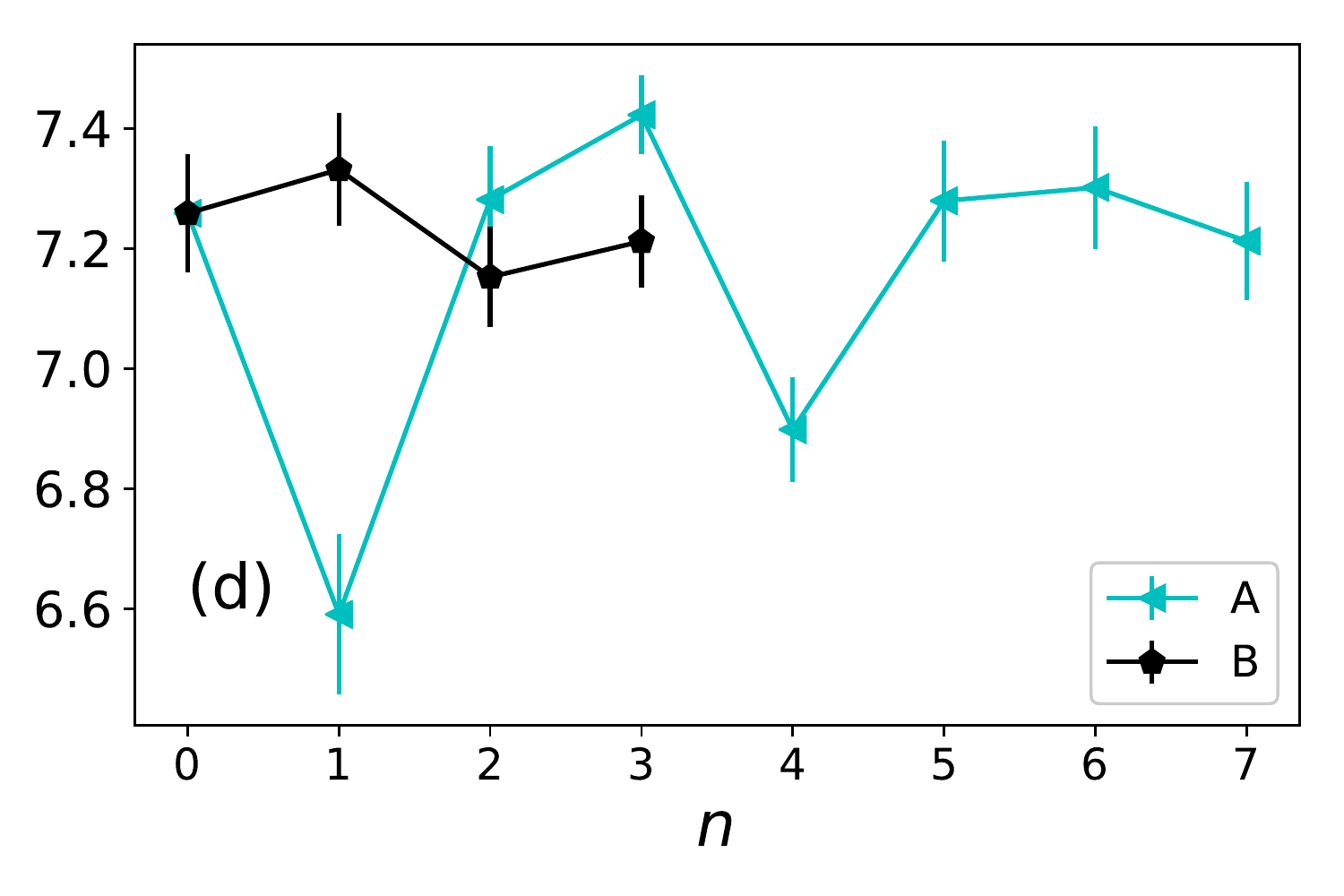}.pdf, height=3.0cm,width=.23\textwidth, clip}
\caption{Comparison between $\chi_{ff}(\mathbf{r}=0)$ at the central impurity (a-b) as a function of the number of droplet rings $N_r$ for various droplets and (c-d) as a function of ring distance $n$ for droplets with a single droplet ring $N_r=1$ in $L=12$ lattices at $T=t/10$.}
\label{center_chiff_L12}
\end{figure}

Finite Size effects are ubiquitous in lattice quantum Monte Carlo calculations. Typically these are most serious when a question concerning long range order is considered since, by definition, one is examining the asymptotic behavior at the large spatial separation. Our adoption of open boundary condition for the lattice without the translational symmetry implies the potentially significant impact of the lattice size. Because the manifestation of the essential features, especially the enhancement of $\chi_{ff}(\mathbf{r}=0)$ upon $N_r$ requires a large enough lattice e.g. $L\geq 8$ while the computational cost limits us to simulate a much larger lattice, the finite size scaling is by and large meaningless in this context. 

Therefore, here we only provide evidence that there are no qualitative changes for a larger lattice system with $L=12$ compared with $L=10$ adopted in Sec. III. Since, after all, the relevant physics is more or less local, e.g. the local repulsion for $f$-conduction electron is strong coupling $U/t=4.0$, we do not expect any significant modification of the local magnetic properties.
Figure~\ref{center_chiff_L12} confirms this expectation via the local susceptibility $\chi_{ff}(\mathbf{r}=0)$ at the central impurity for $L=12$ lattices at $T=t/10$.

Fig.~\ref{center_chiff_L12}(a-b) compare the behavior of $\chi_{ff}(\mathbf{r}=0)$ as a function of the number of impurity rings $N_r$ between two types of systems. All the essential features presented in Fig.~3 maintain in this larger lattice. In particular, the unusual feature of enhanced $\chi_{ff}(\mathbf{r}=0)$ with $N_r$ for some droplets persists. Furthermore, Fig.~\ref{center_chiff_L12}(c-d) displays $\chi_{ff}(\mathbf{r}=0)$ for droplets with a single impurity ring $N_r=1$ as a function of its distance $n$ to the central impurity. The oscillating behavior is reminiscent of that presented in Fig.~4.
The robustness against the lattice size confirms our major conclusion of the tunability of the magnetic properties by the artificial arrangement of the droplet geometry.

\subsection{Effects of hybridization strength $V$}
\begin{figure}
\psfig{figure=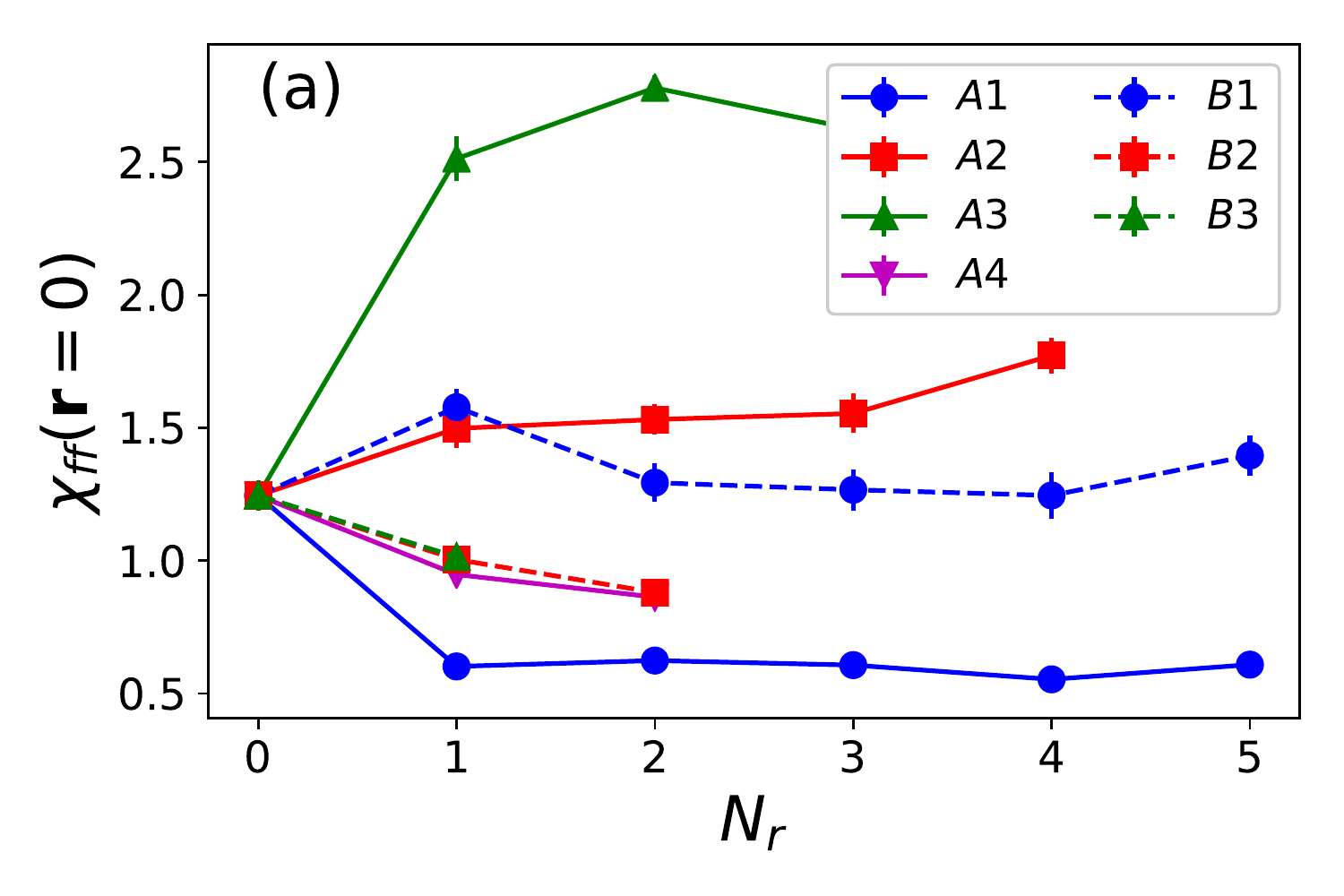}.pdf, height=3.0cm,width=.23\textwidth, clip}
\psfig{figure=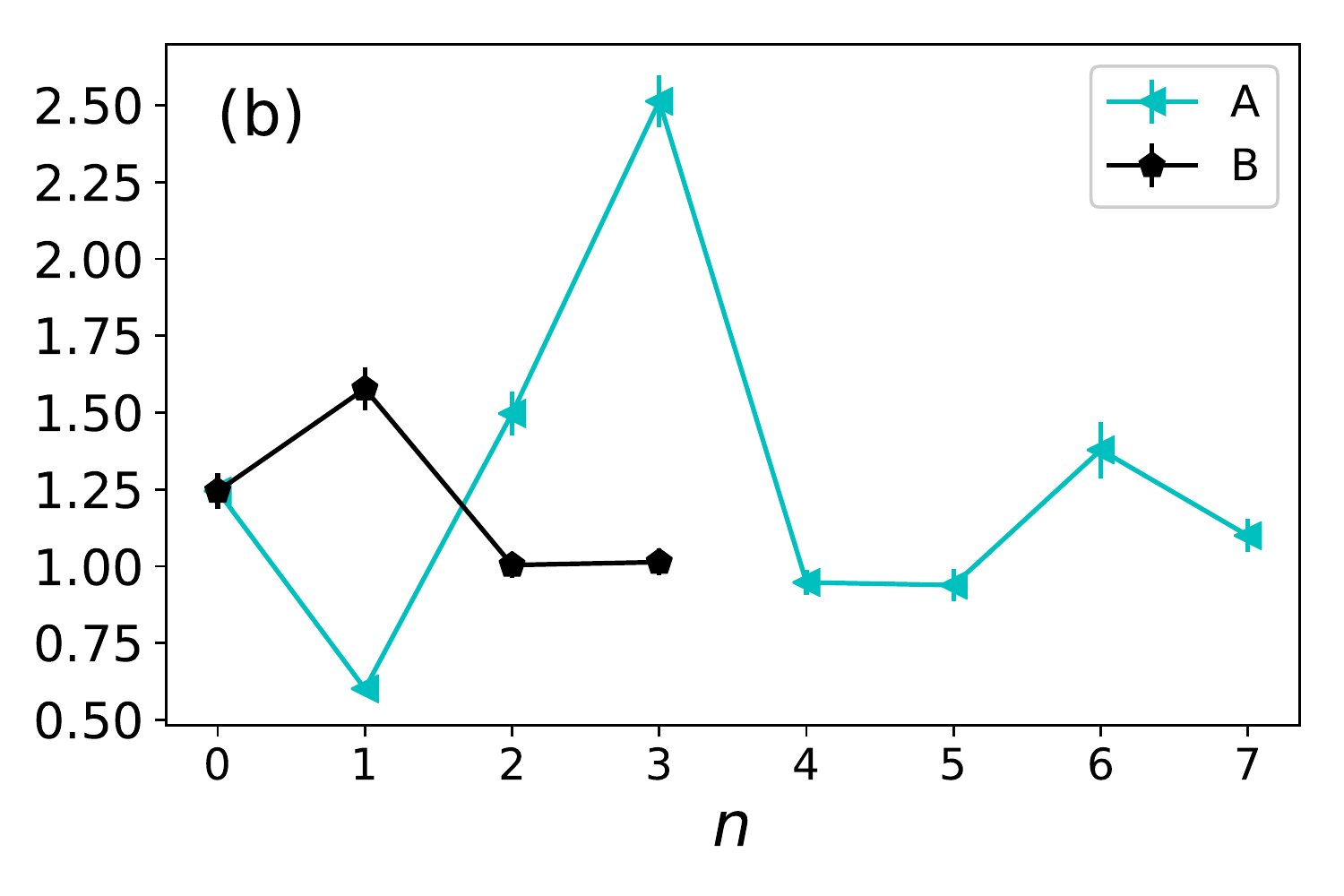}.pdf, height=3.0cm,width=.23\textwidth, clip}
\caption{$\chi_{ff}(\mathbf{r}=0)$ at the central impurity (a) as a function of $N_r$ for various droplets and (b) as a function of $n$ for droplets with a single droplet ring $N_r=1$ for $c-f$ hybridization strength $V/t=2.0$ at $T=t/10$.}
\label{center_chiff_V2}
\end{figure}

In Sec. III, we fixed the $c-f$ hybridization strength as the characteristic $V/t=1.0$. Because the Anderson lattice models describes the competition between inter-impurity antiferromagnetic spin correlation mediated via RKKY interaction and the Kondo screening from conduction electrons, it is natural to ask for the impact of the $c-f$ hybridization strength. Similar to Fig~\ref{center_chiff_L12}, Figure~\ref{center_chiff_V2} demonstrates the the behavior of $\chi_{ff}(\mathbf{r}=0)$ as a function (a) of $N_r$ for various droplets and (b) of $n$ for droplets with a single impurity ring $N_r=1$ with varied distance $n$ to the central impurity. Obviously, there is no qualitative modification of the tunable features of $\chi_{ff}(\mathbf{r}=0)$ compared to those for smaller hybridization (Figs.~3-4) and for larger lattices (Fig.~\ref{center_chiff_L12}). 

\subsection{Effects of chemical potential $\mu$}
\begin{figure}
\psfig{figure=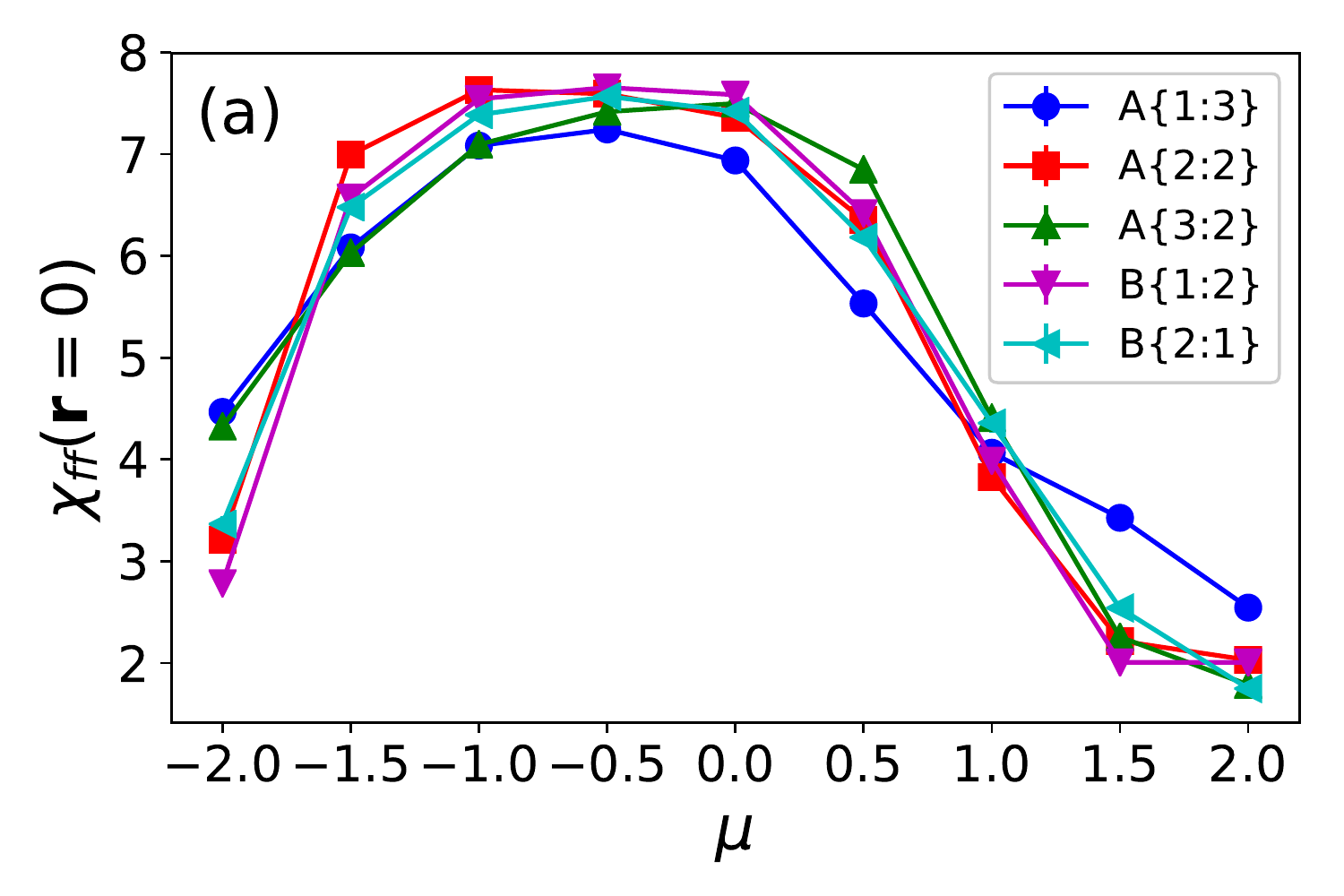}.pdf, height=3.05cm,width=.22\textwidth, clip}
\psfig{figure=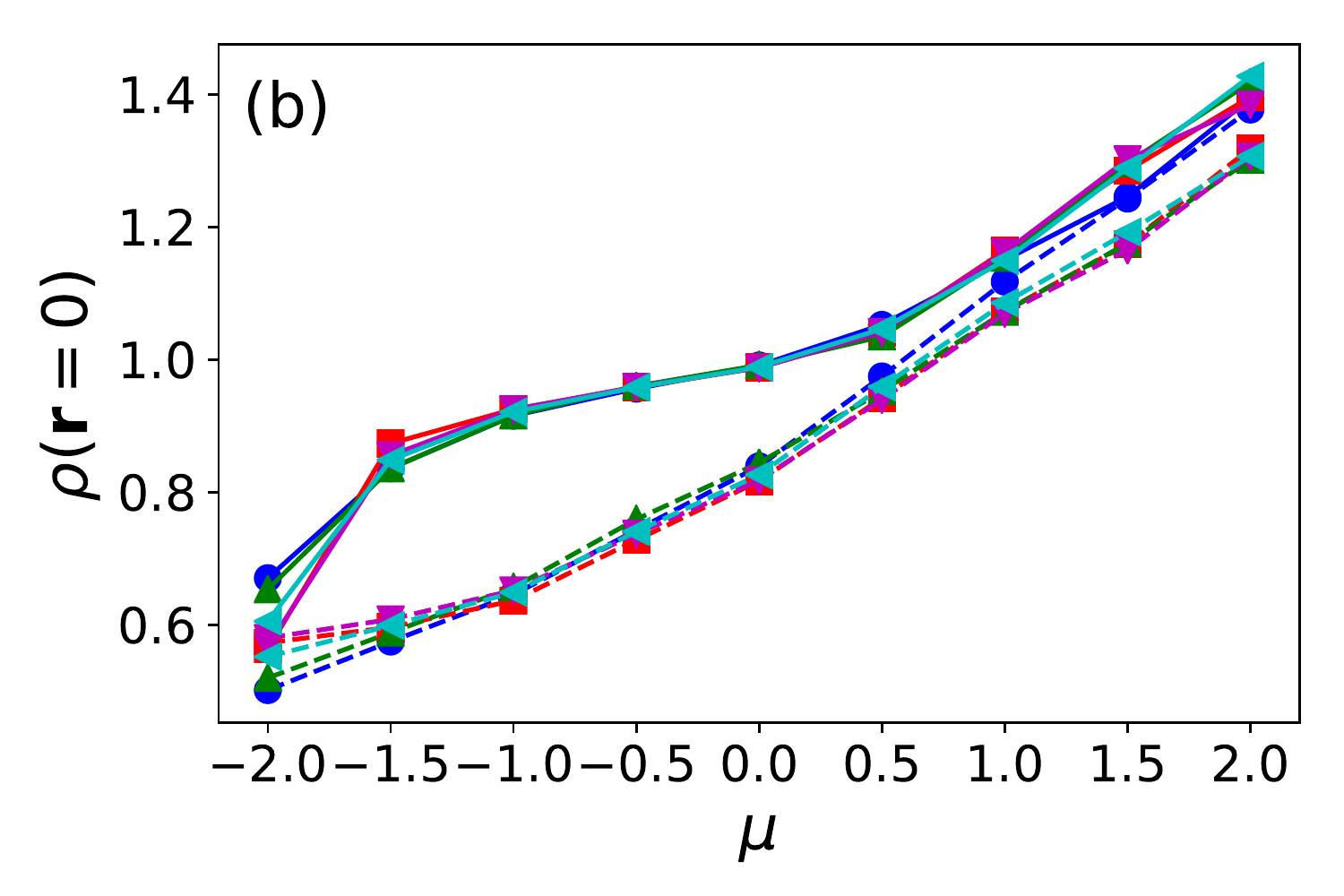}.pdf, height=3.02cm,width=.24\textwidth, clip}
\caption{(a) $\chi_{ff}(\mathbf{r}=0)$ and (b) the local orbital occupancy $\rho(\mathbf{r}=0)$ at the central impurity (solid lines for droplet $f$-orbital and dashed lines for conduction electrons) vs chemical potential $\mu$ for $V/t=1.0$ in $L=10$ lattices at $T=t/10$.}
\label{varymu}
\end{figure}

In Sec. III, we have focussed on two special cases ($\rho=1$ and $\mu=0$) of orbital occupancies by tuning the chemical potential $\mu$. Because the major difference of the Anderson-type models apart from the Kondo-type models is the involvement of the additional charge fluctuations in the determination of their physical properties, the role played by the chemical potential, namely more general cases of orbital occupancies, is worth elaborating upon. 

Figure~\ref{varymu}(a) illustrates the evolution of $\chi_{ff}(\mathbf{r}=0)$ with $\mu$ for some characteristic droplets. Obviously, the dominant feature is the commonly broad peak at $\mu \sim -0.5$, which can be naturally accounted for by the local orbital occupancy $\rho(\mathbf{r}=0)$ (solid lines for droplet $f$-orbital and dashed lines for conduction electrons) shown in Fig.~\ref{varymu}(b). Specifically, at $\mu \sim -0.5$, both the features of (i) almost half-filled occupancy of the droplet $f$-orbital (solid lines) and (ii) their plateau-like evolutions with $\mu$  imply for the strongest magnetic local moment to induce the peak of local magnetic susceptibility. This further indicates the close relation between the local magnetic properties and charge density as discussed in Sec. III.

\subsection{Effects of Coulomb repulsion $U$}
\begin{figure}
\psfig{figure=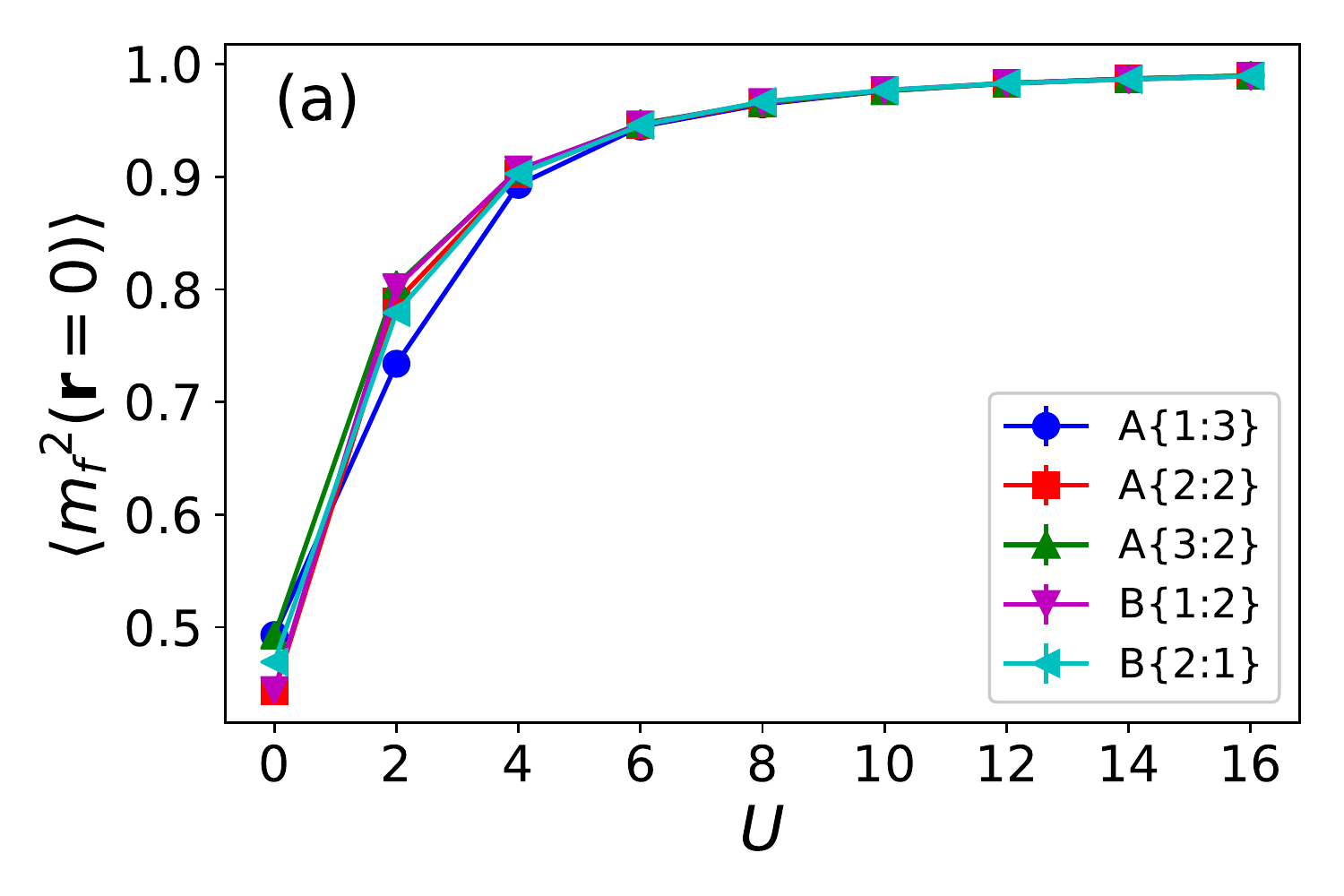}.pdf, height=3.0cm,width=.22\textwidth, clip}
\psfig{figure=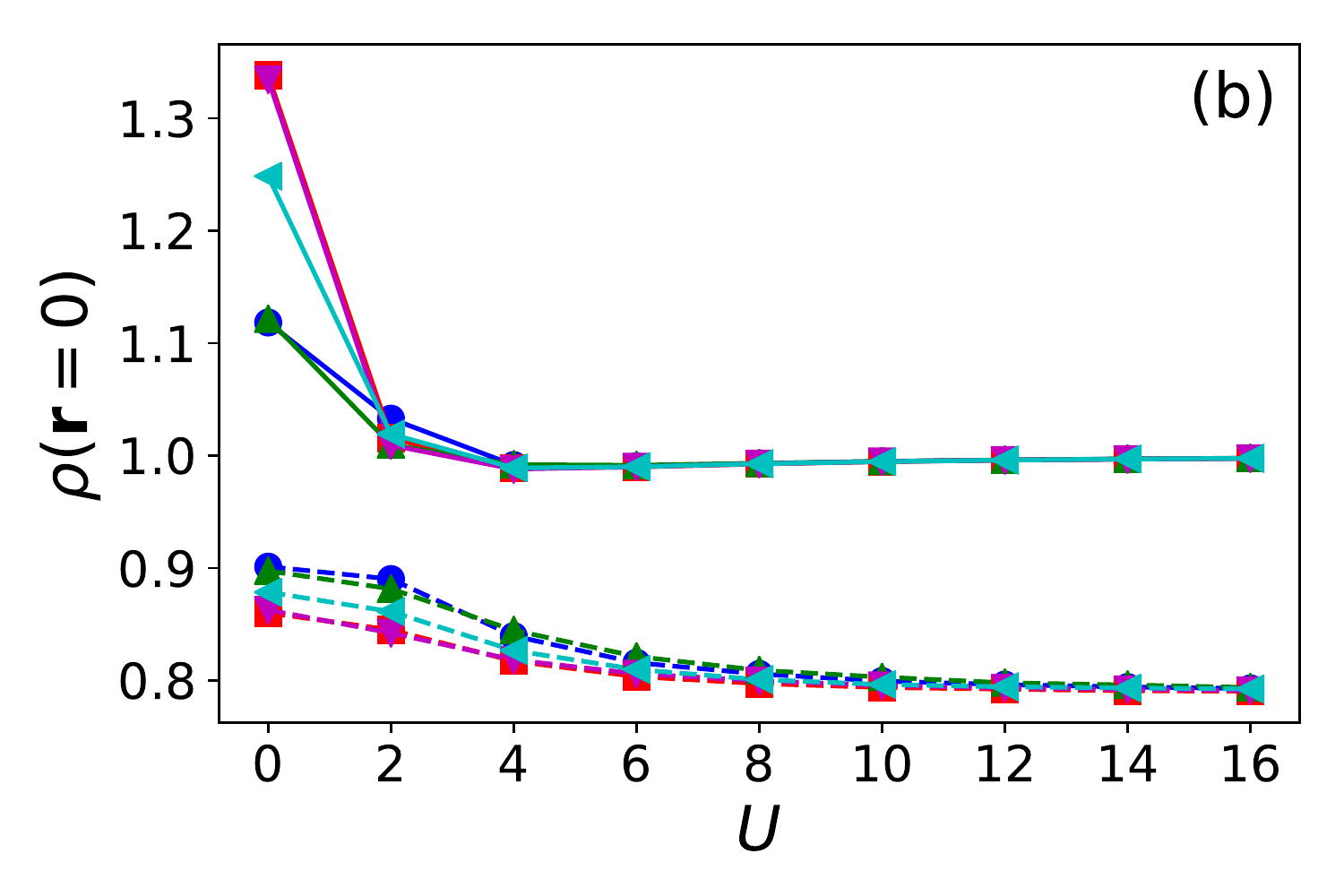}.pdf, height=3.0cm,width=.24\textwidth, clip}
\caption{(a) The local magnetic moment $\langle m^{2}{_f}(\mathbf{r}=0) \rangle$ of droplet center and (b) the local orbital occupancy $\rho(\mathbf{r}=0)$ at the central impurity (solid lines for droplet $f$-orbital and dashed lines for conduction electrons) vs $U$ for $V/t=1.0$ in $L=10$ lattices of $\mu=0$ systems at $T=t/10$.}
\label{varyU}
\end{figure}

Our current work concentrates on the Anderson droplet model with the additional involvement of the charge degrees of freedom instead of the effective Kondo droplet model in the strong coupling limit $U/t \rightarrow \infty$, it is worthwhile to discuss more about the role played by the Coulomb repulsion $U$.

Figure~\ref{varyU}(a) displays the local magnetic moment of the droplet center $\langle m^{2}{_f}(\mathbf{r}=0) \rangle$ versus $U$ for some characteristic droplets and Fig.~\ref{varyU}(b) shows the local orbital occupancy $\rho(\mathbf{r}=0)$ (solid lines for droplet $f$-orbital and dashed lines for conduction electrons). Here we adopt $\mu=0$ to facilitate the half-filled occupancy of the droplet center. Clearly, strong Coulomb repulsion saturates the local charge occupancy. As a result, the forbidden charge fluctuations so that double occupancy at the half-filled droplet center leads to the enhanced and saturated local magnetic moment.  

\section{Conclusion}\label{Conclusion}
In conclusion, we have employed the numerically exact DQMC simulations in the framework of Anderson droplet model to investigate the local magnetic properties associated with the hexagonal droplet embedded in a triangular lattice.
We demenstrated the tunability of the magnetic properties by the evolution of local susceptibility with the droplet geometry for two types of systems with differing local orbital occupancy profile.
Our ADM has the intrinsic intertwined spin and charge degrees of freedom, whose fluctuations largely affects the local magnetic properties, although the charge fluctuation of the droplets can be partially suppressed by enforcing its nearly half-filled occupancy. 

The coincidence between the magnetic properties and the local charge occupancy is manifested by the unusual enhancement of $\chi_{ff}(\mathbf{r}=0)$ with $N_r$ (Fig.~\ref{center_chiff}) presented in Sec. III.A, which is strongly tied to the density redistribution via varying the droplet geometry. This density fluctuation is in turn closely related to the non-periodicity of the lattice due to both the droplet geometry and the open boundary employed. In essence, the possibility of artificial manipulation of the magnetic properties of the Anderson droplet in our current confined lattice system is realized via the potentially controllable density variation enforced by a finite boundary.

It is not straightforward to compare our results to the lastest exploration of Co adatoms on Cu(111) surface~\cite{Morr2019} due to (I) the intrinsic difficulty of extracting the local hybridization strength via the analytical continuation of local interorbital Green's function; (II) the strong charge effects in our ADM; and (III) the finite size effect in contrast to the essentially infinitely large Cu(111) surface~\cite{Morr2019}. Further design of the lattice and/or droplet settings are requisite to perform insightful direct comparison. 

The natural extension of the present work is to investigate the Kondo droplet model~\cite{Assaad2019} with the absence of the droplets' $f$-electron charge fluctuation, and/or with more appropriate lattice band fillings. Besides, the Anderson version of Kondo holes in a droplet reported in the latest investigation deserves further exploration~\cite{Morr2019,kondohole1,kondohole2}. Despite that the direct comparison with the recent progress on quantum engineered Kondo lattice is not straightforward, our simulations complements the exploration of the novel artificial tunability of engineered confined lattice systems.

\section{Acknowledgements}\nonumber
This work was funded by Stewart Blusson Quantum Matter Institute (SBQMI) and Natural
Sciences and Engineering Research Council (NSERC) of Canada. Computational resources  were provided by SBQMI and Compute Canada.


\begin{thebibliography}{40}
\bibitem{Morr2019}
J. Figgins, L. S. Mattos, W. Mar, Y. Chen, H. C. Manoharan, and D. K. Morr, arXiv:1902.04680 (2019).

\bibitem{Morrpaper1}
H. C. Manoharan, C. P. Lutz, and D. M. Eigler, Nature 403, 512 (2000).

\bibitem{Morrpaper2}
C. R. Moon, C. P. Lutz, and H. C. Manoharan, Nat. Phys. 4, 454 (2008).

\bibitem{Morrpaper3}
C. F. Hirjibehedin, C. P. Lutz, and A. J. Heinrich, Science 312, 1021 (2006).

\bibitem{Morrpaper4}
N. Tsukahara, S. Shiraki, S. Itou, N. Ohta, N. Takagi, and M. Kawai, Phys. Rev. Lett. 106, 187201 (2011).

\bibitem{Morrpaper5}
K. K. Gomes, W. Mar, W. Ko, F. Guinea, and H. C. Manoharan, Nature 483, 306 (2012).

\bibitem{Morrpaper6}
M. R. Slot et al., Nat. Phys. 13, 672 (2017).

\bibitem{Morrpaper7}
H. Kim et al., Sci. Adv. 4, eaar5251 (2018).

\bibitem{Steglich}
S. Wirth and F. Steglich, Nat. Rev. Mat. 1, UNSP 16051 (2016).

\bibitem{Assaad2019}
M. Raczkowski and F. F. Assaad, Phys. Rev. Lett. 122, 097203 (2019).

\bibitem{SchriefferWolff}
J. R. Schrieffer and P. A. Wolff, Phys. Rev. 149, 491 (1966).

\bibitem{mapping}
P. Sinjukow and W. Nolting, Phys. Rev. B 65, 212303 (2002).

\bibitem{Kotliar2005}
P. Sun and G. Kotliar, Phys. Rev. Lett. 95, 016402 (2005).

\bibitem{Kotliar2008}
L.D. Leo, M. Civelli, and G. Kotliar, Phys. Rev. B 77, 075107 (2008).

\bibitem{MJ}
M. Jiang, N. J. Curro, and R. T. Scalettar, Phys. Rev. B 90, 241109(R) (2014);
M. Jiang and Yi-feng Yang, Phys. Rev. B 95, 235160 (2017).

\bibitem{TremblayPRX}
W. Wu and A.-M.-S. Tremblay, Phys. Rev. X 5, 011019 (2015).

\bibitem{NFL2008}
A. Amaricci, G. Sordi, and M. J. Rozenberg, Phys. Rev. Lett. 101, 146403 (2008).

\bibitem{Mott2016}
D.E. Logan, M.R. Galpin and J. Mannouch, J. Phys.: Condens. Matter 28, 455601 (2016).

\bibitem{Capone2016}
A. Amaricci, L. de' Medici, and M. Capone, Europhys. Lett., 118, 17004 (2017).

\bibitem{depletedPAM}
N. C. Costa, M. V. Arajo, J. P. Lima, T. Paiva, R. R. dos Santos, and R. T. Scalettar, Phys. Rev. B 97, 085123 (2018).

\bibitem{impPAM1}
A. Benali, Z. J. Bai, N. J. Curro, and R. T. Scalettar, Phys. Rev. B 94, 085132 (2016).

\bibitem{impPAM2}
T. Mendes-Santos, N. C. Costa, G. Batrouni, N. Curro, R. R. dos Santos, T. Paiva, and R. T. Scalettar, Phys. Rev. B 95, 054419 (2017).

\bibitem{triangularAssaad}
M.W. Aulbach, F.F. Assaad, and M. Potthoff, Phys. Rev. B 92, 235131 (2015).

\bibitem{triangularMFT}
S. Hayami, M. Udagawa, and Y. Motome, J. Phys. Soc. Jpn. 81, 103707 (2012).

\bibitem{DQMC}
R. Blankenbecler, D. J. Scalapino, and R. L. Sugar, Phys. Rev. D 24, 2278 (1981).

\bibitem{comment}
The conditions for generating the coherent Kondo lattice behavior have been investigated in Ref.~\cite{Assaad2019} while that is currently not our focus in this work.

\bibitem{kondohole1}
J. Figgins and D. K. Morr, Phys. Rev. Lett. 107, 066401 (2011).

\bibitem{kondohole2}
M. H. Hamidian et al., Proc. Natl. Acad. Sci., 108, 18233 (2011).

\end{thebibliography}
\end{document}